\documentclass[preprint]{aastex631}

\usepackage{graphicx}	
\usepackage{amsmath}	
\usepackage{amssymb}	
\usepackage{booktabs}
\usepackage{enumitem}
\usepackage[caption=false]{subfig}
\usepackage{hyperref}
\usepackage{tabularx}

\usepackage{threeparttable}
\usepackage{threeparttablex}
\usepackage{longtable}
\usepackage{rotating}
\usepackage{natbib}
\usepackage[titletoc,title]{appendix}
\usepackage{xr}
\hypersetup{linkcolor=red,citecolor=green,filecolor=cyan,urlcolor=magenta}

\newcommand{\psqm}{\,m$^{-2}$}

\newcommand{\ncol}{\,$N_{\text{H}_2}$}
\newcommand{\ncold}{\,$N_{\text{H}_2}^{\text{dust}}$}

\newcommand{\ico}{\,$I_{^{12}\text{CO}}$}
\newcommand{\td}{\,$T_{\text{d}}$}
\newcommand{\xco}{\,$X_{\text{CO}}$}
\newcommand{\nco}{\,$N_{\text{CO}}$}

\newcommand{\psec}{\,s$^{-1}$}
\newcommand{\kkms}{\,K~km~\psec}
\newcommand{\coab}{\,[$N_{\text{CO}}/N_{\text{H}_2}$]}

\newcommand{\tex}{$T_{\rm ex}$}

\defcitealias{champ1}{Paper I}
\defcitealias{champ2}{Paper II}
\defcitealias{champ3}{Paper III}
\defcitealias{champ4}{Paper IV}
\defcitealias{mmsignposts}{Paper I-B}
\defcitealias{schap17}{Paper III-B}
\defcitealias{pittsmn}{PBV19}

\graphicspath{{./}{figures/}}

\accepted{May 24, 2021}
\submitjournal{ApJS}

\shorttitle{CHaMP V: On the $X_{\text{CO}}$ Factor}
\shortauthors{Pitts \& Barnes 2021}

\begin{document}

\title{Census of High- and Medium-mass Protostars V.\\ CO Abundance and the Galactic $X_{\text{CO}}$ Factor}

\correspondingauthor{Rebecca Pitts}
\email{rebecca.pitts@nbi.ku.dk}

\author[0000-0002-7937-4931]{Rebecca L. Pitts}
\affiliation{Niels Bohr Institute, Centre for Star \& Planet Formation, 
University of Copenhagen, 
\O{}ster Voldgade 5-7, 
1350 Copenhagen K, Denmark}

\author{Peter J. Barnes}
\affiliation{Space Science Institute, 4765 Walnut St Suite B, Boulder, CO 80301, USA} \affiliation{School of Science and Technology, University of New England, Armidale, NSW 2351, Australia}

\begin{abstract}

We present the second dust continuum data release in the Census of High- and Medium-mass Protostars (CHaMP), expanding the methodology trialed in \citealt{pittsmn} to the entire CHaMP survey area ($280^{\circ}<\ell<300^{\circ}$, $-4^{\circ}<b<+2^{\circ}$). This release includes maps of dust temperature ($T_{\text{d}}$), H$_2$ column density ($N_{\text{H}_2}$), gas-phase CO abundance, and temperature-density plots for every prestellar clump with \emph{Herschel} coverage, showing no evidence of internal heating for most clumps in our sample. We show that CO abundance is a strong function of $T_{\text{d}}$, and can be fit with a second-order polynomial in log-space, with a typical dispersion of a factor of 2--3. The CO abundance peaks at $20.0^{+0.4}_{-1.0}$~K with a value of $7.4^{+0.2}_{-0.3}\times10^{-5}$ per H$_2$; the low $T_{\text{d}}$\ at which this maximal abundance occurs relative to laboratory results is likely due to interstellar UV bombardment in the largest survey fields. Finally, we show that, as predicted by theoretical literature and hinted at in previous studies of individual clouds, the conversion factor from integrated $^{12}$CO line intensity ($I_{^{12}\text{CO}}$) to $N_{\text{H}_2}$, the $X_{\text{CO}}$-factor, varies as a broken power-law in $I_{^{12}\text{CO}}$\ with a transition zone between 70 and 90~\kkms. The $X_{\text{CO}}$-function we propose has $N_{\text{H}_2}\propto I_{^{12}\text{CO}}^{0.51}$ for $I_{^{12}\text{CO}}\lesssim70$~\kkms\ and $N_{\text{H}_2}\propto I_{^{12}\text{CO}}^{2.3}$ for $I_{^{12}\text{CO}}\gtrsim90$~\kkms. The high-$I_{^{12}\text{CO}}$\ side should be generalizable with known adjustments for metallicity, but the influence of interstellar UV fields on the low-$I_{^{12}\text{CO}}$\ side may be sample specific. We discuss how these results expand upon previous works in the CHaMP series, and help tie together observational, theoretical, and laboratory studies on CO over the past decade. 
\end{abstract} 

\keywords{Galaxy: disk --- ISM: abundances --- ISM: dust, extinction --- molecular processes --- stars: formation --- submillimeter: ISM}


\section{Introduction} \label{sec:intro}
The single greatest impediment to the study of star formation is the H$_2$ molecule's lack of a dipole moment. Stars form in dense clouds where most of the hydrogen is in the form of H$_2$ \citep{ladas}, and H$_2$ lacks molecular transitions that can be excited at temperatures (1 to a few $\times10$~K) typical of molecular gas in the prestellar and early protostellar phases of evolution. Astronomers must rely on alternative tracers to probe the physical, chemical, and kinematic conditions of the gas, assuming these tracers are well-coupled to the H$_2$. Without a direct way to test that assumption, a variety of physical and chemical conditions must be surveyed with as many relevant tracers as possible to find the limits of each tracer's utility.  

The Census of High- and Medium-mass Protostars (CHaMP, \citealt{champ1,mmsignposts,champ2,champ3, schap17, champ4}, hereafter Papers I, I-B, II, III, III-B, and IV respectively) 
is one such survey focused on the upper end of the protostellar mass spectrum, where observations are further complicated by short formation timescales ($t\lesssim10^6$~yr, \citealt[e.g.][]{kahn74,schall,kuip11,yusof}), large distances ($>400$~pc) to the nearest objects \citep[see e.g.][]{reid09}, and the crowded fields of the Galactic Plane \citep[see e.g. review by][]{zinnyork}. The CHaMP survey sample consists of about 300 clump- to core-scale ($\sim1$ to $\sim0.1$~pc) condensations in the Carina-Sagittarius Arm within the $280^{\circ}<\ell<300^{\circ}$, $-4^{\circ}<b<+2^{\circ}$ area, mapped with the Mopra\footnote{The Mopra telescope is part of the Australia Telescope, funded by the Commonwealth of Australia for operation as a National Facility managed by CSIRO. The University of New South Wales Digital Filter Bank used for
observations with the Mopra telescope was provided with support from the
Australian Research Council.} 22~m telescope in over 30 molecular species. Aside from questions more specific to massive star formation, like the duration of the prestellar phase, the CHaMP project's goals have expanded to include addressing more fundamental concerns about how common H$_2$ tracers, like CO(J$1-0$) emission, change in their tracing efficacy 
with the physical conditions. Combining the multi-isotopologue CO data released in \citetalias{champ3} and \citetalias{champ4} with archival dust continuum data from emph{Herschel} and other infrared observatories gave us the chance to begin exploring the detailed, temperature-dependent behavior of CO emission in \citealt{pittsmn} (hereafter PBV19). In this study, we expand that work to the rest of the CHaMP sample.

Our previous and current works began, as many continuum surveys do, with fitting modified Planck spectral energy distributions (SEDs) in the manner of \citealt{hild}, followed by detailed examination of the resulting maps of dust temperature (\td) and H$_2$ column density (\ncol). Similar work has been done as part of surveys like HiGAL \citep[][etc.]{higal1,higalirdc,higal2} and MALT90 \citep{malt90}, and our work advances these studies with additional insights.
\td\ trends with \ncol\ or with radial distance from the center of a condensation can indicate whether a clump is actively star forming, and the proportion of clumps with cold versus warm centers is a proxy for the relative amounts of time spent in the prestellar and protostellar phases. Just within our survey, kinematic evidence from \citetalias{champ3} and \citetalias{champ3} in the form of radial velocity shifts of $^{12}$CO relative to $^{13}$CO and C$^{18}$O suggested long infall timescales for prestellar clumps. This is further supported by the tendency of local \ncol\ maxima to more often than not coincide with \td\ minima found in \citetalias{pittsmn}. Most importantly, we used \td\ and \ncol\ from dust continuum emission, in conjunction with CO column densities derived using the three most abundant CO isotopologues (see \citetalias{champ3} for details), to derive gas-phase CO abundance (hereafter simply CO abundance) maps to relate to \td\ and the integrated $^{12}$CO line intensity (\ico).

For most studies that do not focus on the behavior of CO as a function of local conditions, the usual method is to assume that the velocity-integrated intensity of the CO($J=1-0$) line is proportional to the H$_2$ column density by some factor \xco\ \citep{dame01}. Unlike H$_2$, the CO molecule's 3~mm $J=1-0$ rotational transition is easily detectable with ground-based radio telescopes, and CO is the most abundant molecule in the universe after H$_2$. However, with full radiative transfer on multiple CO isotologues \citepalias[see][for details]{champ4}, the $^{12}$CO($J=1-0$) line is seen to be optically thick almost everywhere in the Galactic Plane. Moreover, whenever \xco\ is treated as a constant, typically averaging $2\times10^{24}$~m$^{-2}$~(K~km~s$^{-1}$)$^{-1}$ to within a factor of 2 \citep{bolatto,okamoto2017,gong18,hayashi2019}, it encodes a constant CO abundance of 10$^{-4}$ per H$_2$. The inherent uncertainty in \ncol\ derived from dust, by factor of two to three due to the uncertain gas-to-dust ratio \citep[see e.g.][]{beckw,zubko,reach}, is not nearly enough to explain the $>2$ orders of magnitude variation seen in just the subset of CO abundance maps published in \citetalias{pittsmn}. A wide variety of both observational \citep{caselli99,bacmann,akhern,fontani,ripple,kongco} and laboratory \citep{oberg09,noble12,munoz10,munoz16,cazax17} studies find that the CO abundance must vary with temperature and ambient radiation exposure. In this study we finally have the statistics to quantify the correlation of CO abundance with \td\ in the Galactic Plane. If CO abundance varies systematically with local physical conditions, there is no reason to believe that \xco\ is constant, or that the argument from virialization is applicable.  

Besides the obvious dependency of \xco\ on metallicity, studies by \citealt{desika2}, \citealt{desika}, \citealt{thrumms}, \citealt{wada},  \citetalias{champ4}, and \citealt{sofue20} found additional dependencies of $X_{\text{CO}}$ on CO optical depth. The latter four papers, especially \citealt{thrumms} and \citetalias{champ4}, find $^{13}$CO and C$^{18}$O observations indicate that $N_{\text{CO}}$ is higher than expected given $I_{\text{CO}}$ in both the highest- and lowest-column-density areas, such that the standard $X_{\text{CO}}$-factor relationship can under-predict \ncol\,by up to a factor of three. Simulations by \citealt{wada} suggest the apparent power-law dependence of $X_{\text{CO}}$ on $I_{\text{CO}}$ may be even steeper than that put forth in \citetalias{champ4}. These proposed power-law alternatives to the $X_{\text{CO}}$-factor may reduce the need for so-called CO-dark gas in the diffuse envelopes of molecular clouds, where H$_2$ is dense enough to self-shield, but CO has been photo-dissociated \citep{blitz,reach94}. For instance, the GOTC+ team, who used the standard $X_{\text{CO}}$-factor to calculate molecular gas masses from CO observations, suggest CO-dark percentages may vary from $\sim$20\% for dense, massive clumps, to $\sim$70\% for diffuse H$_2$ filaments \citep{pine,langer}. If the method of \citetalias{champ4} or the fit presented in \S3.3 proves more appropriate, the fraction of H$_2$ in CO-dark gas in dense clump settings may be less significant, given typical uncertainties in total clump mass of $\sim$20\%.\\ 

In this article, we expand upon the results of \citetalias{pittsmn} to show that absolute CO abundance variations are widespread in massive molecular gas clumps and correlate strongly with \td in \S3.1. We present a set of best-fit parametrizations for these effects in \S3.2, which we recommend for future studies of dense molecular clumps. Further, via comparison of \ncol\ derived from dust continuum observations to \ico, we reveal in \S3.3 among the strongest evidence to date that the \xco-factor is systematically dependent on the density (ergo shielding) and excitation conditions of the gas. We conclude with some of the implications of these findings in \S4.

\section{Data and Methods}
\label{sec:meth}
\subsection{CO}
\indent The 2009--2012 phase of Mopra telescope observations for the CHaMP project (Stage II) mapped the $^{12}$CO, $^{13}$CO, C$^{18}$O, and other line data across for the brightest 267 prestellar clumps detected in HCO$^+$ in Stage I \citepalias[see][for Phase I data]{champ1, mmsignposts, schap17}. For the Stage II analysis, integrated $^{12}$CO line intensities (denoted \ico) and CO column densities (denoted \nco) were derived by performing full radiative transfer on all three isotopologues, assuming typical Galactic isotopic ratios. \citetalias{champ3} discusses the Stage II observing plan and conditions, reduction techniques, and radiative transfer equations in great detail.\\

\subsection{Dust Continuum}
\indent We stacked archival far-infrared (FIR) and submillimeter (submm) data from 70 to 870~\micron, with optional data for a second component from 3.4~\micron\ to 24~\micron, and fit pixel-by-pixel color-corrected modified Planck spectral energy distributions (SEDs) using IDL code built around MPFIT \citep{mrkwdt}. We used data from \emph{Herschel}-PACS \citep{pacs}, \emph{Herschel}-SPIRE \citep{spire}, APEX-LABOCA \citep{laboca}, and, where helpful to separate out warmer temperature components, WISE \citep{wise} and MIPS-24~\micron\ \citep{mips}, all obtained from the Infra-Red Science Archive (IRSA). Our anonymous code package is publicly available at \href{https://github.com/rlpitts/Mosaic-Math}{this linked GitHub repository}, and includes several model options and color correction functionality. We assume a gas-to-dust ratio $\gamma$ of 100, dust emissivity index $\beta$ of 1.8, and opacity of $\kappa_0$ of 0.55 m$^2$~kg$^{-1}$ at 250~$\mu$m as per the results of the Planck Collaboration's studies of the Galactic Plane \citep[see e.g.][]{planck23}. Because of the CHaMP sample's proximity to the Carina tangent, we assume negligible foreground and background components except in the cases of Regions 23 and 26, which lie in front of the Dragonfish Nebula. We also do \emph{not} assume optically thin dust emission because there was no significant computational benefit to doing so, and we find that optical depths near 70~\micron\ can approach 1. We refer readers to \citetalias{pittsmn} for a complete description of the SED fitting routine and its features.\\

\subsection{Objects Covered}
\indent The CHaMP survey target fields are divided into 27 Regions with multiple prestellar clumps, plus a number of isolated prestellar clumps, that were mapped by NANTEN2. Of these, 18 Regions and one isolated clump (BYF~123) had complete or nearly complete coverage by both \emph{Herschel} and Mopra. We initially tested the pixel-by-pixel SED-fitting pipeline on Regions 9 and 26 (in and around Gum 31 and RCW 64, respectively), and then expanded our analysis to Regions 10 and 11 (the northern and southern halves of the Carina Nebula Complex respectively) to complete \citetalias{pittsmn}. In the following sections, we expand our sample to cover CHaMP Regions 1--3, 5--8, 12 (marginal), 13, 16, 18, 21, and 23. Table~\ref{tab:pos} summarizes the contents and sky coverage of each Region.
\begin{threeparttable}[htb]
    \centering \caption{Sky coverage and some contents of each Region (see also \citetalias{champ1}, Figure~2)} 
    \begin{tabularx}{\textwidth}{l l l l l}\hline \small
       Region  &  $\ell$ range ($^{\circ}$) & $b$ range ($^{\circ}$) & BYF Clumps & Noted Objects in Region\\\hline
       1  & $280.69^{\circ}$ to $281.55^{\circ}$ & $-1.87^{\circ}$ to $-0.95^{\circ}$ & 2--5,7--9 & IRAS 09578-5649\tnote{a}\\
       2a & $281.48^{\circ}$ to $281.78^{\circ}$ & $-1.17^{\circ}$ to $-0.49^{\circ}$ & 10, 12--14, 16 & MMB G281.710-01.104\tnote{b}\\
       2b+3 & $281.65^{\circ}$ to $282.32^{\circ}$ & $-2.10^{\circ}$ to $-1.35^{\circ}$ & 15, 18--22, 25, 26 & RAFGL 4101\tnote{c}\\
       2c & $282.16^{\circ}$ to $282.35^{\circ}$ & $-0.92^{\circ}$ to $-0.41^{\circ}$ & 23, 24, 27 & THA 35-II-3\tnote{d}\\
       5 & $282.78^{\circ}$ to $283.25^{\circ}$ & $-1.07^{\circ}$ to $-0.92^{\circ}$ & 32, 36, 37 & IRAS 10123-5727\tnote{a}\\
       6 & $283.96^{\circ}$ to $284.21^{\circ}$ & $-1.11^{\circ}$ to $-0.76^{\circ}$ & 40--42 & SEST 39\tnote{e}\\
       7 & $284.64^{\circ}$ to $284.72^{\circ}$ & $-0.71^{\circ}$ to $-0.53^{\circ}$ & 47 & $\ldots$\\
       8 & $284.79^{\circ}$ to $285.43^{\circ}$ & $-0.16^{\circ}$ to $+0.18^{\circ}$ & 50, 54, 56 & Hoffleit 18\tnote{f}\\
       9 & $286.00^{\circ}$ to $286.46^{\circ}$ & $-0.45^{\circ}$ to $+0.24^{\circ}$ & 63, 66--73, 76--79 & Gum 31, NGC 3324 (CNC)\\
       10 & $286.90^{\circ}$ to $287.54^{\circ}$ & $-0.99^{\circ}$ to $-0.05^{\circ}$ & 85--104 & Tr14, Northern Cloud (CNC)\\
       11 & $287.61^{\circ}$ to $288.31^{\circ}$ & $-1.25^{\circ}$ to $-0.42^{\circ}$ & 105--118 & Southern Cloud/Pillar (CNC) \\
       12 & $291.28^{\circ}$ to $291.53^{\circ}$ & $-1.79^{\circ}$ to $-1.59^{\circ}$ & 127, 130 & $\ldots$\\
       13 & $291.19^{\circ}$ to $291.70^{\circ}$ & $-0.81^{\circ}$ to $-0.16^{\circ}$ & 126, 128, 129, 131, 132 & NGC 3576, NGC 3603\\
       15 & $291.97^{\circ}$ to $292.05^{\circ}$ & $-2.00^{\circ}$ to $-1.92^{\circ}$ & 134 & $\ldots$\\
       16 & $293.01^{\circ}$ to $293.38^{\circ}$ & $-1.08^{\circ}$ to $-0.74^{\circ}$ & 141, 142, 144 & Bran 354A\tnote{h}\\
       18 & $293.62^{\circ}$ to $293.84^{\circ}$ & $-1.77^{\circ}$ to $-1.59^{\circ}$ & 149, 150 & Bran 362B\tnote{h}\\
       21 & $294.76^{\circ}$ to $295.23^{\circ}$ & $-1.84^{\circ}$ to $-1.54^{\circ}$ & 161--163, 165, 167 & IC2944/2948\\
       23 & $298.14^{\circ}$ to $298.48^{\circ}$ & $+0.65^{\circ}$ to $+0.80^{\circ}$ & 185, 190 & IRAS 12091-6129\tnote{i}\\
       26a & $298.86^{\circ}$ to $298.94^{\circ}$ & $+0.31^{\circ}$ to $+0.50^{\circ}$ & 199 & $\ldots$ \\
       26b & $298.94^{\circ}$ to $299.58^{\circ}$ & $-0.48^{\circ}$ to $-0.17^{\circ}$ & 201--203, 208 & RCW 64\tnote{j}, Bran 386B--G\tnote{h}\\ \hline
    \end{tabularx}
    \label{tab:pos}
    \begin{tablenotes}\footnotesize
    \item[][a] \citealt{iras1986lrs}; [b] \citealt{green2012}; [c] \citealt{rafgl}; [d] \citealt{theps1966}; [e] \citealt{hlb1998}; [f] \citealt{hoff53}; [g] \citealt{gum31}; [h] \citealt{bran}; [i]\citealt{whrb97}; [j]\citealt{rcw} \end{tablenotes}
\end{threeparttable}

\section{Results and Discussion}
\label{sec:res}
\subsection{Temperature, Density, and Morphological Trends in the Full CHaMP Sample}\label{ssec:morpho}
\begin{figure}
  \begin{center}
    \makebox[\textwidth][c]{\includegraphics[width=1.1\textwidth]{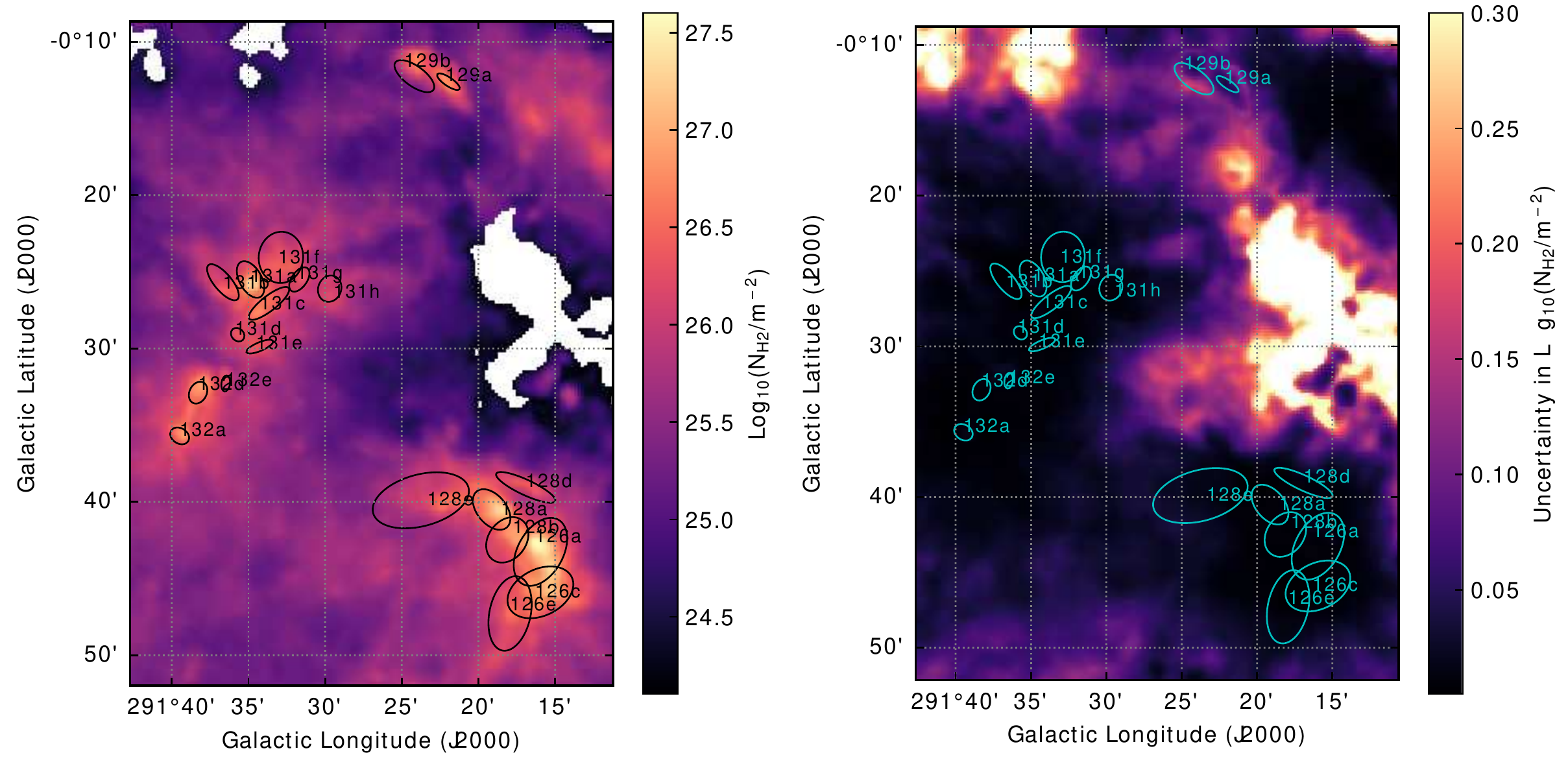}}\\
    \makebox[\textwidth][c]{\includegraphics[width=1.1\textwidth]{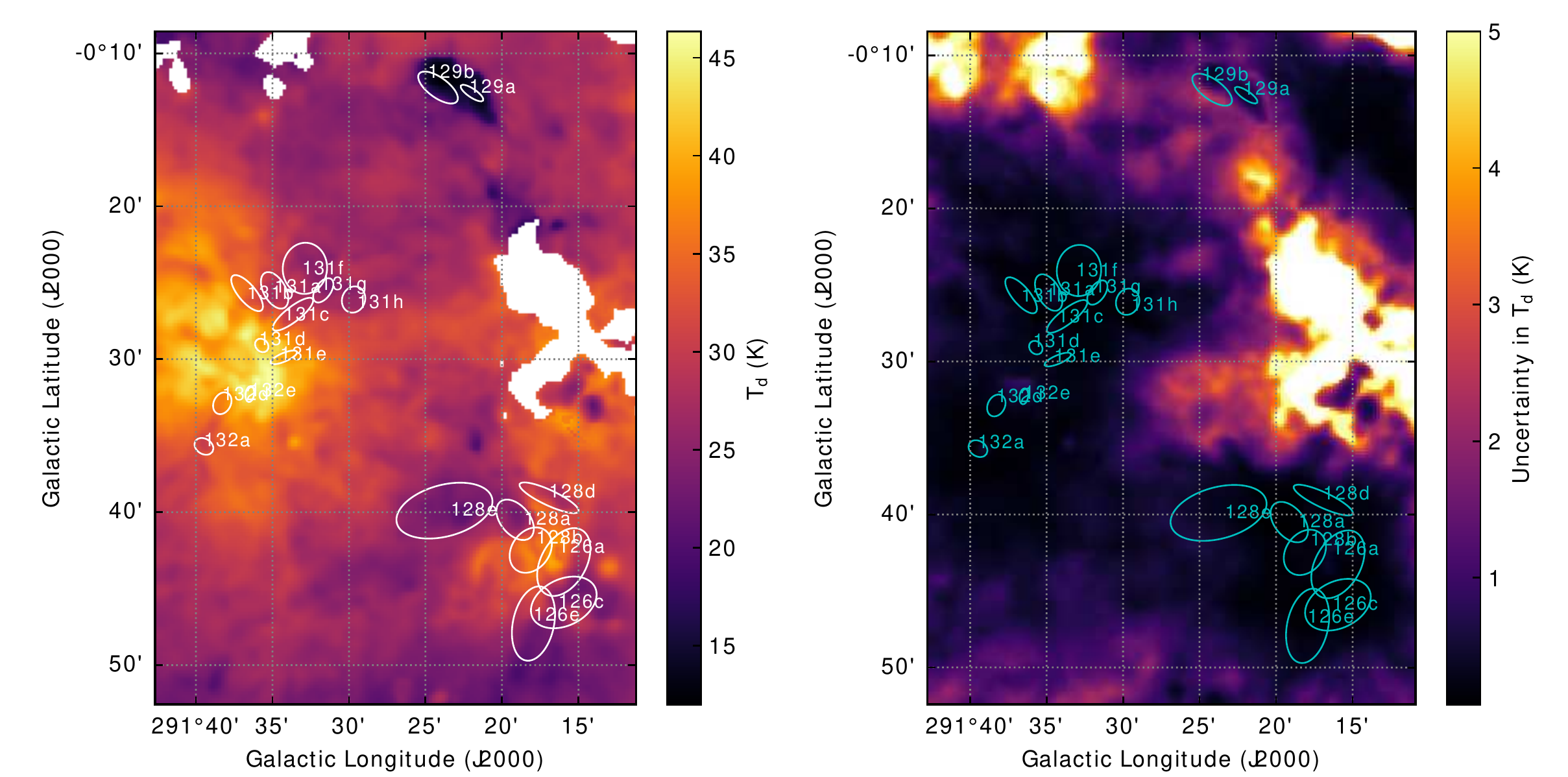}}
    \caption{SED-fitting parameter maps for Region 13. \emph{Top row:} \ncol\, (left) and \ncol\, uncertainty (right) maps. \emph{Bottom row:} \td\, (left) and \td\, uncertainty (right) maps.}\label{fig:pmaps13}
  \end{center}
\end{figure}

Figure~\ref{fig:pmaps13} shows a sample set of dust temperature and H$_2$ column density maps, and their uncertainty maps, for Region 13. Dust temperature, H$_2$ column density, and CO abundance maps by Region, for all other Regions not covered in \citetalias{pittsmn}, are included in the online-only appendix sections~\ref{sec:app0} and \ref{sec:app1}, respectively. Most of Region 13 is unusually warm, but we chose it as the example because it shows the greatest diversity of physical conditions across a Region.

\begin{figure}
\centering
    \includegraphics[height=6in]{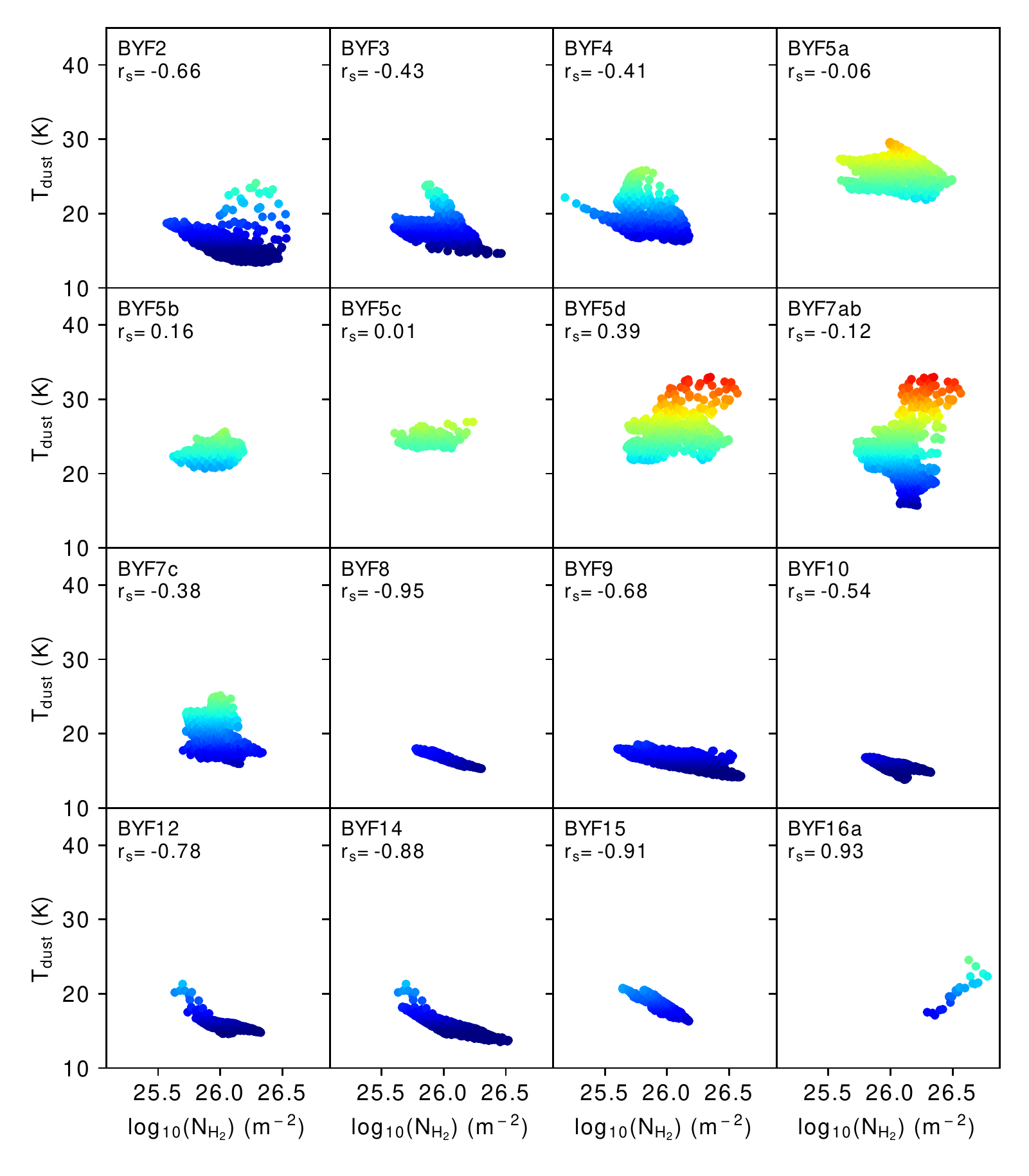}
    \caption{$T_{\text{d}}$ vs. $N_{\text{H}_2}$ plots for clumps in Regions 1 and 2a. Spearman rank correlation coefficients ($r_s$) are given under the BYF catalogue numbers.}
    \label{fig:tvn}
\end{figure}

Most clumps in Figure~\ref{fig:tvn} and Appendix~\ref{sec:app2} show $T_{\text{d}}$ declining, in a variety of functional forms, as $N_{\text{H}_2}$ increases. This broadly inverse relation between temperature and column density is the opposite of the trend one typically observes for $N_{\text{H}_2}$ and $T_{\text{ex}}$ for CO \citep[e.g.][]{kongco,gong18}, but is to be expected if molecular line emission, particularly from CO, is the predominant coolant in molecular clouds (MCs). Even clumps that are known to contain or border on active star-forming regions---like BYF 40b, 109a, 150, and 203---sometimes still show higher $T_{\text{d}}$ at lower column densities. Some particularly warm comet-shaped clumps, like BYF~103a, form a slightly concave-up curve in \td-\ncol\ space suggestive of a compressive front. Not every cometary globule has this shape in the plot of \td\, vs.\, \nco, but every clump that has this functional form is a cometary globule. 

\begin{figure}[tbp]
\centering
    \includegraphics[width=0.5\textwidth]{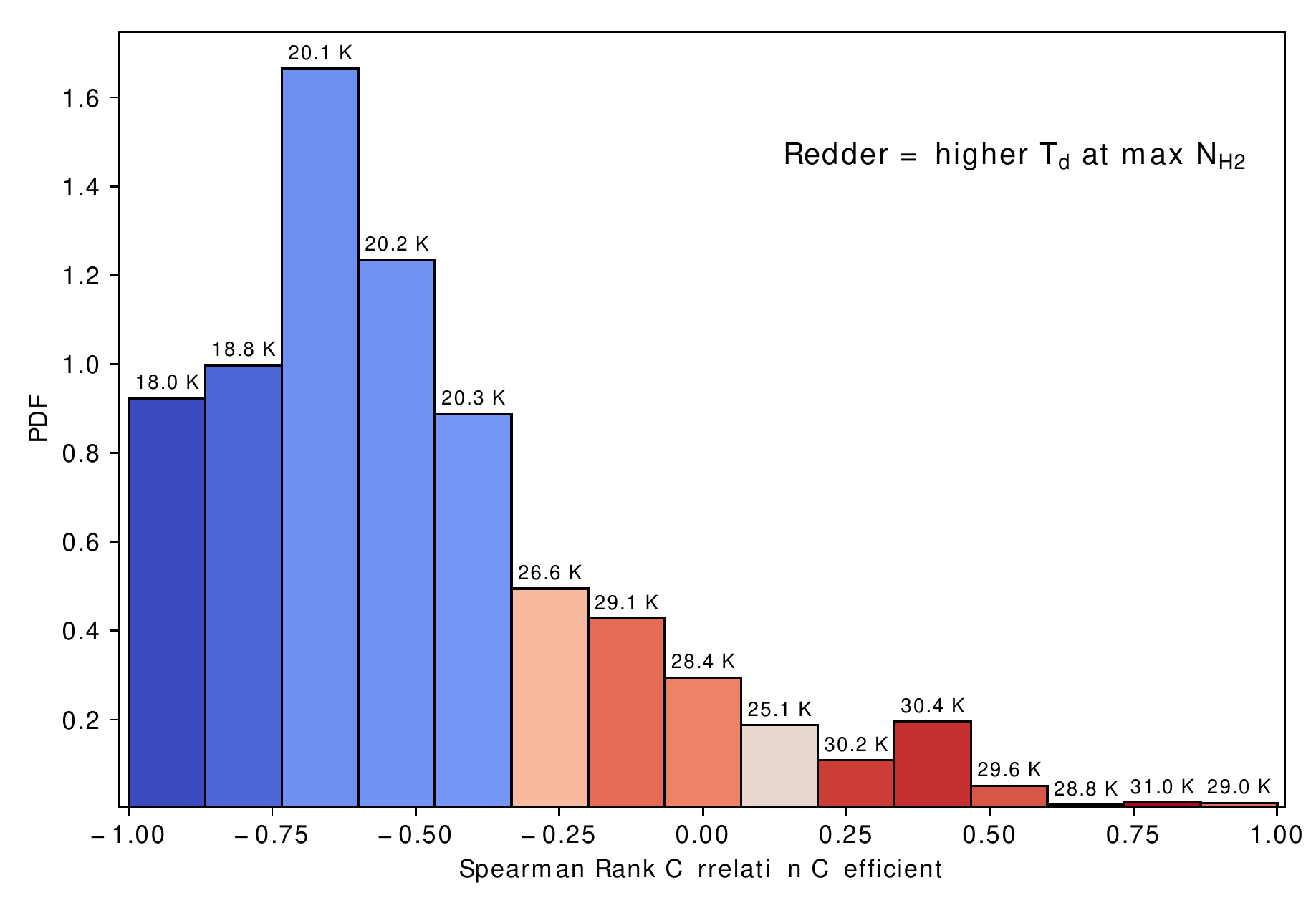}
    \caption{Histogram PDF of Spearman rank coefficients of correlation between \td\, and \ncol. The Spearman coefficients themselves correlate with a clump's \td\, at the position of maximum \ncol\, (indicated by the coloration of the bins, where redder bins have higher \td.)}
    \label{fig:spear}
\end{figure}

Overall, at least 84\% of prestellar clumps in the CHaMP survey that could be separated in dust continuum emission showed clear anti-correlations between \td\, and \ncol,\ (see Figures \ref{fig:tvn} and \ref{fig:spear}), with many of the others too small to judge accurately. We checked for this anti-correlation by evaluating the trend in \td\, versus \ncol\, over 2$\sigma$ in both the major and minor axes around each clump, and calculating the Spearman rank correlation coefficient. We used the Spearman rank correlation coefficient rather than the Pearson correlation coefficient because the latter assumes a linear relationship between parameters, whereas the former only demands that a correlation be monotonic. The 16\% of clumps that had flat or positive Spearman coefficients of correlation between \td\, and \ncol\, were systematically warmer at the location of the peak in \ncol, but were not all associated with other signs of star formation. Some, like BYF~103a and b (see Figure~\ref{fig:tvn9} in Appendix~\ref{sec:app2}), were simply not monotonic. The clumps known to be associated with other indicators of star formation, like masers and 2MASS embedded cluster candidates, showed large scatter in their \td\, versus \ncol\, plots and typically (though not always) had Spearman coefficients $>+0.25$. One known clump around a protocluster, BYF~73, still had a negative Spearman coefficient, but as shown in \citealt{pittsapjl}, this clump is some 98\% gas-dominated. The Spearman correlation coefficient between \ncol\, and \td\, may only be sensitive to star formation at a somewhat later stage and will be dependent on spatial resolution.

These results are generally consistent with prestellar molecular clumps being long-lived structures as previously evinced in \citealt{champ4}. Specifically, it suggests that clumps without a strong internal heat source (``prestellar,'' or more ambitiously, ``starless'') last $\sim10$ times as long as clumps exhibiting active star formation (Figure~\ref{fig:spear}). In other words, most of the gas-phase evolution of a clump happens in the last $\sim10\%$ of its lifetime. This also makes sense in light of simulations showing that the cooling efficiency of CO increases with temperature and density \citep{juve01,stahlpal,gong18}: purely gas-dynamical changes in temperature and pressure should be damped. However, these damping of effects also mean that a clump can contain extremely young protostellar cores while still appearing cold and CO-depleted at clump scales. Indeed, our group wrote about the protostellar content of BYF73 in \citealt{pittsapjl} while documenting its clump-scale appearance of CO depletion and lower-than-ambient temperatures in \citetalias[][Figures 6 and 11]{pittsmn}. Thus, observational applications of these models to verify a clump's evolutionary state demand detailed accounting of heating mechanisms, like embedded protostars. That means catalog matching would have to be done in search of star formation indicators within each clump's angular area--e.g. astrophysical masers, outflows, or K-band point-sources. Besides BYF73, we have done this manually where hot spots and hollows in the column density maps were immediately apparent (see Table~\ref{tab:hotspots}), but follow-up on other sources is ongoing. 

\begin{figure}[tbp]
\centering
    \includegraphics[width=0.5\textwidth]{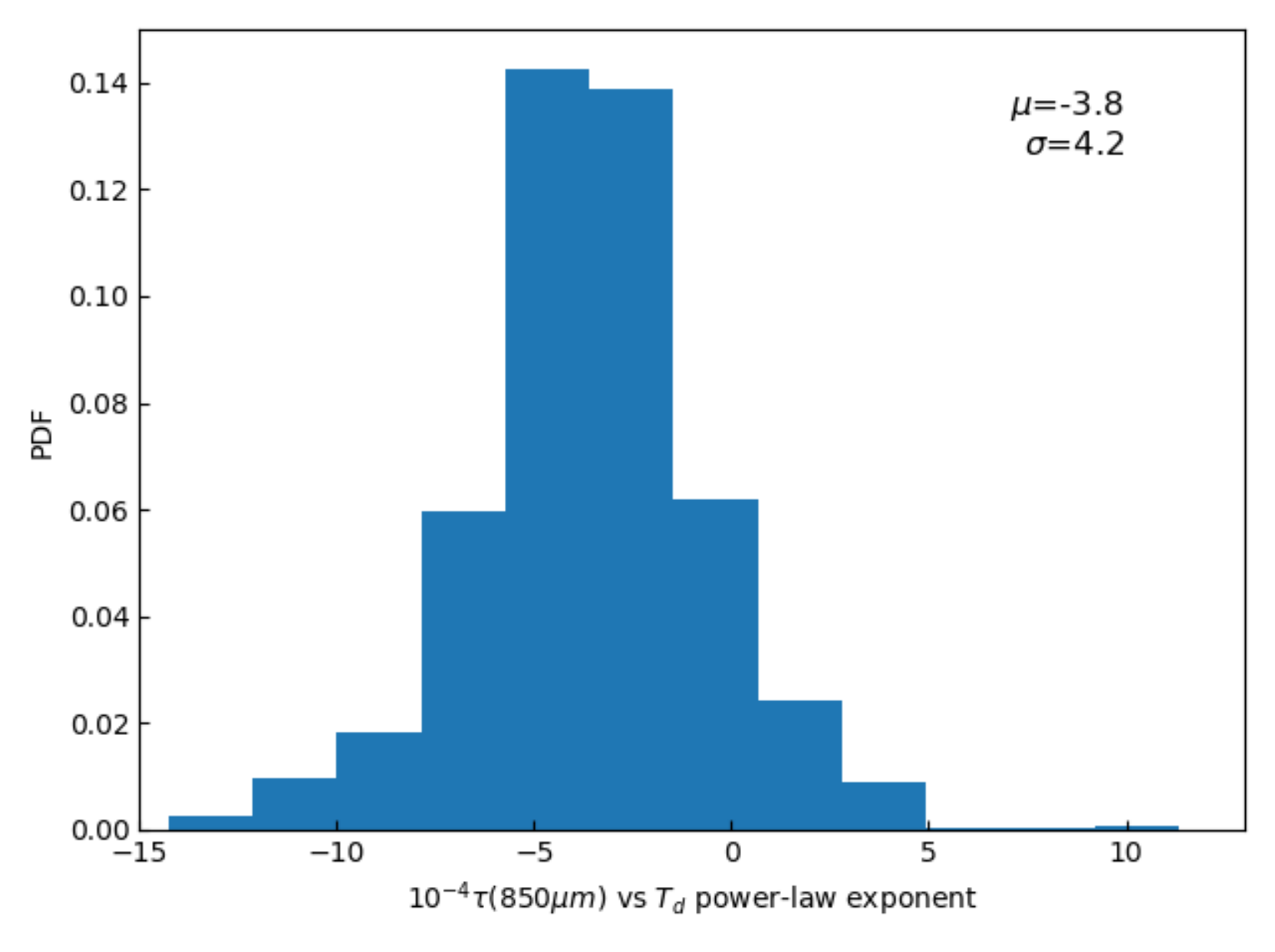}
    \caption{Histogram PDF of power-law exponents for the fits of optical depth $\tau(870\,\mu m)$ versus \td. This was done strictly for comparison with the work of \protect\citealt{hasen}.}
    \label{fig:hasencomp}
\end{figure}

The CHaMP clumps (again $\sim1$~pc in size) are more than ten times further away and sample a much greater diversity of environmental conditions than the cold dense cores ($\sim0.1$~pc) in the Pipe Nebula studied by \citealt{hasen}, but that study is a potentially useful point of comparison. For more direct comparison, we converted \ncol\, to an optical depth at 850~$\mu$m and fit each clump with a line in log[\td]-log[$\tau(850\,\mu m)$] space (a power law in linear space), with $\tau(850\,\mu m)$ multiplied by 10$^4$. The resulting distribution of power-law exponents, shown as a histogram in Figure~\ref{fig:hasencomp}, was 
more symmetric and less purely negative than the distribution of the same parameter in \citealt{hasen}. The power-law exponents for the CHaMP clumps ranged from about -15 to +5 with a mean of about -4, compared to typical values between -5 and -10 and a range of -40 to 0 in \citealt{hasen}. Besides the inclusion of actively star-forming clumps, we suspect at least some of the difference is down to resolution for reasons that should become clearer when we discuss the morphotype scheme developed in \citetalias{pittsmn}. We also see much more substructure in the plots of temperature versus density (or optical depth), at least some of which must be due to how we defined the boundaries of the clumps. Where \citealt{hasen} worked directly with local minima identified in \td\, maps, the CHaMP clumps are defined by CO emission. As shown \citetalias{pittsmn} and in the appendices, CO tends to be under-abundant in the coldest density enhancements, so multiple local \ncol\, maxima sharing a common envelope can look like a single structure in CO. For a truer, apples-to-apples comparison with the physical conditions inside the Pipe Nebula cores, the CHaMP sample would have to be redefined using, e.g., an algorithm like Clumpfind \citep{clumpfind} directly on the \ncol\, maps.


In general we find that the morphotype scheme defined in \citetalias{pittsmn} continues to be a relevant aid in describing and understanding of the spatial variation of CO abundance, both across individual clumps and across their larger Regions. To review, we classified prestellar clumps into five major groups depending on the relative positions or geometries of \ncol\ maxima and CO abundance extrema:
\begin{itemize}
    \item \textbf{Coincident- or C-Type} if \ncol\ and the CO abundance both peaked in the same location, or if the CO abundance appeared constant across the clump;
    \item \textbf{Asymmetric- or A-Type} if there was a CO abundance maximum within the clump boundaries as defined in \citetalias{champ3}, but it was noticeably offset from the \ncol\ maximum;
    \item \textbf{Frozen-out- or F-Type} if the \ncol\ maximum coincided with a local minimum in CO abundance surrounded by a ring or filament of enhanced CO abundance;
    \item \textbf{FA-Type} if the \ncol\ maximum coincides with a CO abundance minimum that is only partly bounded by enhanced CO abundance;
    \item \textbf{Peculiar- or P-Type} in the rare case that there is no visible \ncol enhancement at the location of a clump defined by CO emission.
\end{itemize}
In many cases, it was possible to further classify A- and FA-Type clumps into sublimating (s(F)A-Type) or dissociating (d(F)A-Type) sub-types based on the projected positions of nearby FUV sources relative to the \ncol maxima and CO abundance enhancements surrounding zones of depletion. Clumps classified as sA- or sFA-Type had CO abundance enhancements along the side facing the FUV sources, and clumps classed as dA- or dFA-Type had any proximal CO enhancements behind the \ncol\ maximum relative to the FUV source. Figure~\ref{fig:clumps} (adapted from \citetalias{pittsmn}) qualitatively illustrates the appearances of all the types and sub-types in the CO abundance maps.\footnote{``Undisturbed'' is probably a bit of a misnomer, since the geometries of A- and FA-Type clumps are likely attributable to external irradiation more often than not. The circle on the left more accurately shows how the clumps appear, and are categorized, before assessing the influence of environment.} As discussed at great length in \citetalias{pittsmn}, a prestellar clump's morphotype also generally correlates with its minimum temperature. F- and FA-types systematically reached cooler temperatures than C- or A-types, and more consistently displayed a clear anti-correlation of \td\ with \ncol.

\begin{figure}
    \centering
    \includegraphics[height=0.65\textheight]{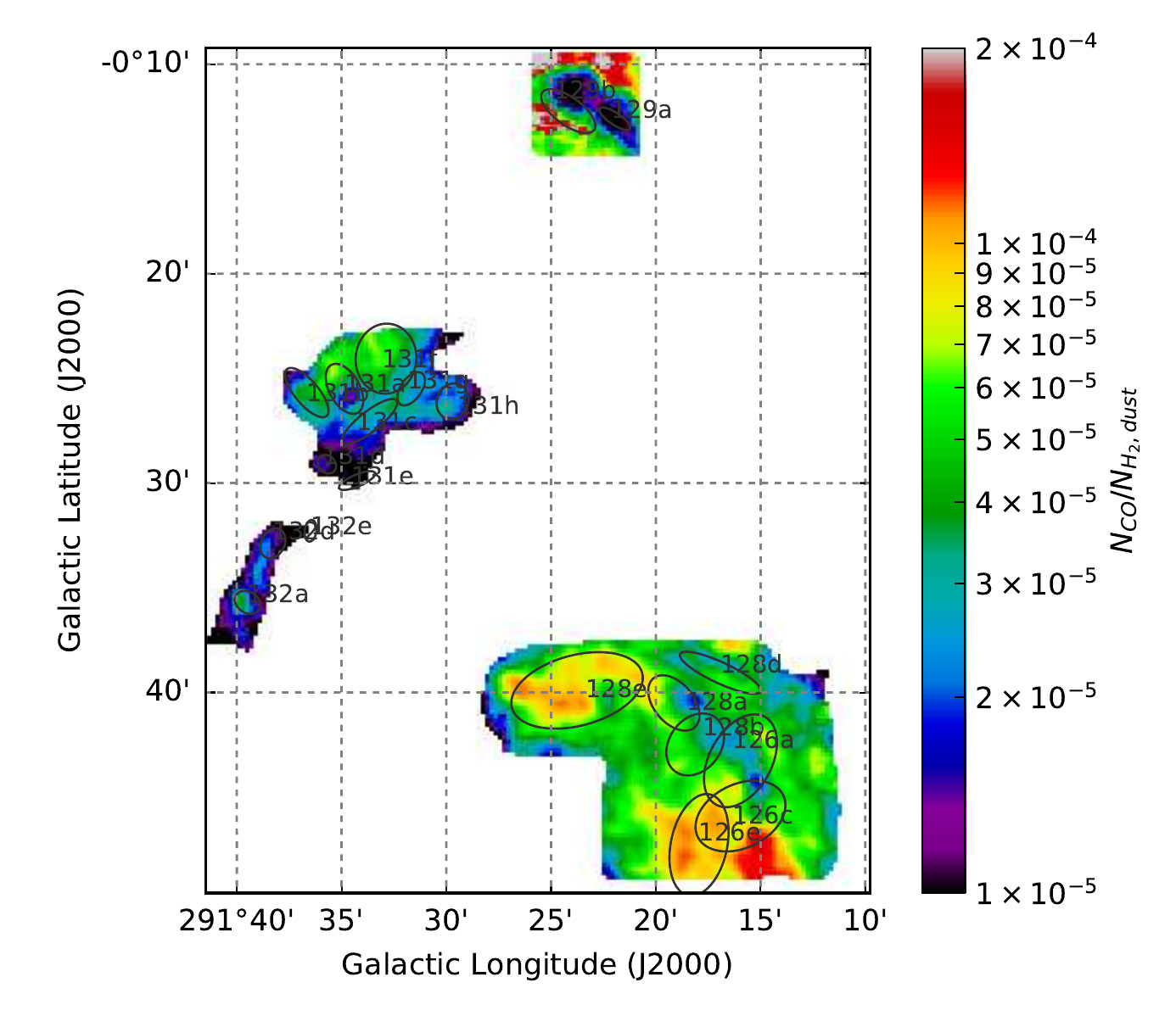}
    \caption{\coab\, map for Region 13, shown as an example.}
    \label{fig:coab13}
\end{figure}

\begin{figure}[htbp]
\centering
    \includegraphics[width=\textwidth]{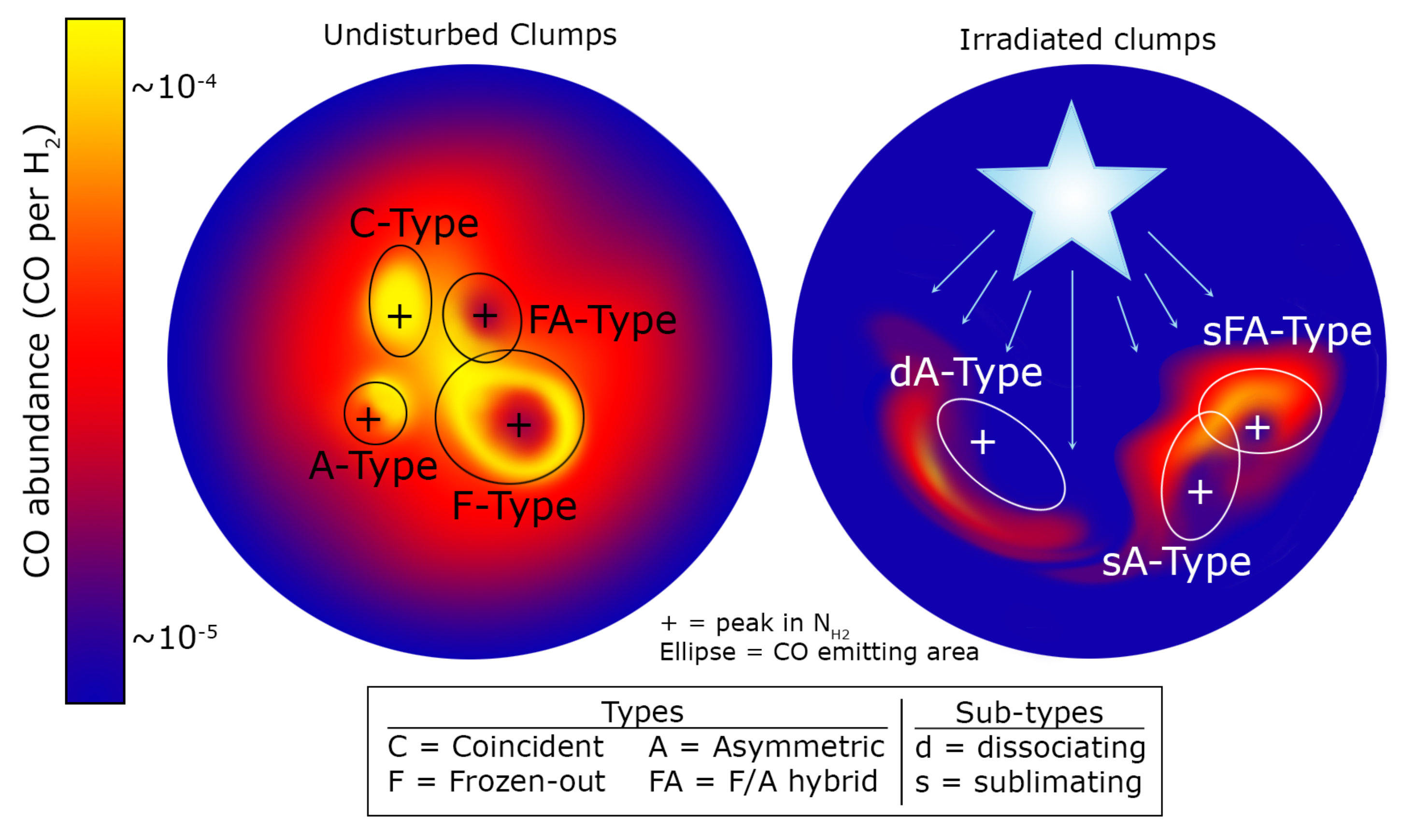}
    \caption{Illustration, adapted from \citetalias{pittsmn} of the clump morphotype scheme defined therein. Mathematical values are only approximate.}
    \label{fig:clumps}
\end{figure}

The abundance map of Region 13 shown in Figure~\ref{fig:coab13} provides a helpful demonstration of the various abundance morphotypes all in one image due to the proximity of several prestellar clump groups in varying stages of development. CO abundance maps for the remaining Regions can be found in Appendix~\ref{sec:app1}. Each of the three clusters of clumps in Region 13 is a product of a different environment viewed at a different distance. BYF~129 (top), by far the closest at about 1.2 kpc away \citep[][and references therein]{champ1}, is an extreme example of an F-Type: cold ($\sim$12~K) and strongly depleted. A quick look at the interactive three-color 2MASS image on SIMBAD shows that BYF~129 is an IRDC. The various components of BYF~126 and 128, both about 2.4 kpc away, trace a dense filament cutting across the (Galactic) southwest of NGC~3576 that appears to be CO-depleted toward the middle and enhanced on the ends. The minima in BYF~128a and 126c may not be the most depleted clumps in the filament: BYF~128b and 126a cover a patch of saturated emission at 250~$\mu$m, so \td\, and \ncol\, for these two clumps are highly uncertain. The most distant grouping at over 6 kpc away, BYF~131 and 132, surround the massive cluster NGC~3603. BYF~131d and g are C-Types, and the rest (where enough data exists to classify them) are A-Types, mostly of the dissociating subtype. Due to their distance, it is reasonable to suspect that the C-Type clumps in the BYF 131 and 132 groups might have a very different classification if viewed at a different angle or closer distance.

This system is not scale-free, and the temperature segregation is less clean than it initially appeared. Small F- and FA-type clumps viewed at sufficient distance may look like C- and A-Type clumps, respectively, which may in part explain why many more of the latter types were identified with lower temperatures. Two clumps originally designated as F-Type were found to be warmer rather than colder toward their dense centers. These were readily identified as known protostellar objects with associated masers and outflows on SIMBAD. Aside from them, the range of temperatures seen in F- and FA-Types did not increase like it did for the other morphotypes, despite a similar increase in the total number of clumps of these two types. This is somewhat unexpected given the range distances probed by the CHaMP survey on average accounts for less of the variation in size than intrinsic differences in the source dimensions. It was probably just luck given that several more clumps initially identified as FA-Types were also found to be associated with embedded objects after inspecting them in the \td\ maps.

Besides requiring adequate separation of clumps, the morphotype classification scheme originally assumed that any CO line emission structure fully enclosing a compact H\textsc{ii} region would be broken up into two or more clumps by virtue of the kinematics or temperature effects. That held true for Regions 9--11 and 26, but between Regions 1--3, 6, 8, and 13, we noted a handful a clumps where that assumption breaks down: BYF~5d, 7ab, 16a, 19b, 27, 40a, 54a, 56a, and 131a. Notably, every one of these clumps appears coincident with one or more known embedded sources, UCHII regions, or astrophysical masers within the clump's CO-emitting area defined in \citetalias{champ3}, as we found on SIMBAD and summarized in Table~\ref{tab:hotspots}. The two misclassified F-Types, BYF~16a and 27, also happened to be the only two of these warm clumps with minimum temperatures exceeding 25~K. Table~\ref{tab:hotspots} lists the clumps containing identified hot spots, the best-known sources therein, and a checklist of star formation tracers detected. Only the reference for the first detection that is not an upper limit is listed if the same source is observed multiple times in the same tracer. Indirect or otherwise unconfirmed detections are indicated with a question mark after the citation.

BYF~16a, 19b, 40a, 54a, and 131a all have coincident $T_{\text{d}}$ and $N_{\text{H}_2}$ maxima, and all are associated with at least one maser. BYF~56a also shows maser activity, but still met all the criteria for an A-Type clump. BYF~27 is centered on the Herbig Be star THA 35-II-3 \citep{tha35ii3}, and the two objects' distances are well within each other's margins of error \citep{pds37,yone}. Both BYF~7ab and BYF~40a would have been designated A-Types, but after accounting for the multiple temperature extrema within each, the published dimensions of these two objects may warrant revision. In addition to the above, BYF~2 and 3 also contained a handful of tiny hot spots, but these were not associated with $N_{\text{H}_2}$ maxima, nor did they seem to substantially affect their clumps' CO abundance distributions. Given these observations, and the fact that BYF~2 and 3 are among the closest clumps to the Carina Tangent, they may be background sources or may simply have not have been bright enough in NANTEN observations to make the cut for Mopra follow-up observations. 

\begin{longrotatetable}
\begin{ThreePartTable}
   \centering
   \begin{TableNotes}\footnotesize
\item[$\dagger$] Hot spot is in clump CO emitting area but not coincident with a local \ncol\ maximum.
\item[$\ddagger$] Object is visible in images in this wavelength range but has no catalog ID in the respective survey.
\item[]Table note marks for each detection correspond to the following references -- [a] \citealt{bnm1996}; [b] \citealt{iras1988psc}; [c] \citealt{wisesup}; [d] \citealt{msx6c}; [e] \citealt{rmssurvey}; [f] \citealt{iras1986lrs}; [g] \citealt{dbs2003}; [h] \citealt{2masscat}; [i] \citealt{gsl2002}; [j] \citealt{green2012}; [k] \citealt{wraycat}; [$\ell$] \citealt{theps1966}; [m] \citealt{PDScat1992}; [n] \citealt{allwise}; [o] \citealt{hlb1998}; [p] \citealt{bcg1980}; [q] \citealt{caswell87}; [r] \citealt{wmg1970}; [s] \citealt{braz83}; [t] \citealt{vdW1995}; [u] \citealt{srm1989}; [v] \citealt{cappa08}; [w] \citealt{pmt2014}; [x] \citealt{povich}; [y] \citealt{amm2003}; [z] \citealt{rath}; [A] \citealt{dutra01}; [B] \citealt{smith05}; [C] \citealt{smith_bally2010}; [D] \citealt{feigel11}; [E] \citealt{mbw2006}; [F] \citealt{caswell04}; [G] \citealt{msxcat}; [H] \citealt{nurn03}; [I] \citealt{sung04}; [J] \citealt{swgm93}; [K] \citealt{bgs1989}; [L] \citealt{nrs2013}; [M] \citealt{brn2015}; 
[N] \citealt{whrb97}; [O] \citealt{cwh2008}; [P] \citealt{akari2010}; 
[Q] \citealt{bubbly}; [R] \citealt{bran}; [S] \citealt{vvv11}; [T] \citealt{rcw}
\end{TableNotes}
   \begin{longtable}{lllllllll}\caption{Identified hot spots and other SF tracers within CHaMP clumps}\label{tab:hotspots}\footnotesize
\\ \hline
BYF	&	Hot Spot ID(s)	&	Masers	&	SiO	&	MIR	&	NIR	&	H\textsc{i} \& H$\alpha$	&	H\textsc{ii}	&	X-rays \\	
No.	&   	&		&	(Outflows)	&	(5-25\micron)	&	(0.7-5\micron)	&		&		&		\\ \hline\endfirsthead\endhead
2	&	IRAS 09568-5619	& $\ldots$		& $\ldots$		&	+\tnote{a}	&	\tnote{$\ddagger$}	&	 $\ldots$	&	?\tnote{a}	& $\ldots$		\\
3	&	IRAS 09581-5607	&	 $\ldots$	& $\ldots$		&	+\tnote{bc}	& $\ldots$		&	 $\ldots$	& $\ldots$		&	 $\ldots$	\\
5a\tnote{$\dagger$}	&	IRAS 09572-5636	& $\ldots$		&	 $\ldots$	&	+\tnote{b}	& $\ldots$		& $\ldots$		& $\ldots$		& $\ldots$		\\
5d	&	MSX6C G281.0472-01.5432	& $\ldots$		& $\ldots$		&	+\tnote{de}	&	\tnote{$\ddagger$}	& $\ldots$		& $\ldots$		& $\ldots$		\\
7ab	&	DBSB 125/IRAS 09578-5649	& $\ldots$		& $\ldots$		& $\ldots$	+\tnote{af}	&	+\tnote{g}	& $\ldots$		&	?\tnote{a}	&	 $\ldots$	\\
9	&	2MASS J10001718-5720342	& $\ldots$		& $\ldots$		&	+\tnote{de}	&	+\tnote{h}	& $\ldots$		& $\ldots$		& $\ldots$		\\
11a	&	IRAS 09563-5743	& $\ldots$		& $\ldots$		&	+\tnote{a}	&	+\tnote{i}	& $\ldots$		&	+\tnote{ia}	&	 $\ldots$	\\
16a	&	MMB G281.710-01.104,	&	CH$_3$OH\tnote{j}	& $\ldots$		&	+\tnote{ade}	&	+\tnote{h}	&	 $\ldots$	&	?\tnote{a}	&	 $\ldots$	\\
&	 2MASS J10050568-5657023?	&		&		&		&		&		&		&		\\
27	&	THA 35-II-3/PDS 37	& $\ldots$		&	 $\ldots$	&	+\tnote{de}	&	+\tnote{h}	&	+\tnote{k$\ell$m}	& $\ldots$		&	 $\ldots$	\\
36c	&	IRAS 10123-5727	& $\ldots$		& $\ldots$		&	+\tnote{f}	&	 $\ldots$	&	 $\ldots$	&	 $\ldots$	&	 $\ldots$	\\
36d	&	AllWISE J101417.36-574316.4	&	 $\ldots$	&	 $\ldots$	& 	+\tnote{n}	&	\tnote{$\ddagger$}	&	 $\ldots$	&	 $\ldots$	&	 $\ldots$	\\
40a	&	GAL 284.0-00.8/SEST 39	&	CH$_3$OH?\tnote{o}	&	+\tnote{o}	&	+\tnote{abde}	& $\ldots$		&	 $\ldots$	&	?\tnote{a}	& $\ldots$		\\
54a	&	Caswell OH 285.26-00.05,	&	CH$_3$OH\tnote{p}, H$_2$O\tnote{q}, OH\tnote{r}	&	\tnote{o}	&	+\tnote{af},	&	+\tnote{gh}	&	+\tnote{s}	&	+\tnote{s} ?\tnote{a}	&	 $\ldots$	\\
&	DBSB 48, GAL 285.3-00.0	&		&		&		&		&		&		&		\\
&	IRAS10295-5746	&		&		&		&		&		&		&		\\
56a	&	IRAS 10303-5746,	&	CH$_3$OH\tnote{jt}	&	 $\ldots$	&	+\tnote{ab}	&	 $\ldots$	&	 $\ldots$	&	?\tnote{a}	&	 $\ldots$	\\
&	MMB G285.337-00.002?	&		&		&		&		&		&		&		\\
73	&	DBSB 127, SEST 44	&	H$_2$O\tnote{u}	&	+\tnote{o}	&	+\tnote{adev}	&	+\tnote{gh}	&	+\tnote{v}	&	?\tnote{a}	&	 $\ldots$	\\
77c	&	MSX6C G286.3773-00.2563,	&	 $\ldots$	& $\ldots$		&	+\tnote{dv}	&	+\tnote{h}	&	+\tnote{v}	&	+\tnote{v}	&	?\tnote{w}	\\
&	  IRAS10361-5830?	&		&		&		&		&		&		&		\\
&	 CXOU J103801.8-584642?	&		&		&		&		&		&		&		\\
103a	&	2MASS J10445816-5931166,	&	 $\ldots$	&	 $\ldots$	&	+\tnote{x}	&	+\tnote{hx}	&	 $\ldots$	&	 $\ldots$	& +\tnote{y}	\\
&	 XMMU J104458.4-593115	&		&		&		&		&		&		&		\\
109a	&	Treasure Chest Cluster	&	 $\ldots$	&	 $\ldots$	&	+\tnote{z}	&	+\tnote{gAB}	&	+\tnote{BC}	&	 $\ldots$	&	+\tnote{D}	\\
115a	&	2MASS J10481308-5958516	&	$\ldots$	&	$\ldots$	&	+\tnote{x}	&	+\tnote{hx}	&	$\ldots$	&	$\ldots$	&	$\ldots$	\\
126a	&	2MASS J11115198-6118374,	&	H$_2$O\tnote{p}	&	+\tnote{o}	&	+\tnote{de}	&	+\tnote{hE}	&	$\ldots$	&	$\ldots$	&	$\ldots$	\\
&	 SEST 47	&		&		&		&		&		&		&		\\
131a	&	Caswell OH 291.57-00.43,	&	CH$_3$OH\tnote{F}, H$_2$O\tnote{q}, OH\tnote{r}	&	+\tnote{o}	&	+\tnote{cGH}	&	+\tnote{H}	&	?\tnote{H}	&	$\ldots$	&	?\tnote{I}	\\
&	 SEST 49	&		&		&		&		&		&		&		\\
163a	&	MSX5C G294.9709-01.7271,	&	CH$_3$OH\tnote{J}, H$_2$O\tnote{K}	&	+\tnote{o}	&	+\tnote{G}	&	$\ddagger$	&		&	(in IC2948)	&	?\tnote{L}	\\
&	 RAFGL 4134	&		&		&		&		&		&		&		\\
163b	&	MMB G294.990-01.719,	&	CH$_3$OH\tnote{j}	&	$\ldots$	&	+\tnote{M}	&	+\tnote{M}	&	$\ldots$	&	(in IC2948)	&	?\tnote{L}	\\
&	 La Serena 2?	&		&		&		&		&		&		&		\\
185	&	IRAS 12091-6129	&	CH$_3$OH\tnote{N}	&	$\ldots$	&	+\tnote{GO}	&	+\tnote{P}	&	$\ldots$	&	?\tnote{N}	&	$\ldots$	\\
190b$\dagger$	&	Bran 382/GN 12.10.3,	&	$\ldots$	&	$\ldots$	&	+\tnote{bMQ}	&	+\tnote{MR}	&	$\ldots$	&	$\ldots$	&	$\ldots$	\\
&	IRAS 12102-6133	&		&		&		&		&		&		&		\\
203a	&	IRAS 12175-6236,	&	$\ldots$	&	$\ldots$	&	+\tnote{bM}	&	+\tnote{S}	&	$\ldots$	&	$\ldots$	&	$\ldots$	\\
&	VVV CL012	&		&		&		&		&		&		&		\\
203d$\dagger$	&	Bran 386E/RCW 64	&	$\ldots$	&	$\ldots$	&	+\tnote{Q} ?\tnote{M}	&	+\tnote{R} ?\tnote{M}	&	+\tnote{T}	&	+\tnote{T}	&	$\ldots$	\\
&	 La Serena 31?	&		&		&		&		&		&		&		\\\hline
\insertTableNotes
\end{longtable}
\end{ThreePartTable}
\end{longrotatetable}

\subsection{Temperature and CO Abundance}\label{ssec:tvab}
\begin{figure}
\centering
	\includegraphics[width=\textwidth]{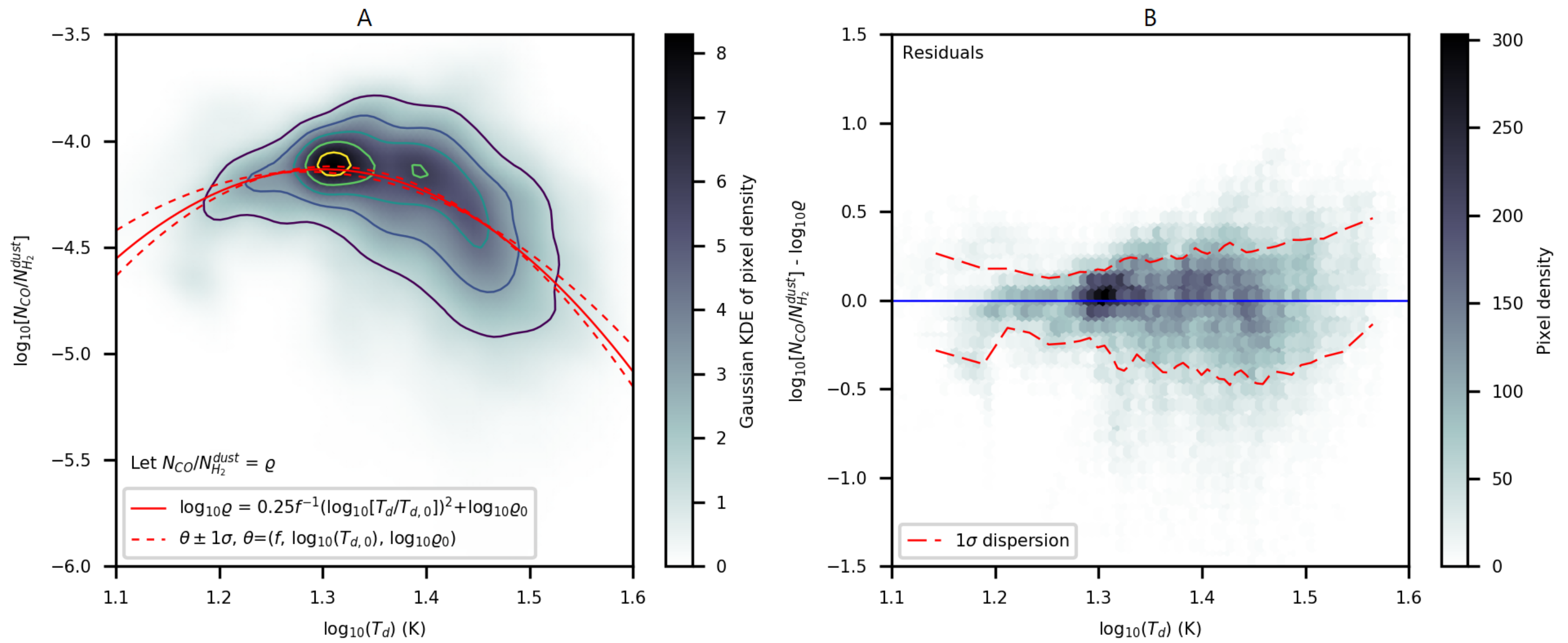}
    \caption{Log-log plot of \coab\, vs. \td\,, with fitted curve and residuals. A) Gaussian-kernel-smoothed pixel density plot of CO abundance (denoted $\varrho$) vs.\ $T_\mathrm{d}$, fitted empirically with a second-degree polynomial in log-space (red solid line). The dashed red lines show fits where the parameters are adjusted by 1$\sigma$ in their mutual posterior probability distributions as estimated by MCMC. B) Pixel density distribution of residuals in the fit of $\varrho$ vs.\, $T_\mathrm{d}$, with the $1\sigma$ dispersion around the binned average indicated by red dashed lines.}
    \label{fig:abvt}
\end{figure}

Useful as the morphotype scheme is, one of the real strengths of the full CHaMP sample is that the number of pixels and the variety of environments is sufficient to start quantifying the correlation between \td\, and \coab\, directly, for the first time outside of the laboratory. Plotted as a two-dimensional histogram, the fragments first shown in Figure~18 of \citealt{pittsmn} merge into the distinct concave-down trend shown in panel (a) of Figure~\ref{fig:abvt}. Panel (a) of Figure~\ref{fig:abvt} is the Gaussian kernel-density smoothed distribution of the CO abundance versus dust temperature in log-log space, which we fit empirically with a parabola in log$T_{\text{d}}$ so that the parameters of the fit could be sanity-checked by eye. Kernel density smoothing was required to make the fit converge at the highest density of points. The residuals of the fit are shown in panel (b), with the 1-$\sigma$ dispersion marked by red dashed lines. Higher-order polynomials did not improve the fit near the limits the temperature range that our data probe. For brevity, hereafter we use $\varrho$ to denote the \emph{fitted} \coab\, ratio as a function of $T_{\text{d}}$, which has the form
\begin{equation}\label{eq:coabvt}
    \mathrm{log}_{10}\varrho = \mathrm{log}_{10}\left[\frac{N_{\mathrm{CO}}}{N^{\mathrm{dust}}_{\mathrm{H}_2}}\right] = 0.25f^{-1}\mathrm{log}_{10}^2\left(\frac{T_d}{T_{d,0}}\right) +\mathrm{log}_{10}\varrho_0
\end{equation} 
in the log. The coefficients $f$, log$_{10}T_{d,0}$, and log$_{10}\varrho_0$ respectively represent the focus of the parabola, the log of the $T_{\text{d}}$ where \coab\, is maximized, and the log of the maximal \coab\, ratio. We find that.
$0.25f^{-1}=-10.0^{+0.8}_{-2.4}$, log$_{10}T_{d,0}=1.30^{+0.01}_{-0.02}$~K ($20.0^{+0.4}_{-1.0}$~K), and log$_{10}\varrho_0=-4.13^{+0.01}_{-0.02}$ ($7.4^{+0.2}_{-0.3}\times10^{-5}$ per H$_2$).  

\begin{figure}
\centering
	\includegraphics[width=\textwidth]{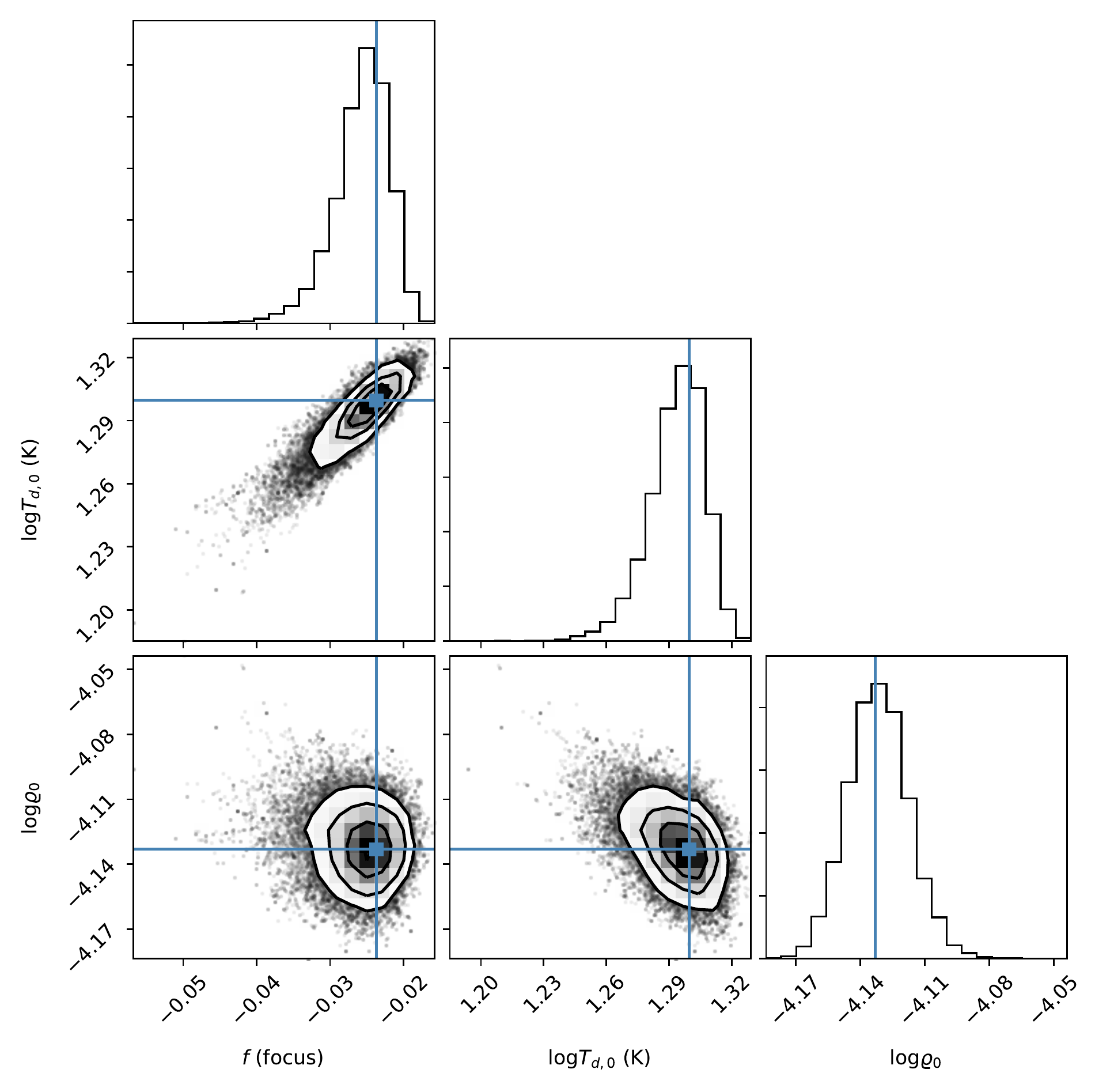}
    \caption{Corner plot of the posterior distributions of the variables in equation~\ref{eq:coabvt}. It is unclear if the focus parameter $f$ has any physical significance, and if so, what it might be.}
    \label{fig:cornabvt}
\end{figure}

Since we fitted a kernel-smoothed resampling of the data rather than the data themselves, simple curve-fitting routines did not provide usable error estimates. 
The best way to estimate the uncertainty was to minimize a negative log-likelihood function, and use the Python MCMC package \texttt{emcee} \citep{emcee} to estimate the posterior distributions of $f$ (the parabolic focus, with unclear physical significance), log$_{10}T_{d,0}$, and log$_{10}\varrho_0$. A Gaussian noise term, $\epsilon$, was also included as a stand-in for unquantified factors expected to influence the dispersion of the data, chief among them the local ISRF. However, the fitted value was $10^{-12}$, so it was removed with no effect on the other parameters. The posterior distributions are plotted in Figure~\ref{fig:cornabvt}. We used uniform priors with the following loose restrictions based on what was apparent from the plot: $-10<f<0.0$, $0.5<$log$_{10}T_{d,0}<2.0$ (that is, $T_{d,0}$ must be between 3~K and 100~K), and $-10.0<$log$_{10}\varrho_0<0.0$. The log-likelihood function used assumes the posterior distributions are close enough to Gaussian that the estimated closed form of the likelihood function for a skew-normal distribution would not have offered a significant improvement. The dispersion in CO abundance as a function of $T_{\text{d}}$ is about 0.3 dex over most of the data, with the smallest dispersion between about 15 and 20~K, and the largest between 25 and 30~K.

To be clear, Equation~\eqref{eq:coabvt} is a purely empirical model intended for observational astronomers with large data sets and limited need to delve into astrochemistry. The benefit of this approach lies in the simplicity of the method: the model is sufficiently simple and geometric that, provided there are enough data on the plot to form a coherent structure (we find the trend requires at least half of all the CHaMP data or a few~$\times10^4$ pixels to see easily), the parameters of the fit can be estimated to within a factor of a few by eye for easy initialization. The most obvious missing piece right now is the strength of photodissociating interstellar radiation, the FUV part of the ISRF. The greatest Region-to-Region variability is almost always in the high-temperature side of the distribution, where trends for smaller individual regions suggest the relative symmetry with the low-temperature side may be due to averaging over many much steeper trends. 

One popular method of estimating the FUV field strength from FIR/submm data is to follow \citealt{kramer} in assuming that nearly all FIR emission is reprocessed FUV emission, and so the ISRF is proportional to the integral of the SED. There are two problems with this approach as it pertains to our study. First, it assumes that internal heating is negligible---not a terrible assumption when close to 90$\%$ of prestellar clumps are effectively quiescent (at this resolution) and as likely to be dispersing as accreting \citep{champ4}---but where that assumption fails, the internally-heated gas falls in the part of the \coab-\td\, plot where the ISRF is most needed to explain the dispersion. Second and more importantly, much like $L/M$ (see discussion in \citetalias{pittsmn}, \S4.2), the ISRF field computed this way is not an independent quantity. All of the variables used to compute it are already in Equation~\eqref{eq:coabvt} in some form.

Comparison of this temperature-abundance distribution to laboratory measures of ion current through CO (proportional to the partial pressure or column density of CO in the gas phase) versus substrate temperature reveals several possible features of astrophysical interest. First, as detailed previously in \citealt{pittsmn}, laboratory experiments show that the amount of gas-phase CO peaks for substrate temperatures between 27 and 30~K in the absence of an externally-applied UV field \citep{oberg09,noble12,munoz10,munoz16}. As one might expect for a range of environments with nontrivial ISRFs, in Figure~\ref{fig:abvt}, the peak abundance in CO occurs at a substantially lower temperature, 20.0$^{+0.4}_{-0.9}$~K. Twenty Kelvin also looks at first glance like a sort of preferred temperature, but as most of this concentration is occurring in just three Regions, we are skeptical of the significance. Next, we refer readers to Figure~10 in \citealt{munoz10} and Figure~9 in \citealt{cazax17} to visualize how the gas-phase CO concentration varies with substrate temperature (the temperature-programmed desorption or TPD diagrams). These figures show that the distribution has a sharp peak in CO concentration between 27 and 30~K atop a much broader, lower distribution---except where the figure from \citealt{cazax17} shows that the sharper desorption peak can nearly disappear into the broader distribution if the CO ice is deposited at relatively high ($\sim$27~K) temperatures. On the other hand, the figure from \citealt{munoz10} shows that a lower substrate temperature upon CO ice deposition allows thermal CO desorption to start at lower temperatures, with escaping H$_2$ further encouraging CO desorption to begin around 15~K rather than somewhere between 20 and 25~K. If the physical quantities in Figure~\ref{fig:abvt} are in fact comparable to the quantities plotted in these two publications, the shape of the \coab-\td\, curve is a surprisingly good qualitative reflection of how one might expect natural CO ice to be be deposited, and later sublimate, given the temperature-density anticorrelation discussed last chapter. The quick rise with flat or downward concavity in CO abundance between 15 and 20~K suggests that the surface layers of CO ice in depleted clumps are weakly bound and infused with trapped H$_2$. The rest of the temperature-abundance distribution lacks the sharp peak near 30~K indicative of CO ice deposited either at fixed temperatures much below 27~K, or decreasing temperatures from inside out as in \citealt{cazax17}, Figure~8. However, this peak seems unlikely to be observed in nature regardless of \td\, at the point of CO ice deposition---it appears to be an artifact of the lack of UV photons or cosmic rays that would stochastically push some of the CO molecules over the desorption activation barrier before the substrate (dust) temperature rises to meet the activation barrier, releasing all but the deepest monolayer(s) of CO in quick succession. All of this suggests that the gas-phase CO abundance should fall toward the center of a prestellar clump, and rise again as embedded protostars form.

Despite the apparent goodness of fit in Figure~\ref{fig:abvt}, the parameter log$_{10}\varrho_0$ puts a floor under \ncol\, derived from CO. Equation~\eqref{eq:coabvt} is well suited for data in the regime where CO is depleted due to freeze-out, but it is not recommended it for use in tandem with the power-law \ico-\ncol\, relationship in \citealt{champ4} in gas with \ncol$\lesssim3\times10^{25}$ \psqm. Its greatest utility will likely be in giving observers in the submm-to-mm range a new method to estimate dust temperatures when observations sampling the peak of the SED are not available.

\subsection{Rethinking the X\texorpdfstring{$_\textrm{CO}$}{CO}-Factor}
Laboratory studies and simulations of how strongly the CO abundance varies as a function of temperature and radiation field strength suggest it should vary by upwards of three orders of magnitude (see \S~\ref{sec:intro}). Moreover, regional red- and blue-shifts of the $^{12}$CO line relative to the $^{13}$CO, C$^{18}$O, and HCO$^+$ lines observed in Papers III and IV suggest that the $^{12}$CO line is so universally opaque that it merely traces the surface of last scattering, rather than the encoding the optical depth through the whole cloud in the line wings. That means physical conditions derived from $^{12}$CO observations may not be consistent with $^{13}$CO or C$^{18}$O or applicable to the cloud interior. At least insofar as H$_2$ column density is concerned, optically thin thermal dust emission offers the chance to directly calculate an \xco-factor expression that compensates for the tendency of low-excitation $^{12}$CO lines to saturate or become optically thick.

\subsubsection{Analysis}
\begin{figure}
\centering
	\includegraphics[width=\textwidth]{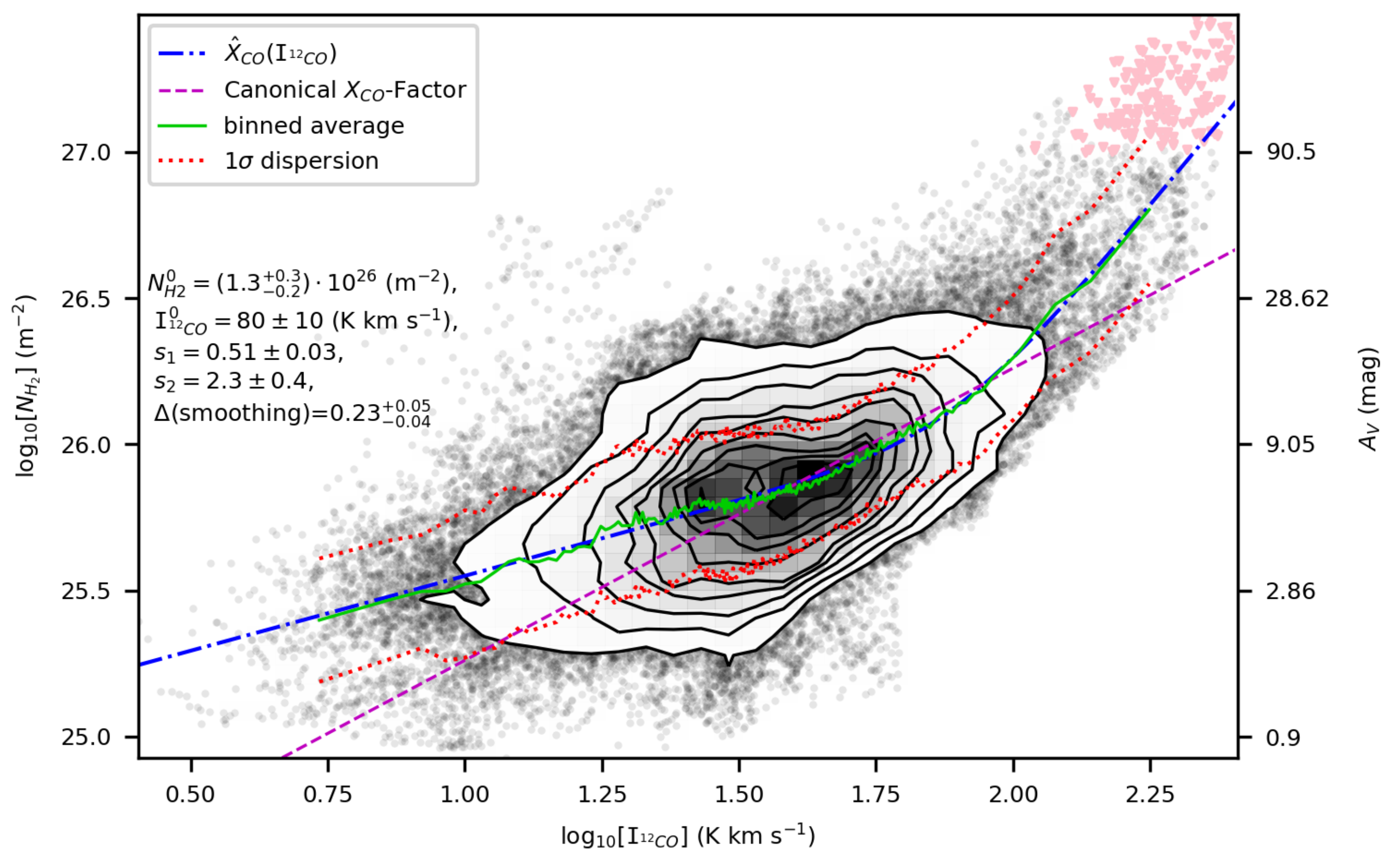} 
    \caption{Log-log 2D histogram of \ncol\ as a function of \ico, the ratio of which is the \xco-factor. Overlaid are the binned average (solid green line), 1-$\sigma$ dispersion (dotted red lines), classical \xco-factor (dashed magenta line), and the fit to the Equation~\eqref{eq:sbplaw} (dash-dotted blue line). To distinguish our fitted model from the classical constant \xco, we denote the model \protect{$\hat{X}_{CO}$(\ico)}. Pink downward triangles are points from a saturated patch of Region 13 where column densities are likely to be upper limits.} 
    \label{fig:xco}
\end{figure}

After computing pixel-by-pixel maps of \ncol\, from dust SED-fitting (hereafter denoted \ncold\, to distinguish it from \ncol\, calculated using CO) as per \citep{pittsmn}, we simply aggregated all the data with naive uncertainties of 15$\%$ or less and plotted them against \ico\ as shown in Figure~\ref{fig:xco}. The pink triangles are points from BYF~128b and 126a where saturation at 250~$\mu$m made the SEDs poorly sampled. Ordinarily the slope of the three SPIRE data points drives the fitted SED to slightly overshoot the 160~\micron\ data and slightly undershoot the 70~\micron\ data due to non-thermal contaminant emission in the latter filter; without the 250~\micron\ point, the fitting routine tends to down-weight the 70~\micron\ data further and make the SED drop more steeply at 160~\micron, nudging the amplitude of the curve higher while lowering the fitted temperature. The net effect is an increase in the reported \ncol; hence the data marked by pink triangles down are upper limits.

To this distribution we fit a broken power law (dash-dotted blue line in Figure~\ref{fig:xco}, labeled $\hat{X}_{CO}(I^{0}_{^{12}CO})$) with a smoothed transition of the form
\begin{equation}\label{eq:sbplaw}
N_{\mathrm{H}_2} = N^0_{\mathrm{H}_2}\mathtt{I}_{\star}^{s_1}\left[\frac{1}{2}+\frac{\mathtt{I}_{\star}^{1/\Delta}}{2}\right]^{\Delta(s_2-s_1)}
\end{equation} where
\begin{equation}\label{eq:istar}
\mathtt{I}_{\star} = \frac{\mathtt{I}_{^{12}\mathrm{CO}}}{\mathtt{I}^0_{^{12}\mathrm{CO}}}
\end{equation}
is the integrated CO line intensity divided by the transition point (indicated by the naught subscripts) between power laws in \ico, $N^0_{\text{H}_2}$ is the H$_2$ column density at $I^0_{^{12}\text{CO}}$, and $s_1$ and $s_2$ are the power law exponents to the left and the right of the transition point, respectively. $\Delta$ is a smoothing parameter defined such that, if $x$ and $y$ denote the abscissa and ordinate respectively, then $\Delta\approx0.5\mathrm{log}_{10}(x_2-x_1)$, where $y(x\lesssim x_1)\propto x^{s_1}$ and $y(x\gtrsim x_1)\propto x^{s_2}$. In other words, $\Delta$ describes the half-width of the data range, in log-space, inside of which neither power law component dominates enough to ignore the other. The equations and definitions above are essentially the same as the Astropy modeling module used to fit them \texttt{SmoothlyBrokenPowerLaw1D} \citep{astropy1,astropy2}, but in all of our equations and figures, we change the signs of the power law exponents to be more intuitive. Visually, the goodness of fit can be judged by comparing the fit (again, the dash-dotted blue line) with the binned average (solid green line). The binned average and the 1-$\sigma$ dispersion around it (dotted red lines) were  computed such that each bin has $\lceil2n^{2/5}\rceil$ data points (assumed equiprobable), where $n$ is the total number of points, and the bins are positioned so that the leftover data points are roughly equally distributed between each end of the distribution. Equation~\eqref{eq:sbplaw} is technically not an expression of \xco---to get an expression for \xco\, in the usual units, the entire equation must be divided by \ico. For comparison of the power law exponents to those of other studies, it is enough to simply subtract 1 from $s_1$ and $s_2$. we fit the expression as above because when we tried to fit \ncol/\ico\, directly, $s_2$ was consistently much too small.

\begin{figure}
\centering
	\includegraphics[width=\textwidth]{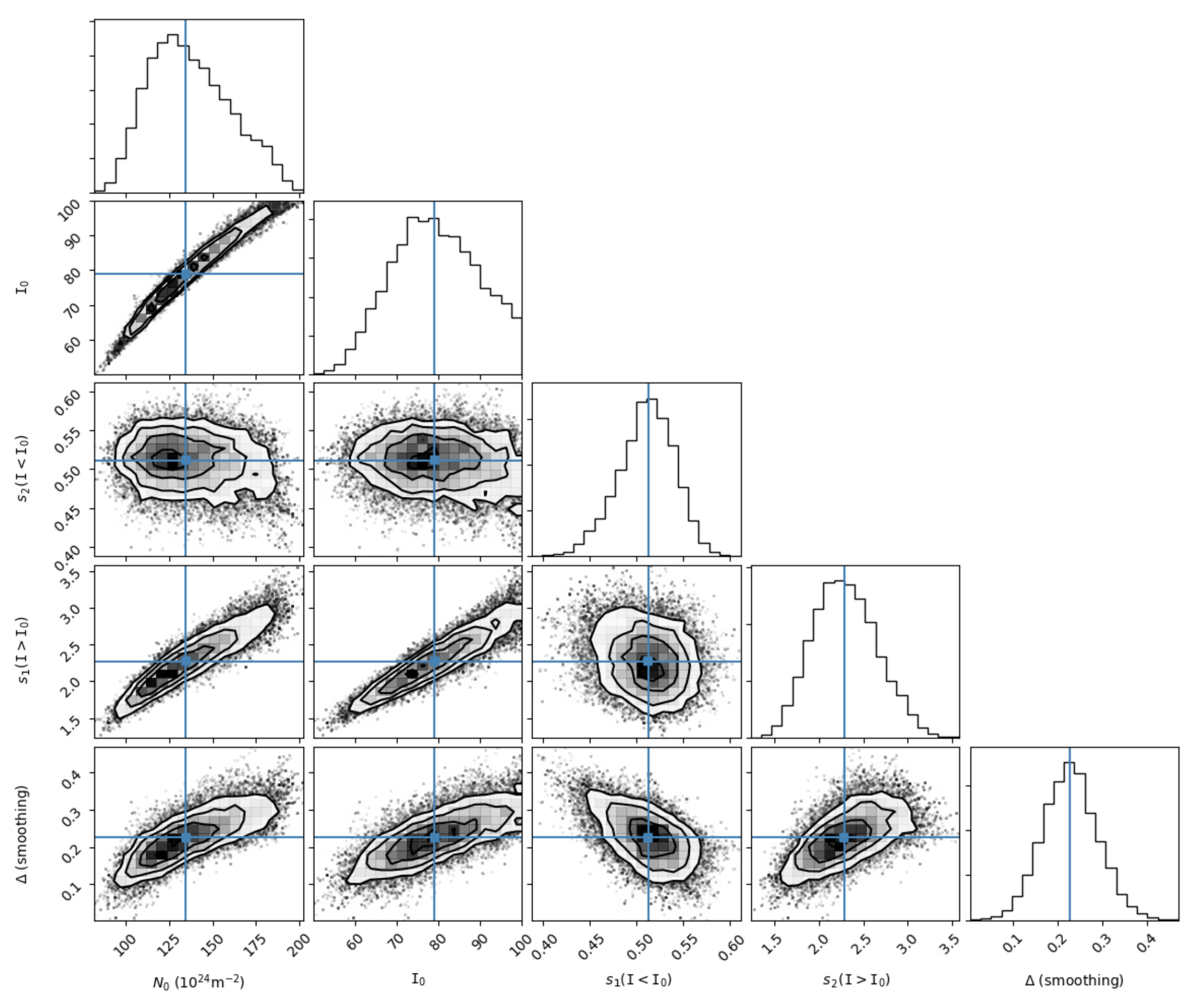}
    \caption{Posterior distributions of the free parameters of Equations \eqref{eq:sbplaw} and \eqref{eq:istar}.}
    \label{fig:cornx}
\end{figure}

As with the \coab-\td\, relation, we minimized the negative log-likelihood function over the parameters of Equation~\eqref{eq:sbplaw}, assuming Gaussian uncertainties and flat priors, and produced the posterior distributions shown in Figure~\ref{fig:cornx} using \texttt{emcee}. We find that $N_{H2}^{0}=(1.3^{+0.3}_{-0.2})\times10^{26}$ (m$^{-2}$), $I^{0}_{^{12}CO}=80\pm10$ (K km s$^{-1}$), $s_1=0.51\pm0.03$, $s_2=2.3\pm0.4$, and $\Delta=0.23^{+0.05}_{-0.04}$. We used inverse-$\sigma$ weighting in the minimization routine, as opposed to the more typical inverse-variance weighting used in the fit of the \coab-\td\, relationship, because the points far from the mean can still be highly significant. It is also worth noting that the range of variation in intrinsic clump size is larger than the range of distances, so resolution is not expected to play a major role in the shape of the distribution overall.

\begin{figure}
\centering
	\includegraphics[width=0.8\textwidth]{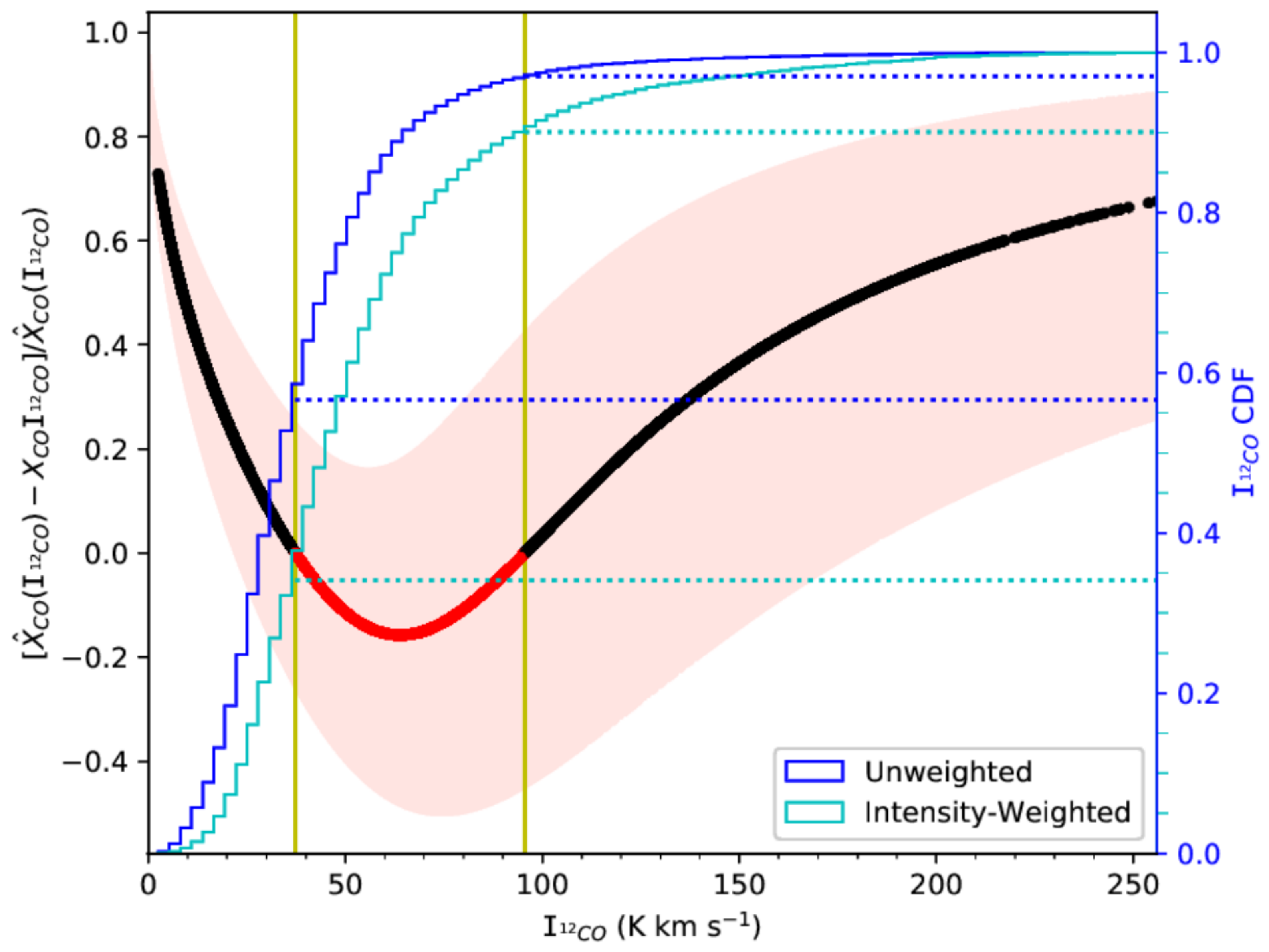}
    \caption{Fractional difference in derived \ncol\, using the constant \xco-factor and our smoothed broken power law, $\hat{X}_{CO}$(\ico). The blue and cyan stepped histograms are the unweighted and \ico-weighted CDFs of \ico\, to give an idea of what fraction of \ncol\, calculations will be over- or under-estimated by using the constant \xco-factor, and at what ranges of of \ico. The pink filled range spans the 1-$\sigma$ variation in all of the fitted parameters of $\hat{X}_{CO}$. The vertical gold lines bracket the range of \ico\, where the constant \xco-factor overstates \ncol, and the dotted blue and cyan lines help show what fraction of pixels fall in that range.}
    \label{fig:xhatcdf}
\end{figure}

Figures~\ref{fig:xco} and \ref{fig:xhatcdf} make abundantly clear that the relationship between \ico\, and \ncol\, is not linear, though there is a fairly wide range of intermediate \ico\, values where \ncol\, calculated with a constant \xco-factor are well within the 1-$\sigma$ dispersion. Note the exquisite agreement of the fit of Equation~\eqref{eq:sbplaw} with the binned average along the latter's entire length in Figure~\ref{fig:xco}. 
To better show what is being missed by computing \ncol\ using a constant \xco-factor, \ref{fig:xhatcdf} shows the difference between that and \ncol\ derived using Equation~\eqref{eq:sbplaw}, normalized by the latter and plotted as a thick black and red curve. The pink filled area represents the combined uncertainty in the difference given the uncertainties of the five free parameters of Equation~\eqref{eq:sbplaw}. At the highest and lowest values of \ico, the traditional \xco-factor may miss upwards of 60$\%$ of the total mass in H$_2$, whereas for intermediate values of \ico, the constant \xco-factor may actually overestimate the mass by up to 20$\%$. On the same plot we have also included intensity-weighted (cyan) and equiprobable (blue) cumulative histograms of \ico\, on the secondary axis for all good pixels in the CHaMP sample, and have drawn dotted lines to benchmark the percentiles (for lack of a better word) against the difference curve on the primary axis so readers can see for what fractions of the data the traditional \xco-factor overstates and understates \ncol.

Sixty percent of all pixels in the CHaMP sample have \ncol\, greater than predicted by the standard \xco-factor. Of these 60$\%$ of pixels, more than 90$\%$ are in the $s_1=0.51\pm0.03$ regime, where the difference between using a constant \xco-factor and using Equation~\eqref{eq:sbplaw} is most statistically significant. The remaining $\lesssim10\%$ of pixels with \ncol\ greater than predicted by the standard \xco-factor are in the high-\ico\ $s_2=2.3\pm0.4$ regime of the broken power law. In the $s_2$ regime, where large increases in \ncol\ produce only small changes in \ico, optical depth increases proportionally with excitation temperature (\tex), which approaches \td\ at these column densities. 
The $s_1=0.51\pm0.03$ regime can only be a consequence of how radiative transfer works in moderately dense gas where CO is present but in a very low-excitation state. Where CO is sufficiently subthermal, its optical depth will rise as the excitation temperature falls, at least until the gas is too diffuse to effectively shield the CO from destruction \citep[see also, e.g.][]{penaloza17}. If CO dissociation was the main cause of deviation from the standard \xco-factor, one would expect that \ico\ would drop off with falling column density super-linearly instead of sub-linearly (indeed, as we discuss in \S\ref{ssec:lit}, this is a predicted third regime that occurs below the column densities we chose to consider). To be fair, observations of low-excitation, low-opacity CO will always be subject to sensitivity limits, which bias observations of low-excitation CO toward areas where the CO is still optically thick. However, per \citetalias{champ4}, CO observations should be complete down to 10~\kkms, and we were aggressive in our continuum background subtraction to limit inclusion of incompletely-sampled diffuse gas.

That the form of the \ico-\ncol\, relationship has (at least) two power-law regimes with a transition region meshes well with what \citep{champ4} found and similar analysis of ThrUMMS data (Barnes et al. in prep.) with respect to the \ico-\nco\, relationship. The range of physical conditions in which $^{12}$CO is optically thin is narrow: the only CHaMP data with $\tau_{^{12}\text{CO}}\lesssim1$ had 1$\lesssim N_{\text{CO}}\lesssim2\times10^{20}$~\psqm, and the minimum reliable $\tau_{^{12}\text{CO}}$ is about 4 \citep{champ4}. Much less than 10$\%$ of the CHaMP sample has \nco\, in that range; $^{13}$CO and sometimes C$^{18}$O were needed to recover \nco\, in the overwhelming majority of pixels. Most of the data follow two trends in \nco\, vs \ico\, depending on the $^{12}$CO excitation temperature, denoted $T_{\text{ex}}$. In the strongly subthermal regime (typically with $T_{\text{ex}}<10$~K), as $T_{\text{ex}}$ falls toward the $^{12}$CO brightness temperature minus the temperature of the CMB, the optical depth spikes and \nco\, becomes nearly independent of \ico\, (equivalently, \ncol$\propto$\xco$^s$ where $s$ goes to -1 for low \tex). For excitation temperatures well in excess of about 10~K, \nco\, approaches an \ico$^2$ dependence. The similarity of both power law exponents in Equation~\eqref{eq:sbplaw} to those of the velocity-resolved \nco\, vs \ico\, plots in \citealt{champ4}, within their mutual margins of uncertainty, seems unlikely to be coincidence, and will be explored in future work by this collaboration. 

Our findings are of imminent import to the question of how much molecular gas in a cloud is CO-dark, only accountable via other tracers like dust. CO-dark gas is thought to be relatively diffuse gas in the envelopes of molecular clouds where gas densities are high enough for H$_2$ to self-shield, but not high enough for CO to be shielded \citep{blitz,wolf}. As mentioned in the introduction, estimates of the CO-dark gas fraction vary wildly from 20$\%$ to upwards of 70$\%$ \citep{pine,langer}. Unfortunately, to truly put this consideration to the test, observers must be able to resolve the width of a CO-dark envelope, not just the width of a molecular cloud. This has been done in Taurus \citep{xu16}. However, without the benefit of an accurate and appropriately varying CO abundance, the equation for the dark gas fraction is dubious. 

The implications of this and similar studies go far beyond the obvious effects on measurements of CO-dark gas and star formation efficiency (SFE). The density of a parcel of molecular gas is the chief determinant of any physical or chemical evolutionary timescale. Take for example the rate at which either an individual molecular cloud or an entire galactic disk is converted into stars---intuitively, the rate of H$_2$ consumption is determined by how fast the gas collapses, which, for an extended mass like a cloud, must be dictated by the gas density. The Kennicutt-Schmidt (KS-) Law \citep{schmidt59,schmidt63,kenn98} quantifies these expectations---the eponymous studies and their successors show that the star formation rate (SFR) per unit area is proportional to a slightly super-linear power law in H$_2$ column density. These studies rarely include other isotopologues of CO. If the \xco\ factor is actually a broken power law, that could change the exponent of the KS-Law or even its entire functional form. It would be interesting to see if the recent case made in \citealt{kenn2020} for a bimodal or broken KS-Law could be tied to the broken power-law X-factor in the sense that the high-density, high-\ico\ gas is analogous to the bulk of the CO-bright gas in starburst galaxies. However, that is beyond the scope of this paper. 

\subsubsection{Comparison with Literature}\label{ssec:lit}
\begin{figure}[htbp]
	\includegraphics[width=\textwidth]{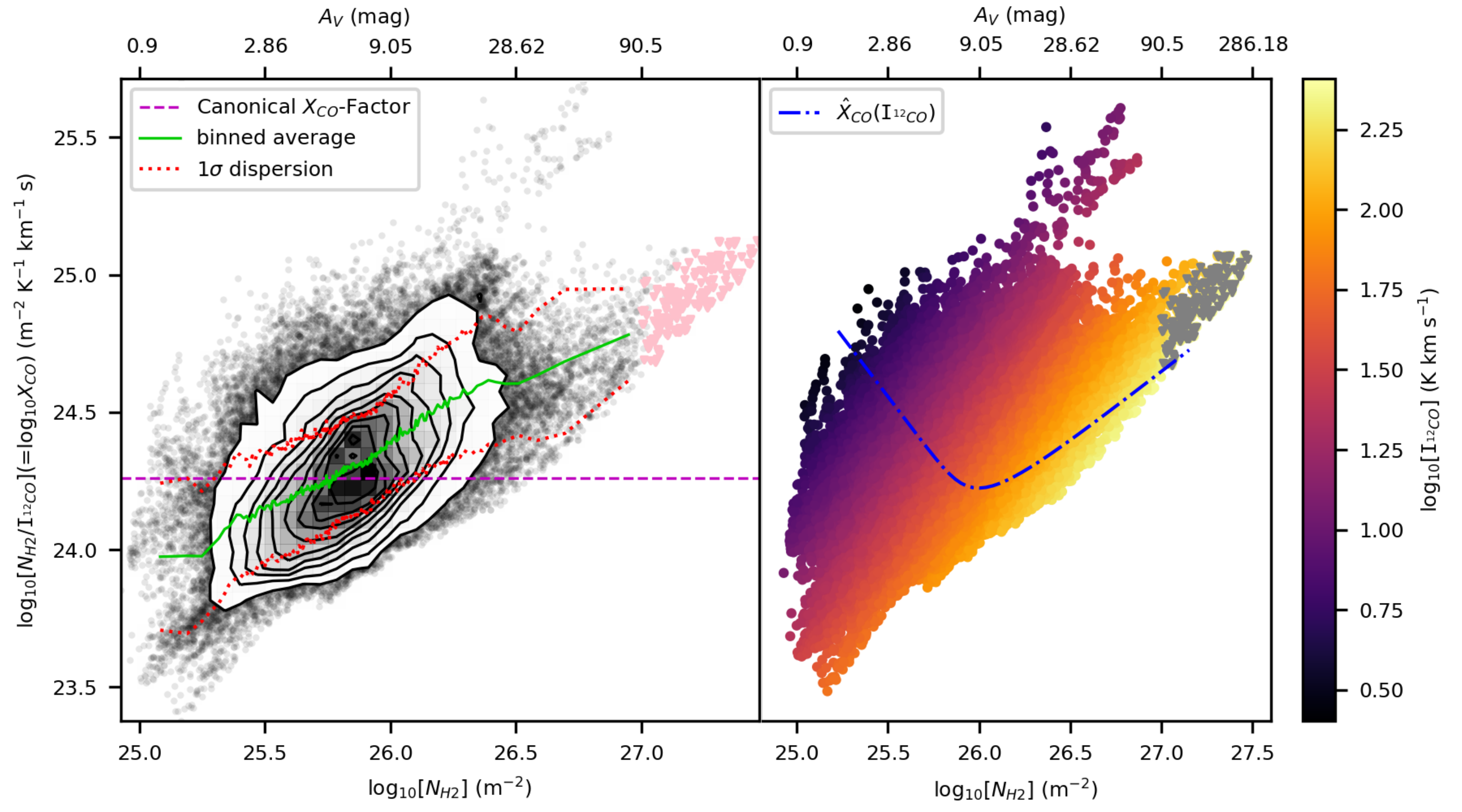}
    \caption{Plots of the \xco-factor as a function of \ncol\ for comparison to synthetic models, one color-coded by \ico. The left panel shows a 2D histogram of the CHaMP data overlaid with the binned average (solid green line), 1-$\sigma$ dispersion (dotted red lines), and classical \xco-factor (dashed magenta line), so the slope can be measured by eye. The right-hand panel shows the CHaMP data again as a scatter plot color-coded by \ico, with the $\hat{X}_{CO}$ curve overlaid as a dash-dotted blue line, to show how orientating and normalizing the data this way obscures the \ico\, dependence of \xco. The blue $\hat{X}_{CO}$ curve was generated by dividing the ordinate (\ncol) in Figure~\ref{fig:xco} by the abscissa (\ico), and plotting the quotient ($\hat{X}_{CO}$) as a function of the ordinate.} 
    \label{fig:xcoshet}
\end{figure}

\citealt{shetty11a} and \citealt{shetty11b} collectively describe perhaps the most thorough investigation of the \xco-factor in synthetic Galactic molecular clouds, incorporating MHD simulations with non-LTE radiative transfer and the non-LTE chemical modeling described in \citealt{glover10} and \citealt{glover11}. This series focuses on a range of densities that top out in the middle of the range that CHaMP probes, about 10$^{26}$~\psqm\, and extend down to column densities about two orders of magnitude lower than any of the CHaMP data that we trust. Where the CHaMP data overlap in \ncol\, (or $A_V$) with the models in \citealt{shetty11a}, the model initialized with a H$_2$ volume density $n\geq3\times10^6$~m$^{-3}$ had roughly the same slope in the log for \ncol$\gtrsim10^{26}$~\psqm\, as the running mean in the left panel of Figure~\ref{fig:xcoshet}: 1.2 to our 1.3. The normalization $N_{H2}^{0}/I^{0}_{^{12}CO}$ (the minimum in $\hat{X}_{CO}$(\ico) and also the power-law transition point) from our model is within about 0.1 dex of the inflection point of their minimum in \xco, as well. In the right-hand pane of Figure~\ref{fig:xcoshet}, where we color-code the data by \ico\, and overlay the model $\hat{X}_{CO}$(\ico), it becomes apparent that the data distribution is hiding the turnover to the regime where the slope of $\hat{X}_{CO}$(\ico) trends negative. Plotting the \xco-factor as a function of \ncol\ is natural for theoretical calculations, but it obscures a very real concern for observational astronomers that the right-hand panel of Figure~\ref{fig:xcoshet} is meant to demonstrate. That is, for most of the range of H$_2$ column densities one would want to probe, there is nearly a factor of 30 difference between the largest and smallest values of \ncol\ that could be derived from a given value of \ico. 

The agreement of our results with \citealt{shetty11a} and \citealt{shetty11b}, at least with the solar-metallicity model initialized $n\geq3\times10^6$~m$^{-3}$, is remarkable for several reasons, not least of which is that Equation~\eqref{eq:sbplaw} incorporates no other physics than the two LTE radiative transfer analyses required to do (1) the SED-fitting and (2) the multi-CO line fitting. Their models fix the ISRF at 1.7 Habing units, far less than what is expected to be impinging on the CHaMP clumps around the Carina Nebula, NGC 3576, NGC 3603, and a number of other H\textsc{ii} Regions in our sample. Their analysis of \xco\, as a function of CO abundance in \citealt{shetty11b} allows for photodissociation, but apparently not depletion; however, only one model occupies the parameter space above \ncol$\gtrsim3\times10^{26}$~\psqm, and they only explore the \xco-\coab\, relationship for \ncol$\leq3\times10^{26}$~\psqm. Still, it is worth noting that whether \coab\, was allowed to vary, fixed at 10$^{-4}$, or fixed at 10$^{-5}$, the normalization may have changed but the slope of log\xco\, versus log\ncol\, stayed roughly the same for \ncol$\gtrsim3\times10^{25}$~\psqm\, ($A_V\sim3^m$). That suggests the change of power-law exponent is an effect of shielding, and that while the normalization may change, the $s_2$ regime of Equation~\eqref{eq:sbplaw} is robust to more than an order of magnitude variation in CO abundance. For the Milky Way-type galaxies at least, then, it should be generally safe to assume \ncol\ is approximately proportional to the square of \ico\ where $^{12}$CO is bright (a few $\times10$~\kkms), and other CO isotopologues are not available. The slope of the $s_1$ regime, by contrast, appears highly sensitive to the CO abundance, whose characterization is aided by dust temperature, but is ultimately incomplete without understanding the ISRF dependence. The effects of sensitivity limits are also under investigation (V$\acute{\textrm{a}}$zquez-Semadeni et al., private communication), but our sampling should be complete for \ico$\gtrsim10$~\kkms.

The Shetty et al. studies did not cover the effects of varying the ISRF on the \xco-factor or the effects of CO line velocity width. These issues were tackled in \citealt{feld121}, who used essentially the same underlying chemical models and simulation techniques to expand on the works by the Shetty et al. series above. \citealt{feld121} found that at large \ncol\ ($A_V$), the \xco-factor becomes directly proportional to \ncol\, if the CO line width is constant, or proportional to the square root of \ncol\, if the CO line width if the cloud is bound and in virial equilibrium. A look at the slope of the blue dot-dashed $\hat{X}_{CO}$(\ico) line in Figure~\ref{fig:xcoshet} (right) suggests that a virial distribution of CO line velocity widths more accurately describes reality: over the range $10^m<A_V<100^m$, $\hat{X}_{CO}$(\ico) increases by about half a dex. However, the trends in the data and the scatter essentially rotate 90$^{\circ}$ in this plot just shy of $A_V\sim10^m$ or log[\ncol]$\sim$26 \psqm. The scatter around the low-\ncol\, side of the function spans almost the entire range of \ncol\, and is parallel to the trend on the high \ncol\, side---the rise in the \xco-factor toward low column densities would have gone unseen had we not modeled \xco\, as a function \ico\, instead of \ncol. Not only is the anti-correlation between the \xco-factor and \ncol\, at low column densities real, the enormous scatter appears to be reproduced in the \citealt{feld121} study with a variety if ISRF strengths ranging from 0.1 to 100~$G_0$ in \citealt{habing68} units. However, after they divided out the metallicity and ISRF dependencies, we observed some statistically significant differences between their plot of \xco\, vs. \ico\, \citep[][Figure 4, bottom right]{feld121} and ours (Figure~\ref{fig:xco}, subtracting one from both exponents). Their power-law transition point is closer to 30 \kkms, compared to our 80$\pm10$~\kkms, and while their high-\ico\ power-law exponent ($s_2$) is within 1$\sigma$ of ours, their low-\ico\ power-law exponent ($s_1$) is almost twice as steeply negative as ours. It seems highly likely that these differences are explainable by the fact that we did not have ISRF information to divide out: averaging over several steep trends with varying vertical-axis intercepts will tend to flatten the slope. Moreover, as the \citealt{feld121} study makes clear, stronger ambient UV irradiation pushes the power-law break point to higher \ncol\, and larger \ico. Therefore, while the differences between our model and theirs are statistically significant, the findings of our studies seem physically consistent.

Previous observational studies of \xco-factor variance have mostly relied on dust extinction and heavily favored nearby, lower-mass star forming regions, like the Pipe Nebula \citep{lombardi06}, Perseus molecular cloud \citep{pine08, lee14, tafalla21}, the Taurus molecular cloud \citep{pine10}, and the California Molecular Cloud \citep{kongco}. Even including similar analysis of the Orion molecular cloud complex by \citealt{ripple}, none of these studies probe \ncol\ values even half as high as the CHaMP sample. Still, only data for highly-irradiated portions of Orion and the California MC fail to show the \ico-\ncol\ relationship entering the equivalent of our fit's $s_2$ regime, i.e. \ico\ becoming less sensitive to increasing $A_V$ starting between 1$^m$ and 10$^m$. Moreover, \citealt{lee14} and \citealt{ripple} also find that, as with our $s_1$ regime, the trend at low \ncol\ ($A_V$ of 1 to a few magnitudes) is for \ico\ to be more sensitive to changes in \ncol\ than the canonical \xco-factor would suggest, whereas \citealt{lombardi06} find the opposite to be true in the Pipe Nebula for $A_V\lesssim1$ and \ico$<40$~\kkms. In the Perseus cloud, both \citealt{pine08} and \citealt{lee14} find that all parts of the cloud enter the $s_2$ regime at $A_V\gtrsim4^m$ starting at $20<$\ico$<60$~\kkms. However, unlike \citealt{lee14}, $^{12}$CO line emission data from \citealt{pine08} appear to follow the classical \xco-factor exactly, suggesting optically thin emission. This seems to be due to their use of curve-of-growth analysis on $^{12}$CO to derive \nco\ and variations in the ratios of $^{13}$CO and $^{18}$CO to $^{12}$CO, rather than performing simultaneous radiative transfer on the three isotopologues assuming constant isotope ratios. In \citealt{kongco}, 11 out of 17 equal-sized segments of the California cloud have roughly the same measured \ico\ over the entire range of $A_V$ ($<5^m$ to $35^m$) with \ico\ hovering between 10 and 20 \kkms; two more segments closer to the active star forming part become optically thick in $^{12}$CO emission at $A_V$ between 5$^m$ and 10$^m$ (\ncol\ of 6--10$\times10^{26}$ \psqm), and \ico\, just shy of 40 \kkms; and the remaining four segments abutting the embedded cluster fan out in $A_V$-\ico\ space such that the standard \xco-factor might actually overestimate \ncol in a majority of pixels. Most notably, \citealt{kongco} find a slope $s_2\sim2.4$ among the quiescent parts of the California MC, well within the margin of uncertainty for our value of $s_2$.

Most of the qualitative discrepancies between our data distribution in \ico-\ncol-space and data from the aforementioned observational studies lie in the width and position of the transition zone, and the behavior of the data at low values of \ico\ and \ncol\ (extinction). Virtually all the aforementioned studies find narrower transition zones than ours, which is to be expected since our data set is aggregated over a larger area than the previous studies. Most of the other studies' data also enter the high-\ico, high-\ncol\ regime at lower \ncol than our data for high-mass prestellar clumps, also as expected given that high-mass star-formation tends to occur in clouds that are both denser and more strongly irradiated. 
Aside from \citealt{kongco} and \citealt{tafalla21}, the latter of whom fit each CO isotopologue separately, most of the studies lacked the statistics to fit slopes to their $s_2$ regimes, even if the data clearly suggested a turnover.  


Other attempts have been made by \citealt{okamoto2017} and \citealt{kalber2020} to combine H\textsc{i} data with CO and dust emission. Of these, the findings of  \citealt{okamoto2017} from the Perseus Cloud most closely approach our results: in the \ico\ regime that overlaps our data (\ico$<30$~\kkms), they too find that \xco\ is higher at lower \ico\ values (recall that dividing \ncol\ by \ico\ yields the same power-law form except all the exponents are reduced by 1; thus our derived exponent on the low-\ico\ side is $-0.49$). However, their data, taken as it was from a single giant molecular cloud at high resolution, shows significant substructure in the \ico-\xco\ plane for \ico$<15$~\kkms, and mostly plateaus for \ico$\gtrsim15$~\kkms\ where they indicate the CO data are optically thick. The substructure at \ico$<15$~\kkms\ and relative sparseness of data at \ico$\gtrsim15$~\kkms\ makes it difficult to determine if the apparent change of slope is significant. If one shifted our power-law \xco\ fit along the dashed magenta line representing the constant \xco-factor in Figure~\ref{fig:xco} until the slope transition starts around 15~\kkms, the fit would still be within the 1$\sigma$ dispersion, so we do not consider our results to be inconsistent with those of \citealt{okamoto2017} at this time.

While \citealt{kalber2020} mostly worked in a drastically different temperature regime, they bring up an alternative interpretation of the data that should be mentioned for the sake of completeness. They assumed constant ratios of CO to H$_2$ and $A_V$ to H\textsc{i}, and used CO, $A_V$, and H\textsc{i} to map the distribution of CO-dark gas and fit a temperature-dependent ``correction factor'' for the relationship between extinction and hydrogen column density. This correction factor is essentially a variable gas-to-dust ratio, which is the main alternative explanation for our CO abundance maps. To be fair, if CO is depleting onto grains, grain growth is occurring by definition, although increases in grain size do not necessarily imply a decrease in grain number density. However, we still believe that the increase in \xco\ at the highest densities is predominantly due to CO depletion, not a falling gas-to-dust ratio, because of how well changes in CO abundance with temperature track with those observed in the lab. Moreover, the distance to the Carina Tangent is such that Herschel images should not resolve the scales of individual protostellar disks, where dust settling and coagulation would occur.

\begin{figure*}
    \centering
    \includegraphics{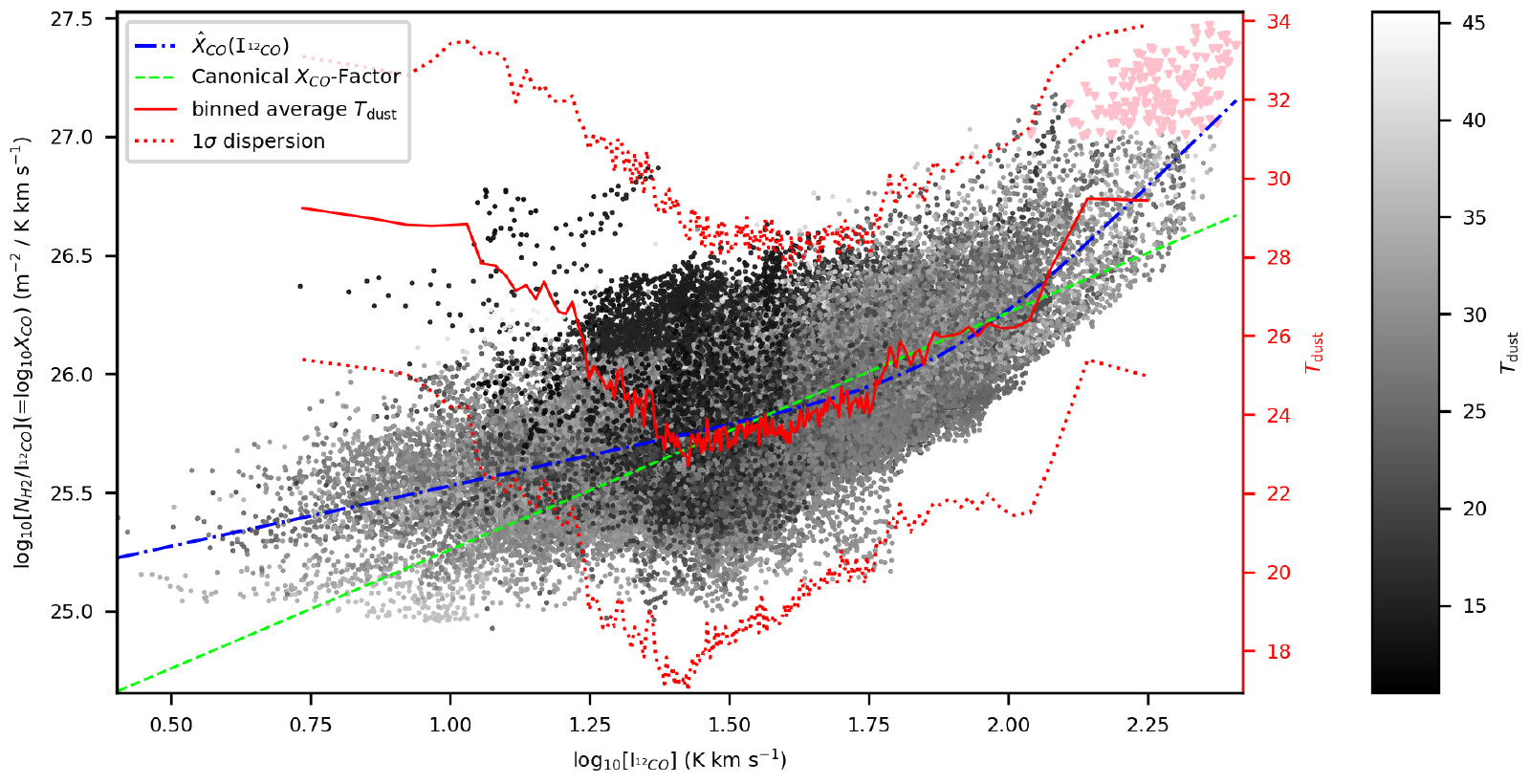}
    \caption{Similar to Figure~\ref{fig:xco}, except all individual points are shown value-coded by \td, and with the binned average \td\ and its $1\sigma$ dispersion plotted as solid and dotted red lines, respectively. The dashed line showing the canonical \xco-factor is recolored in green for visibility. Sampling is complete down to \ico$\;\sim10$~\kkms; data below that should be taken with caution.}
    \label{fig:xcotdist}
\end{figure*}

The effects of CO excitation were partly addressed in \citetalias{champ3} and \citealt{thrumms}, and will be addressed further in a forthcoming paper by Vasquez-Semadini and Barnes (in prep.). However, in light of a recent paper by \citealt{sofue20}, it seems worthwhile to clarify what we think is happening in regions where both \ncol\ and \ico\ are low, given the benefit of \td\ information. Shading the data in Figure~\ref{fig:xco} by \td\ yields the distribution shown in Figure~\ref{fig:xcotdist} (note that data may be incomplete for \ico$\;<10$~\kkms). The binned average temperature and its $1\sigma$ dispersion are plotted as red solid and dotted lines, respectively, against the second vertical axis. Note that the lowest average dust temperatures occur at middling values of \ico and moderately high values of \ncol, rising toward both higher and lower values of \ico. This seems to be a result of averaging over prestellar clumps that cool toward their centers and protostellar clumps that are both warmer toward their centers and systematically denser overall than their prestellar counterparts. The fact that \td\ rises toward lower values of \ico\ as well as higher values suggests the low \ico\ values can be explained by one of two possibilities: either CO is beginning to dissociate as gas becomes warmer and less dense, or the dust and excitation temperatures are decoupling and the CO becoming subthermal, which tends to raise the CO optical depth \citealt[see also the discussion in \S3.1.2 of][]{gong18}. In either case, CO should underestimate H$_2$, contrary to the interpretations of \citealt{sofue20}, and that is indeed what the slope of the low-\ico\ $s_1$ regime indicates. A more direct comparison of the dust and CO excitation temperatures is, however, outside the scope of this paper.

\section{Conclusions}
\label{sec:conc}
SED-fitting parameter maps for the complete CHaMP sample show that H$_2$ column-density maxima usually coincide with dust temperature minima, implying that density enhancement/collapse and cooling must be contemporaneous phases in the evolution of prestellar clumps. However, protostellar clumps and externally-irradiated clumps overlap prestellar clumps in the \ncol-\td\, plane. Eighty-four percent of clumps have anticorrelated \ncol\, and \td, and a few percent of clumps with positively-correlated \ncol\, and \td\, are observably being heated and compressed by external radiation sources. Catalog-matching is needed to determine whether these results are consistent with roughly $90\%$ of prestellar clumps being genuinely quiescent, or only effectively so at the resolution of our observations.

We compared \nco\ maps from \citetalias{champ4}, derived using multiple CO isotopologues, to \ncol\ derived from dust emission and published here, and mapped the computed CO abundance for the rest of the CHaMP sample not covered previously in \citetalias{pittsmn}. We find that not only is there no single CO to $H_2$ abundance ratio appropriate for all physical conditions possible in Galactic molecular clouds, but that the CO abundance is a strong function of dust temperature, in line with laboratory experiments. The relationship has a dispersion of a factor of a few, and the dispersion seems to increase with temperature, as expected given that warmer pixels are dominated by gas in outer clump envelopes subject to varying degrees of external irradiation. Our empirical temperature-abundance relationship is parabolic in log-space, with a peak CO abundance of $7.4^{+0.2}_{-0.3}\times10^{-5}$ per H$_2$ at $20.0^{+0.4}_{-1.0}$~K.

We show that the \xco-factor is significantly non-linear: the standard constant \xco-factor yields H$_2$ column densities outside the 1$\sigma$ dispersion of the CHaMP data for \ico$<10$~\kkms and \ico$>170$~\kkms, as well as for $A_V\lesssim2$ and $A_V\gtrsim28$. Our broken power-law fit to the \xco-factor has \ncol$\propto$\ico$^{0.51}$ for \ico$\;\lesssim70$~\kkms, \ncol$\propto$\ico$^{2.3}$ for \ico$\;\gtrsim90$~\kkms, and a smooth transition between. The larger of the two exponents is consistent with the simulations of \citealt{shetty11a} and \citealt{feld121}, but the smaller exponent is somewhat smaller than expected for a fixed UV field strength. However, the simulations of \citealt{feld121} show that the transition point is a function of local UV field strength, so averaging over many environments with a variety of UV field strengths is expected to broaden the transition region and reduce the exponent in the low-\ico, regime. Compared to the constant \xco-factor, the broken power-law ($\hat{X}_{CO}$(\ico) for short) predicts \ncol\, values as much as 20$\%$ lower for $37<$\ico$<95$~\kkms, but higher by up to 60$\%$ for \ico\, outside of this range. More than half of all CHaMP data pixels are in the low-\ico\ regime where \ncol$\propto$\ico$^{0.51}$ and the increase in predicted \ncol\, is most statistically significant.

\section*{Acknowledgements}
 The CHaMP project was funded by the National Aeronautics and Space Administration from 2015--19 through grant ADAP-NNX15AF64G, which we gratefully acknowledge. We would also like to thank Dr. Erik Deumens of the UF chemistry department for his guidance. This research makes use of the NASA/IPAC Infrared Science Archive, which is operated by the Jet Propulsion Laboratory, California Institute of Technology, under contract with the National Aeronautics and Space Administration. This research has also made extensive use of the SIMBAD database, operated at CDS, Strasbourg, France; as well as the Spanish Virtual Observatory (http://svo.cab.inta-csic.es) supported by the Spanish MICINN / MINECO through grants AyA2008-02156, AyA2011-24052.
\vspace{5mm}
\facilities{Herschel(PACS and SPIRE), APEX(LABOCA), Spitzer(MIPS), WISE, MSX, Mopra}

\bibliographystyle{aasjournal}
\typeout{}
\bibliography{ppr3bib}



\appendix

\section{Plots of Dust Temperature and H\texorpdfstring{$_2$}{2} Column Density}\label{sec:app0}
Here we present SED-fitting parameter maps for CHaMP Regions 1 through 8, 12, 16, 18, 21, and 23. The maps for Region 13 are presented in the text in \S~\ref{sec:meth}. Similar maps for Regions 9 through 11 and 26 can be found in \citealt{pittsmn}.
\begin{turnpage}
\clearpage
\begin{figure}
  \begin{center}
    \includegraphics[width=9in]{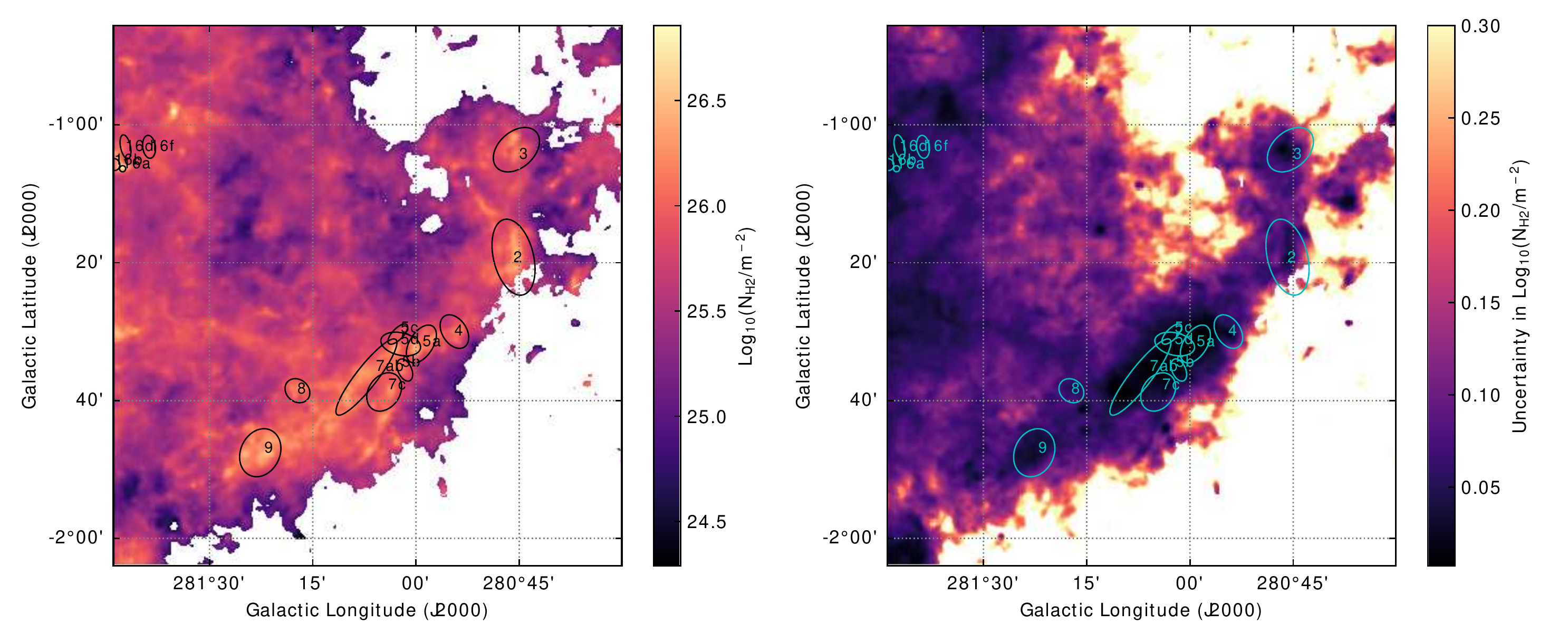}
    \caption{Region 1 \ncol\, (left) and \ncol\, uncertainty (right) maps.}
  \end{center}
\end{figure}
\pagebreak
\begin{figure}
  \begin{center}
    \includegraphics[width=9in]{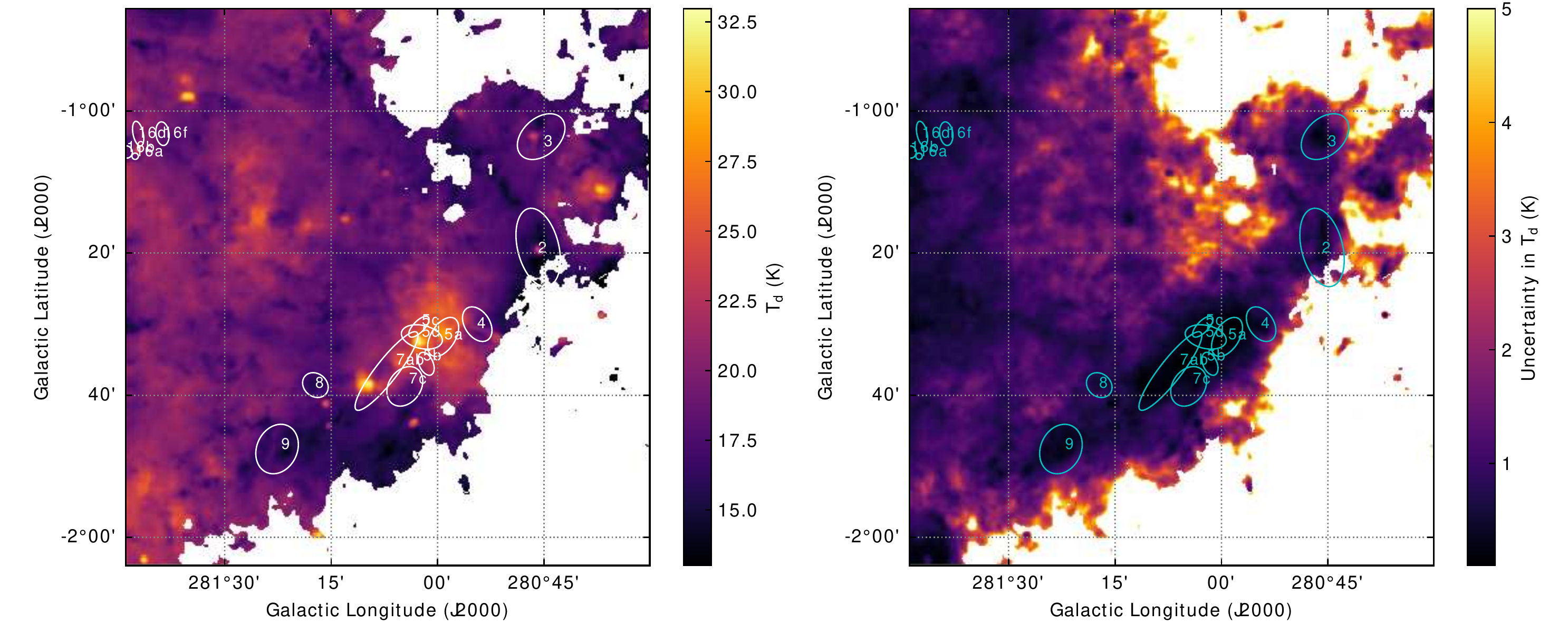}
    \caption{Region 1 \td\, (left) and \td\, uncertainty (right) maps.}
  \end{center}
\end{figure}
\pagebreak
\begin{figure}
  \begin{center}
    \includegraphics[width=9in]{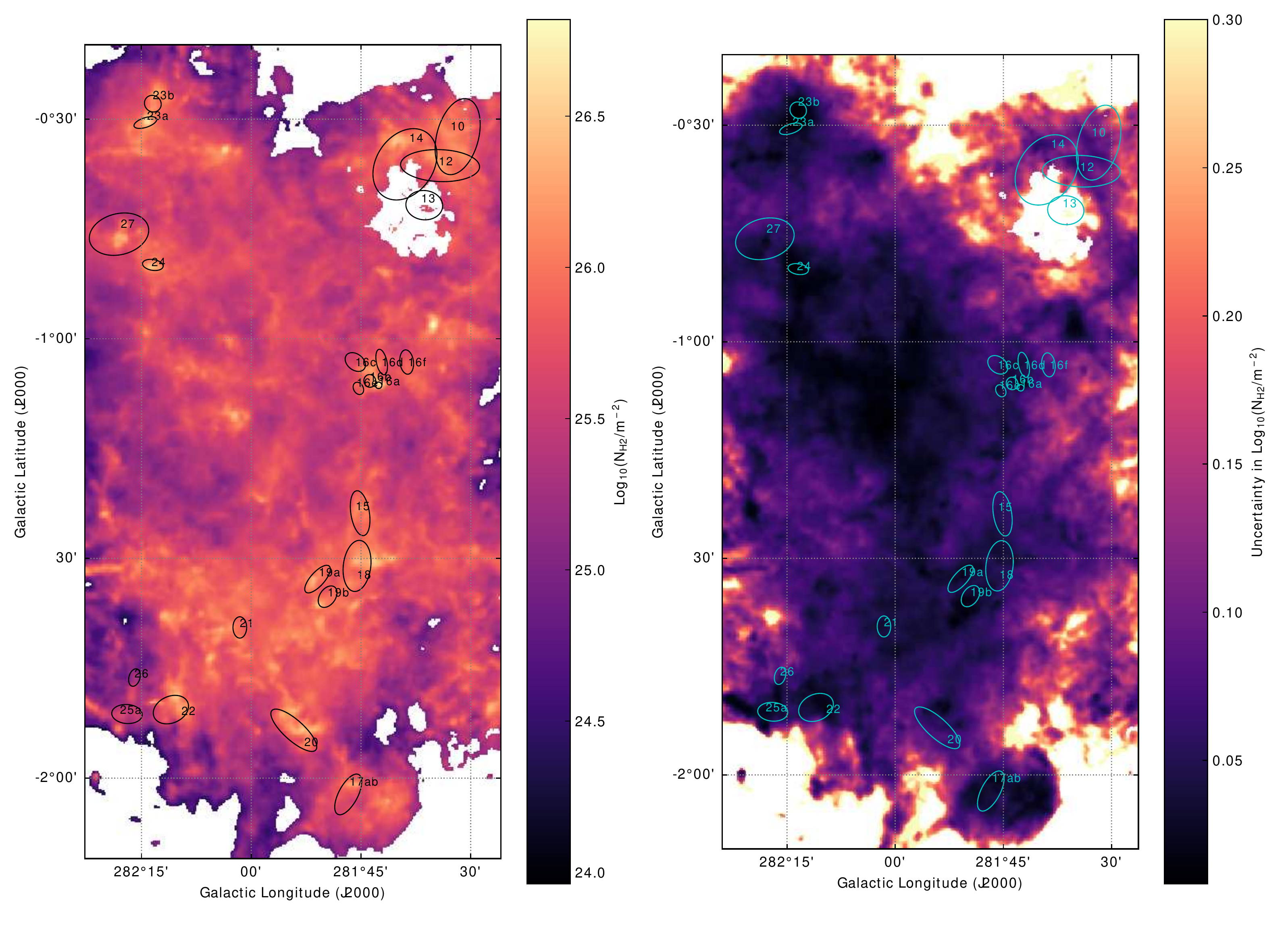}
    \caption{Regions 2 and 3 \ncol\, (left) and \ncol\, uncertainty (right) maps.}
  \end{center}
\end{figure}
\pagebreak
\begin{figure}
  \begin{center}
    \includegraphics[width=9in]{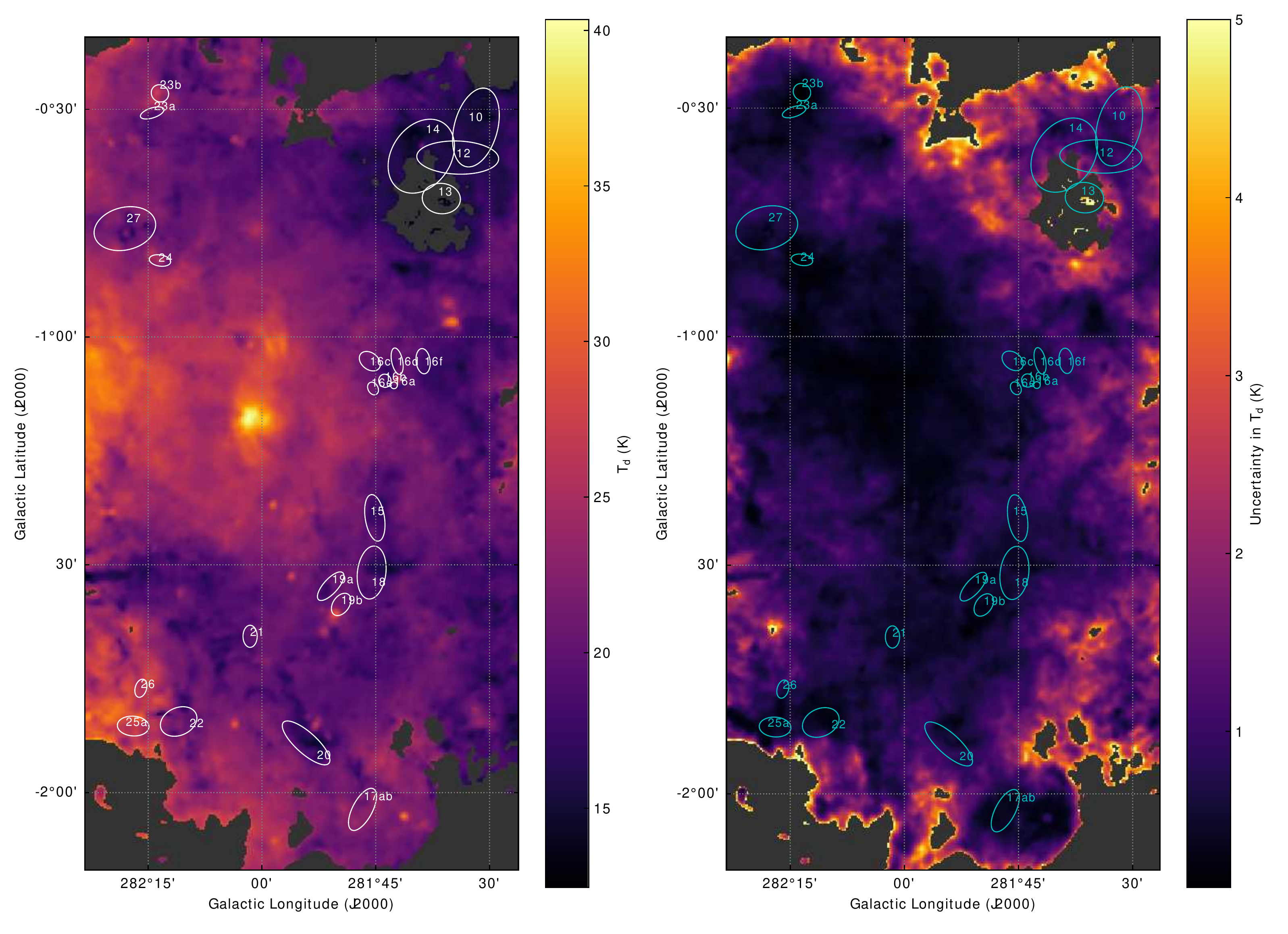}
    \caption{Regions 2 and 3 \td\, (left) and \td\, uncertainty (right) maps.}
  \end{center}
\end{figure}
\pagebreak
\end{turnpage}

\begin{figure}
  \begin{center}
    \includegraphics[width=\textwidth]{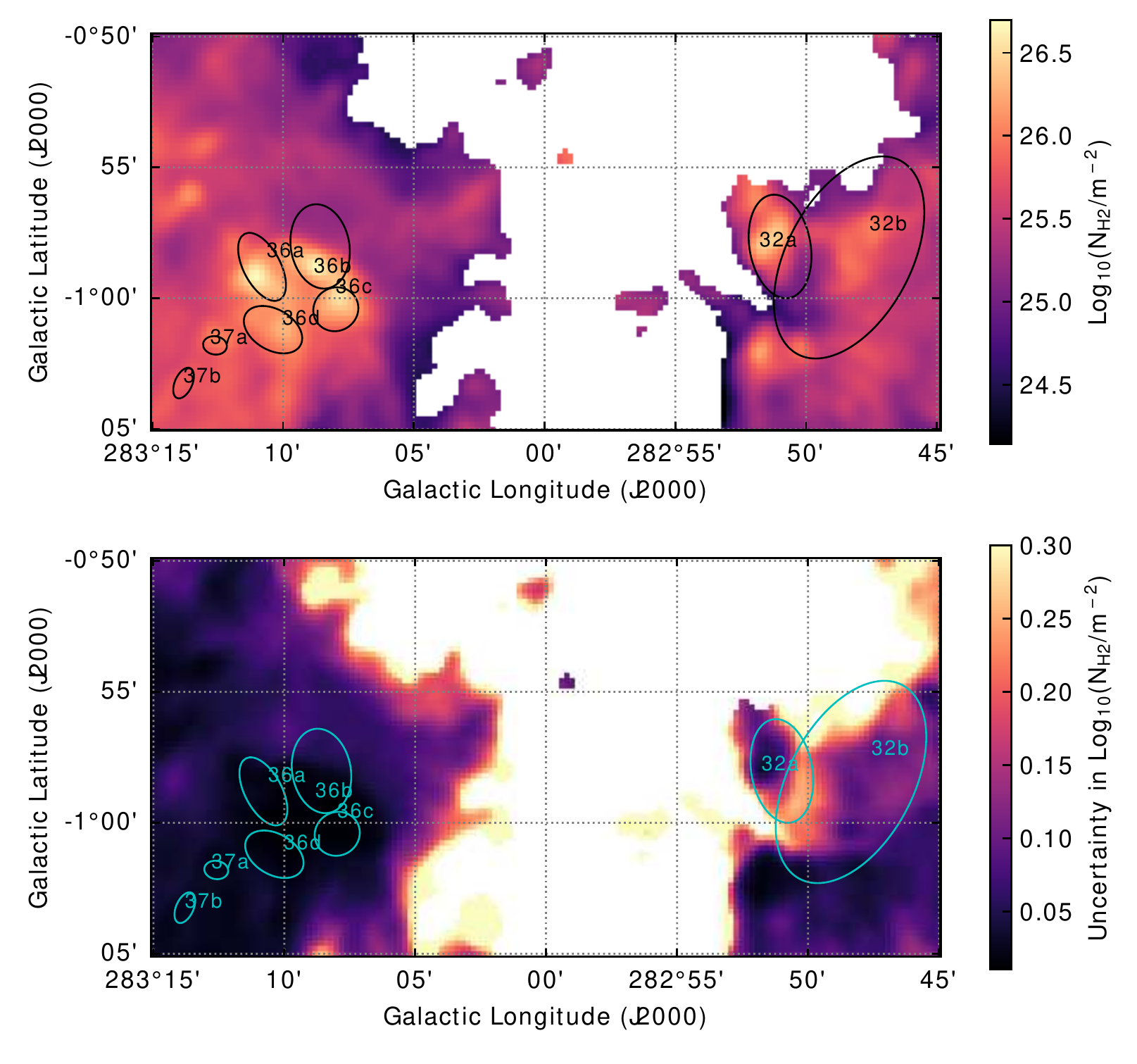}
    \caption{Region 5 \ncol\, (left) and \ncol\, uncertainty (right) maps.}
  \end{center}
\end{figure}

\begin{figure}
  \begin{center}
    \includegraphics[width=\textwidth]{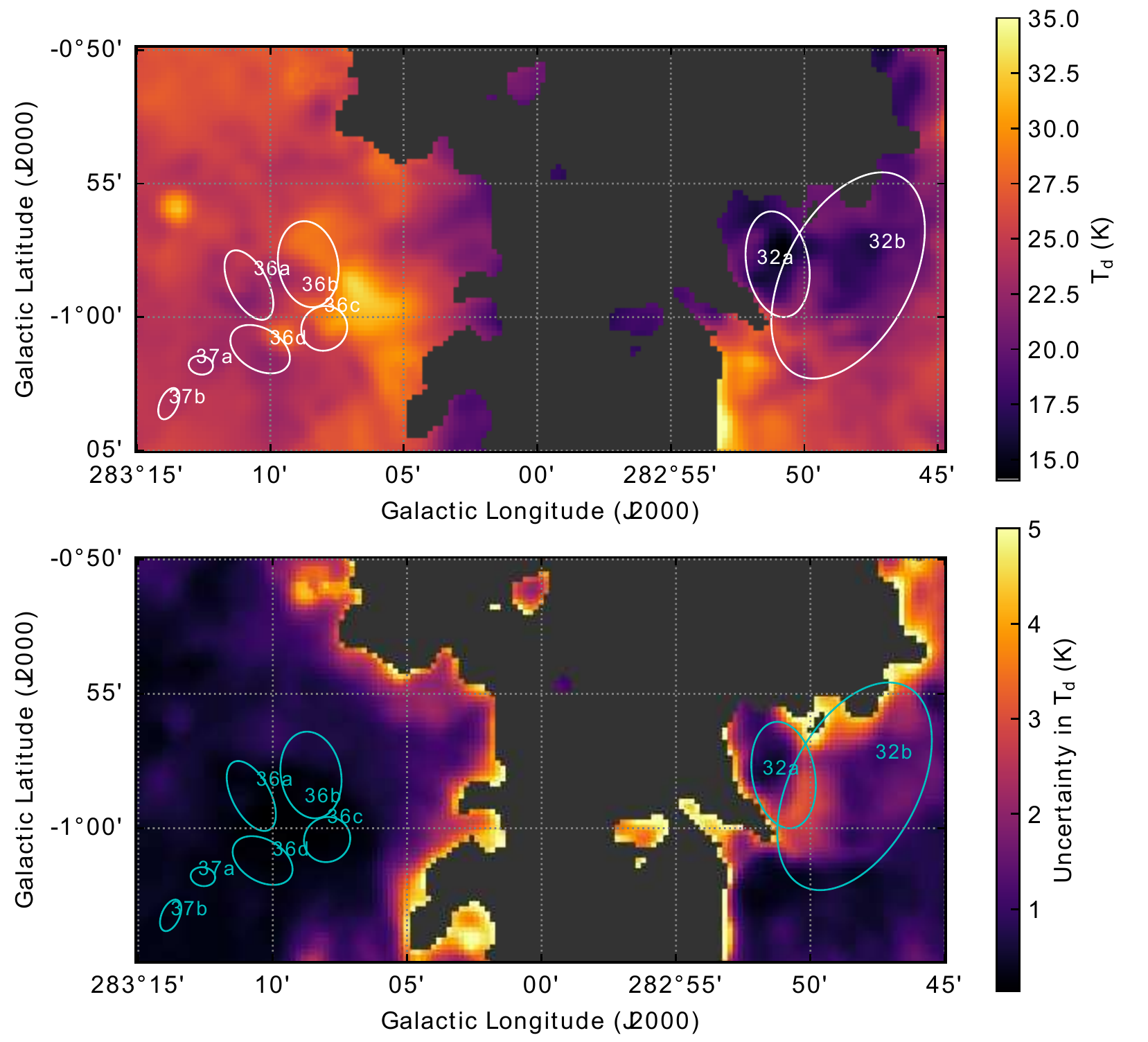}
    \caption{Region 5 \td\, (left) and \td\, uncertainty (right) maps.}
  \end{center}
\end{figure}

\begin{turnpage}
\pagebreak
\begin{figure}
  \begin{center}
    \includegraphics[width=9in]{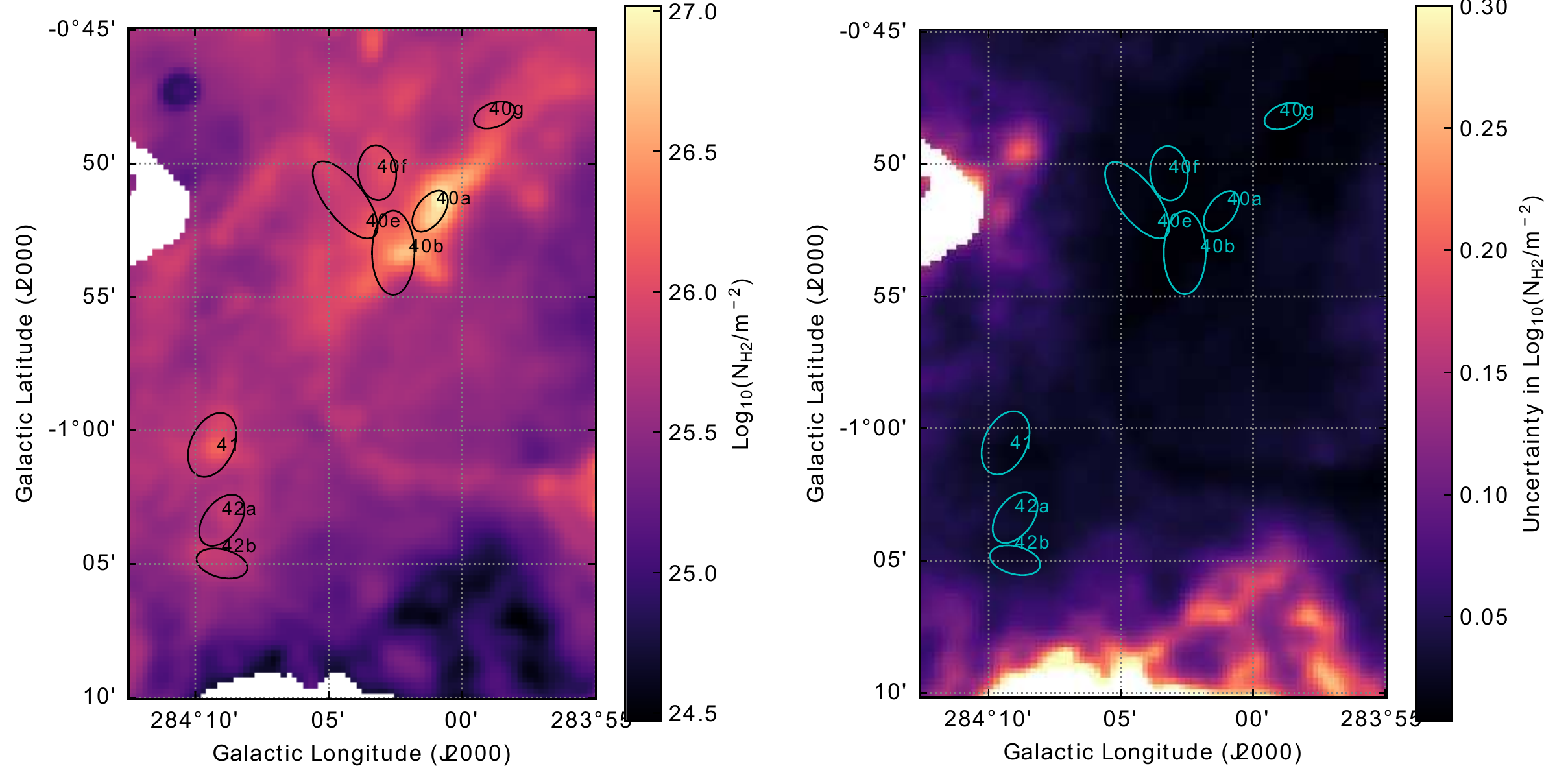}
    \caption{Region 1 \ncol\, (left) and \ncol\, uncertainty (right) maps.}
  \end{center}
\end{figure}
\pagebreak
\begin{figure}
  \begin{center}
    \includegraphics[width=9in]{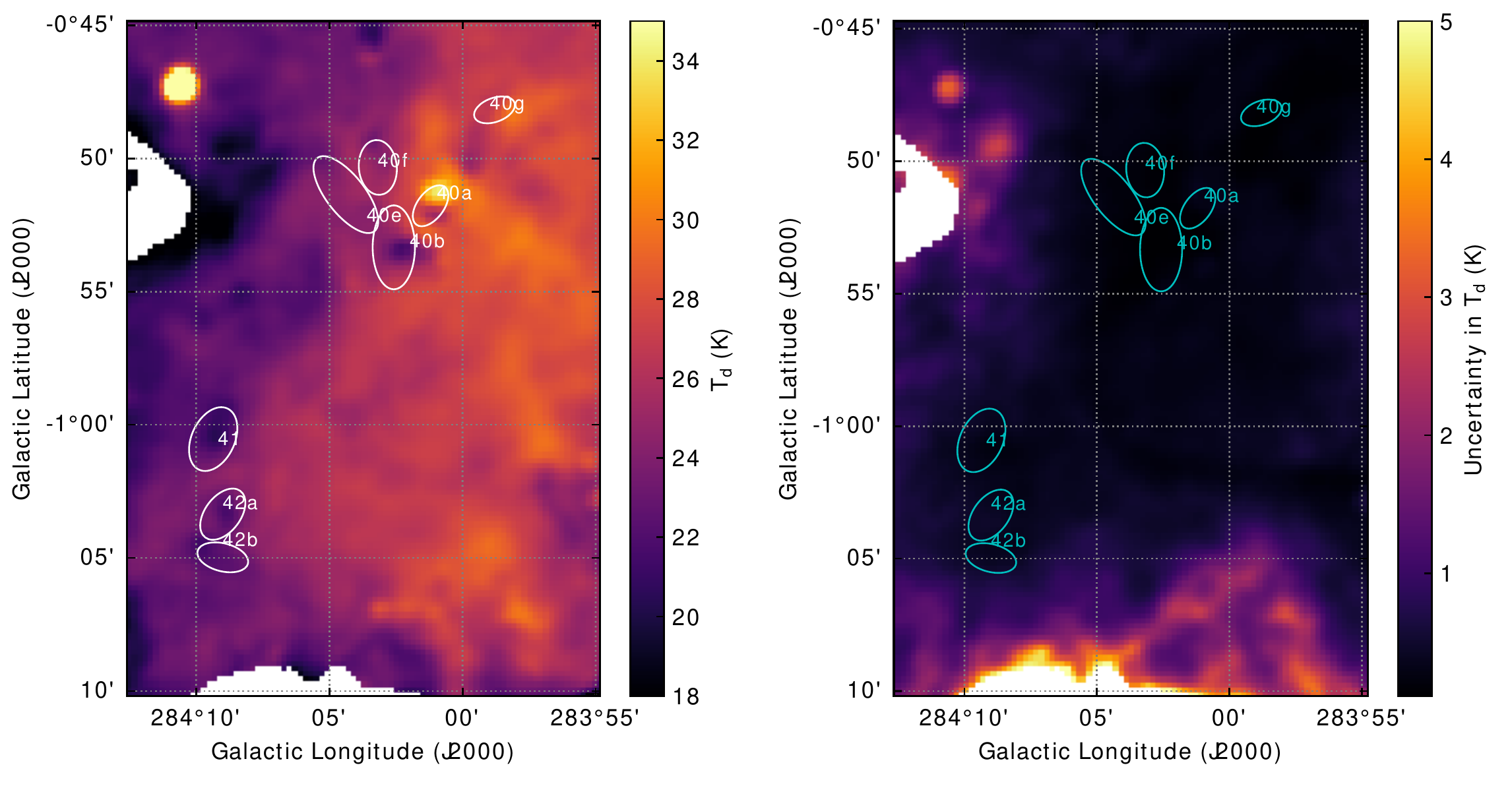}
    \caption{Region 6 \td\, (left) and \td\, uncertainty (right) maps.}
  \end{center}
\end{figure}
\pagebreak
\end{turnpage}

\begin{figure}
  \begin{center}
    \includegraphics[height=4in]{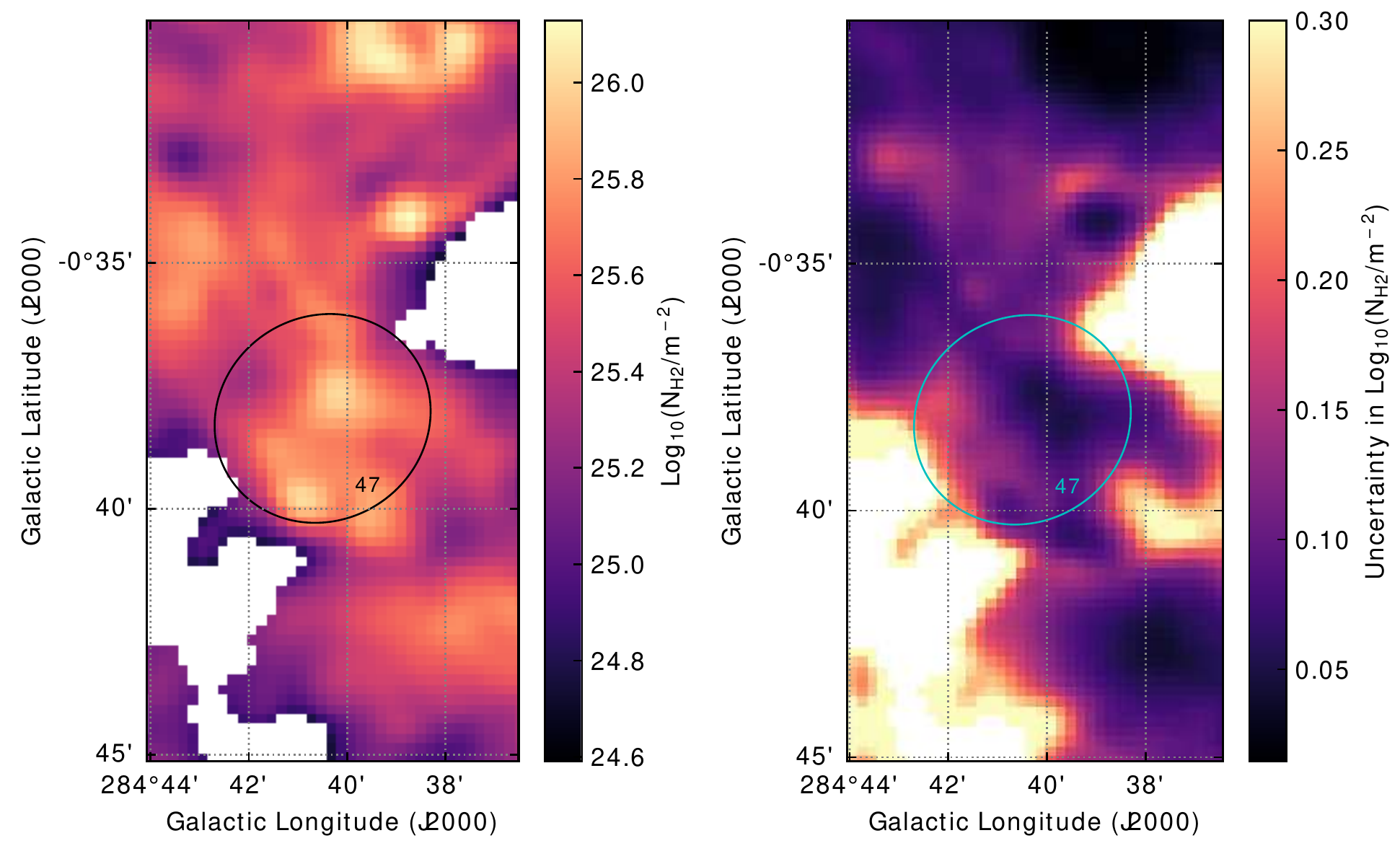}
    \caption{Region 7 \ncol\, (left) and \ncol\, uncertainty (right) maps.}
  \end{center}
\end{figure}

\begin{figure}
  \begin{center}
    \includegraphics[height=4in]{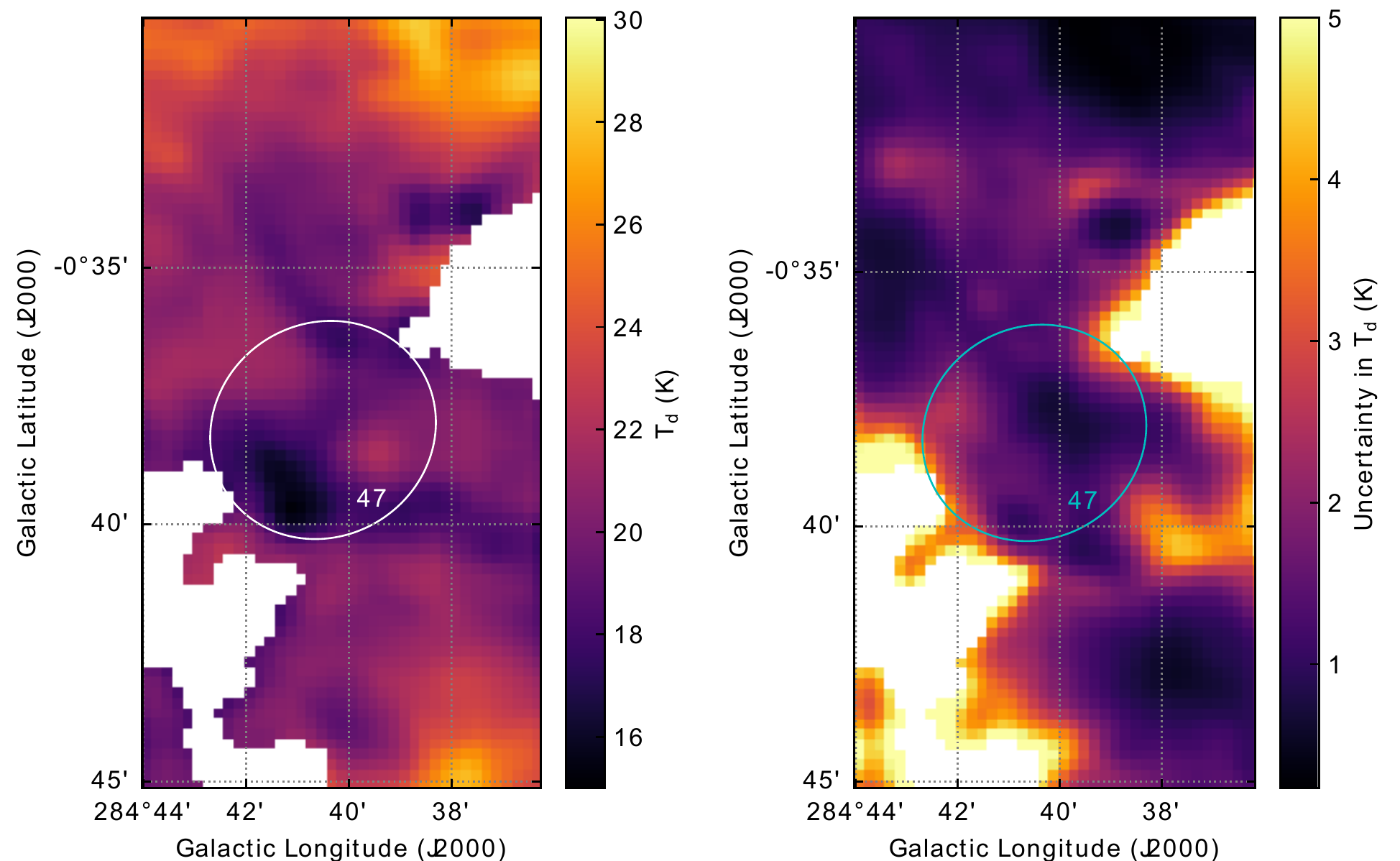}
    \caption{Region 7 \td\, (left) and \td\, uncertainty (right) maps.}
  \end{center}
\end{figure}

\begin{figure}
  \begin{center}
    \includegraphics[width=\textwidth]{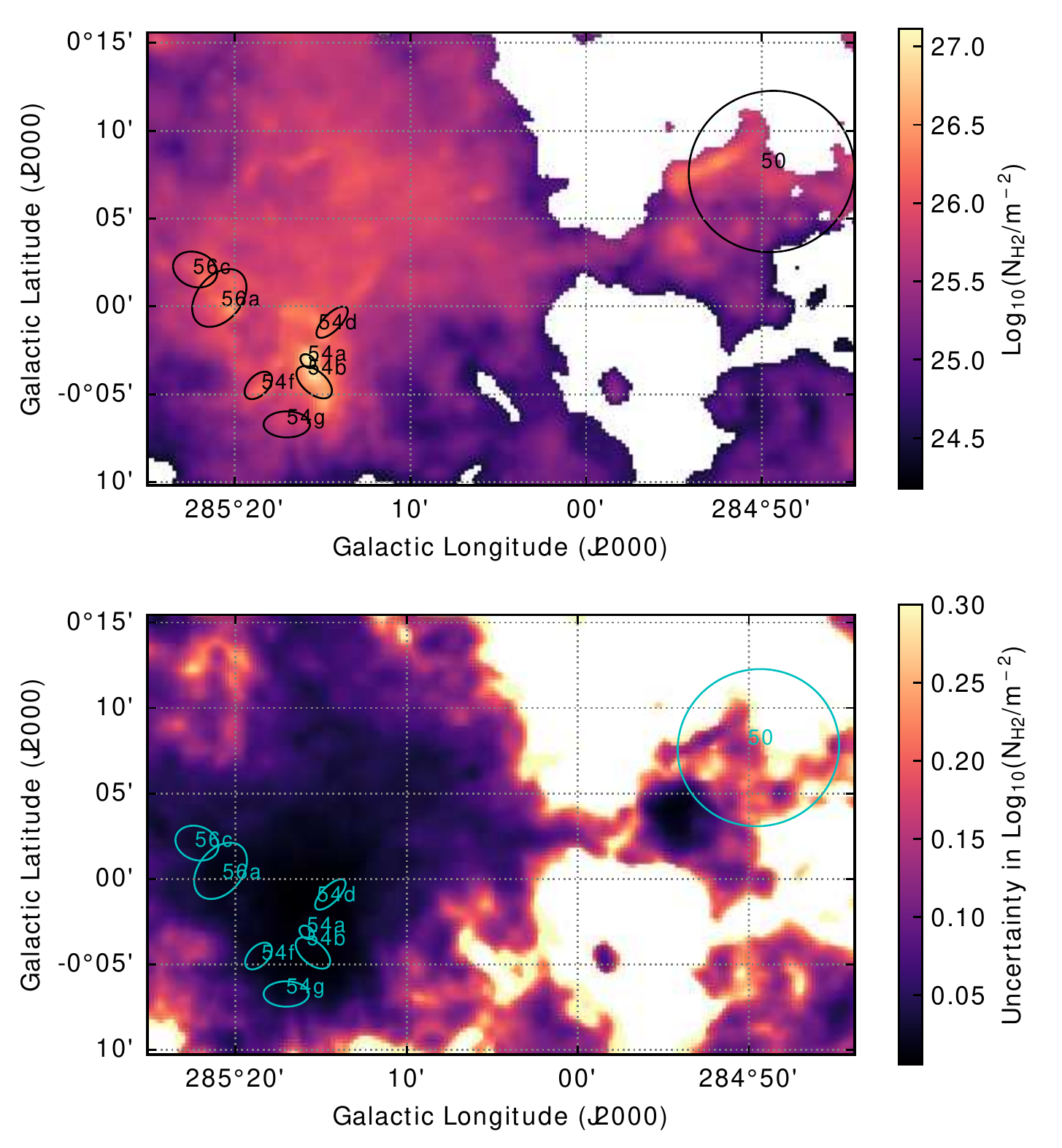}
    \caption{Region 8 \ncol\, (left) and \ncol\, uncertainty (right) maps.}
  \end{center}
\end{figure}

\begin{figure}
  \begin{center}
    \includegraphics[width=\textwidth]{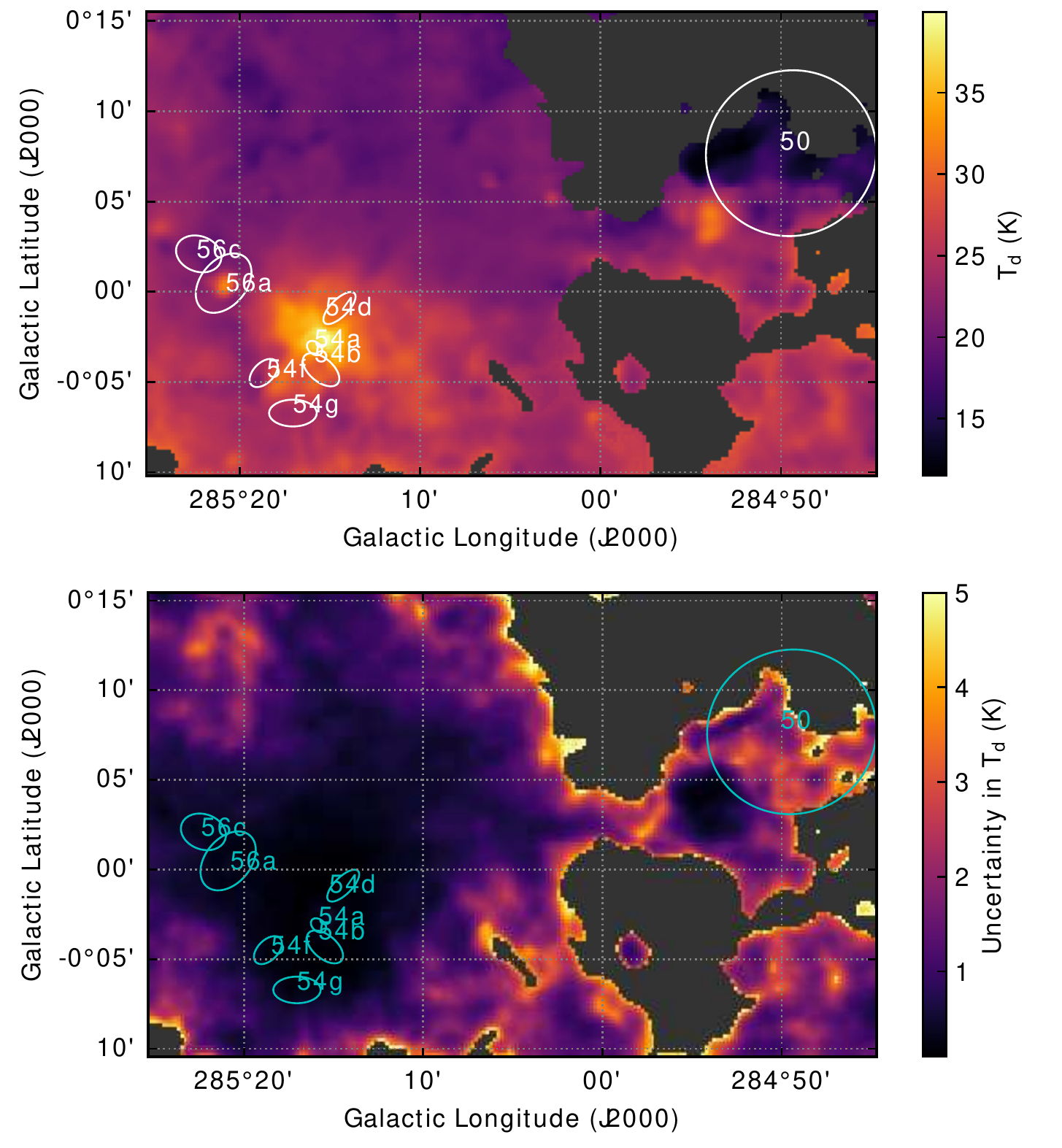}
    \caption{Region 8 \td\, (left) and \td\, uncertainty (right) maps.}
  \end{center}
\end{figure}

\begin{turnpage}
\begin{figure}
  \begin{center}
    \includegraphics[width=9in]{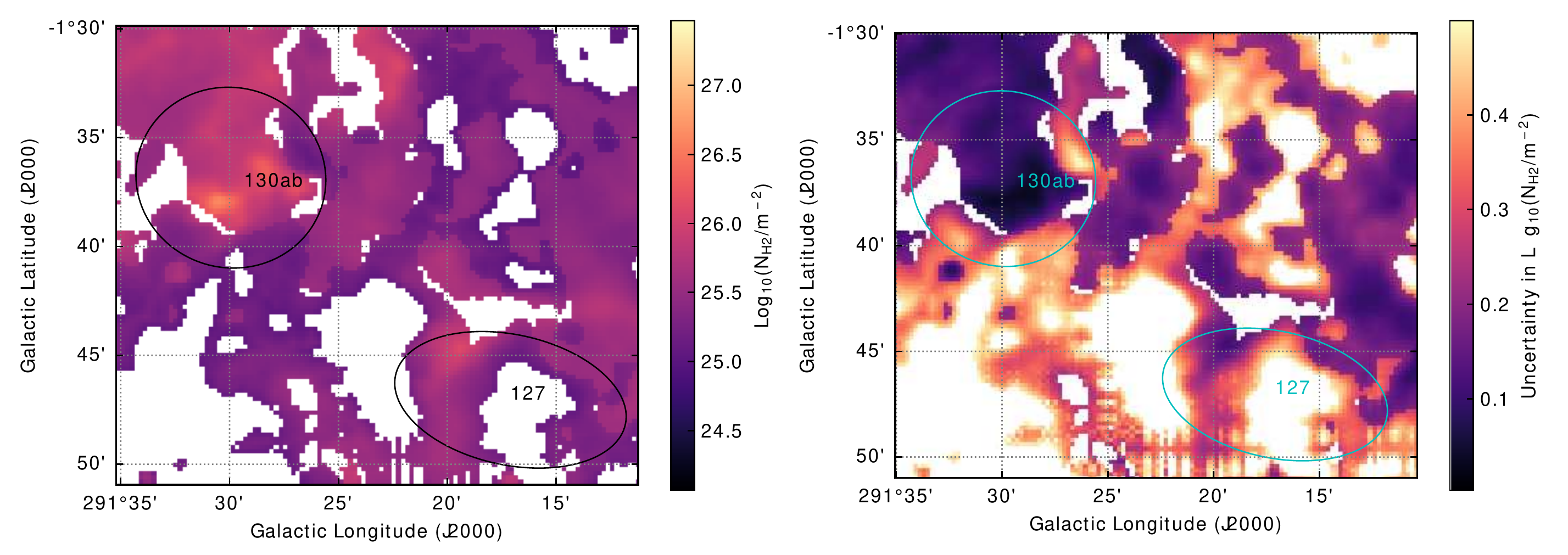}
    \caption{Region 12 \ncol\, (left) and \ncol\, uncertainty (right) maps. The paucity of usable data in this Region leads us not to trust the few good pixels in these clumps.}
  \end{center}
\end{figure}

\begin{figure}
  \begin{center}
    \includegraphics[width=9in]{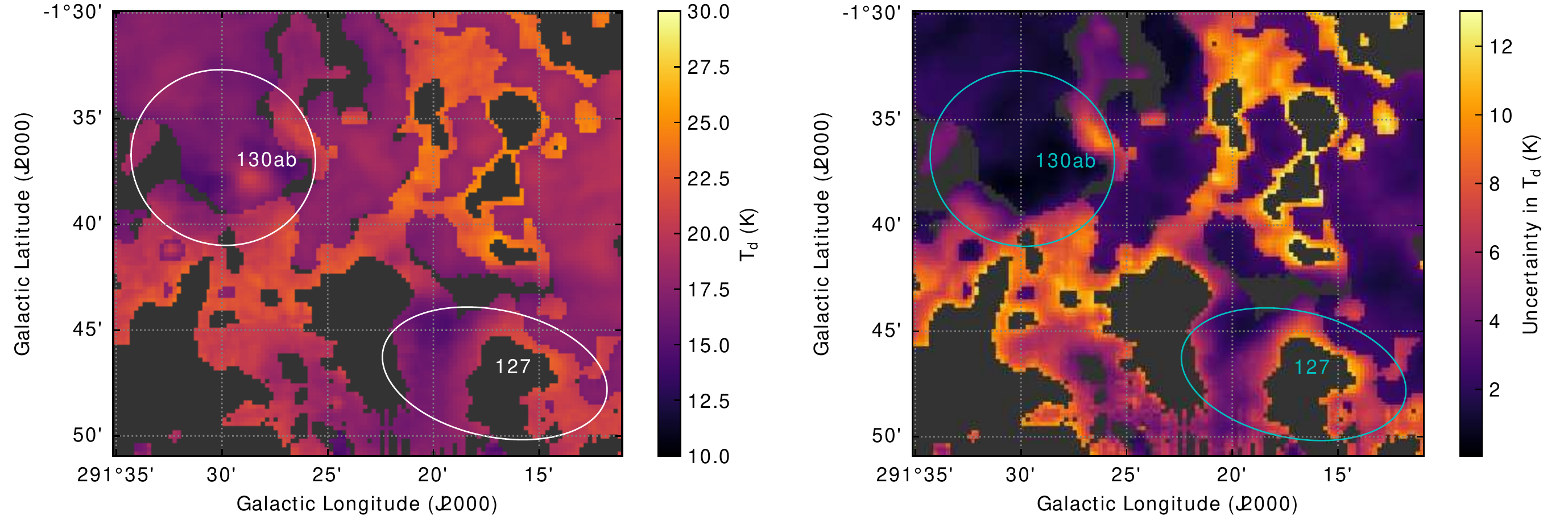}
    \caption{Region 12 \td\, (left) and \td\, uncertainty (right) maps.}
  \end{center}
\end{figure}
\end{turnpage}

\begin{figure}
  \begin{center}
    \includegraphics[width=\textwidth]{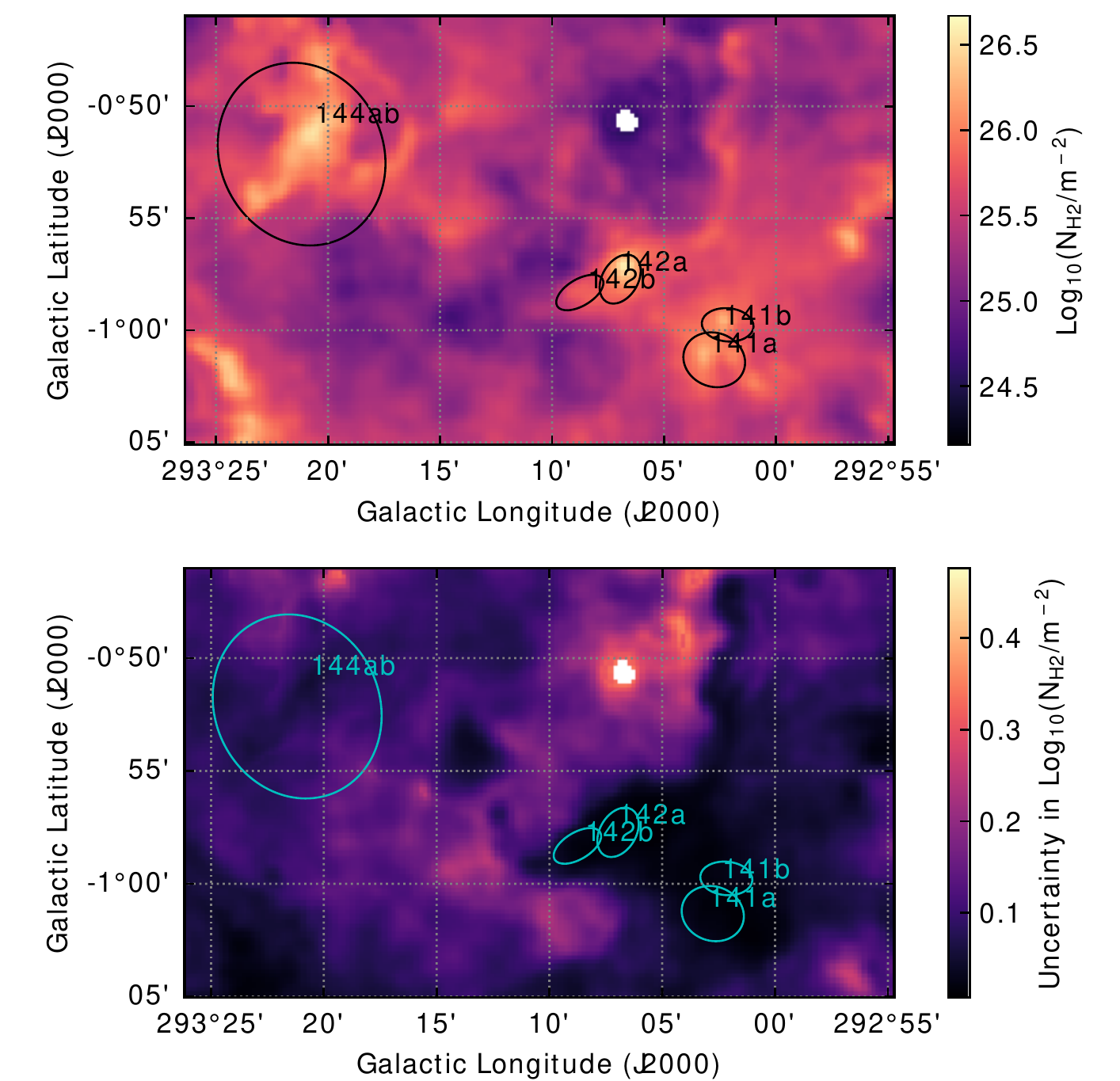}
    \caption{Region 16 \ncol\, (top) and \ncol\, uncertainty (bottom) maps.}
  \end{center}
\end{figure}

\begin{figure}
  \begin{center}
    \includegraphics[width=\textwidth]{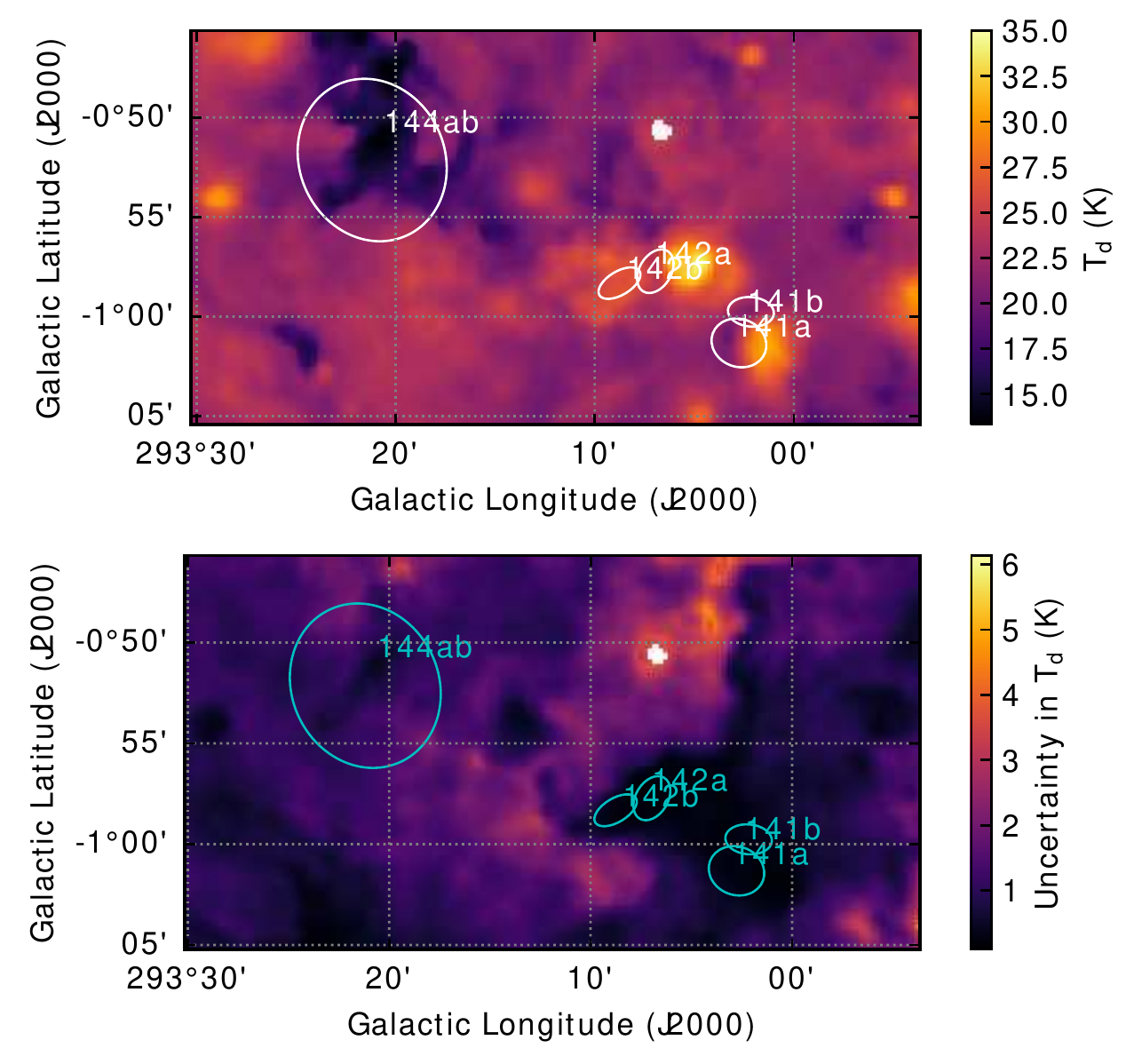}
    \caption{Region 16 \td\, (top) and \td\, uncertainty (bottom) maps.}
  \end{center}
\end{figure}

\begin{figure}
  \begin{center}
    \includegraphics[width=\textwidth]{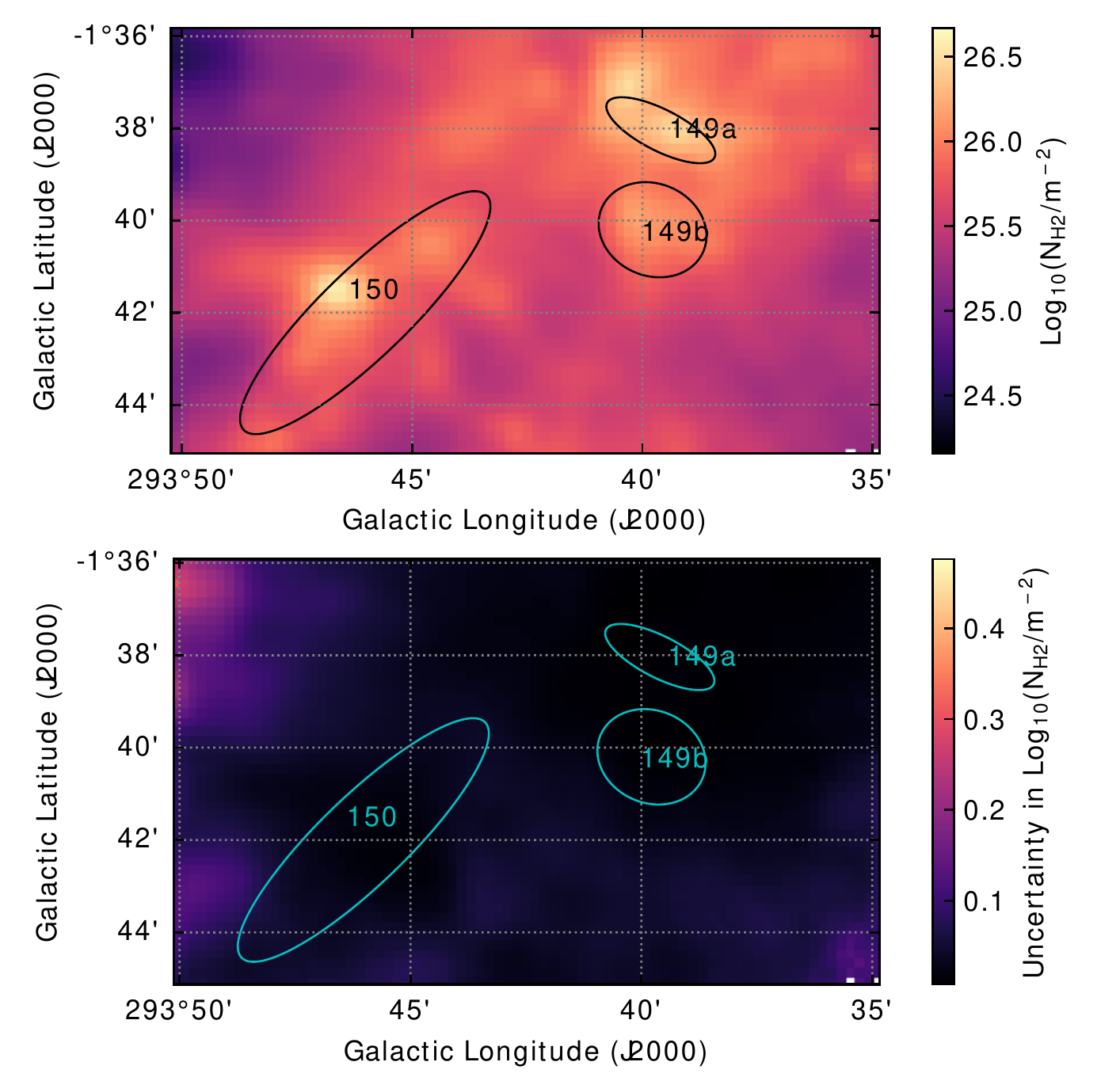}
    \caption{Region 18 \ncol\, (top) and \ncol\, uncertainty (bottom) maps.}
  \end{center}
\end{figure}

\begin{figure}
  \begin{center}
    \includegraphics[width=\textwidth]{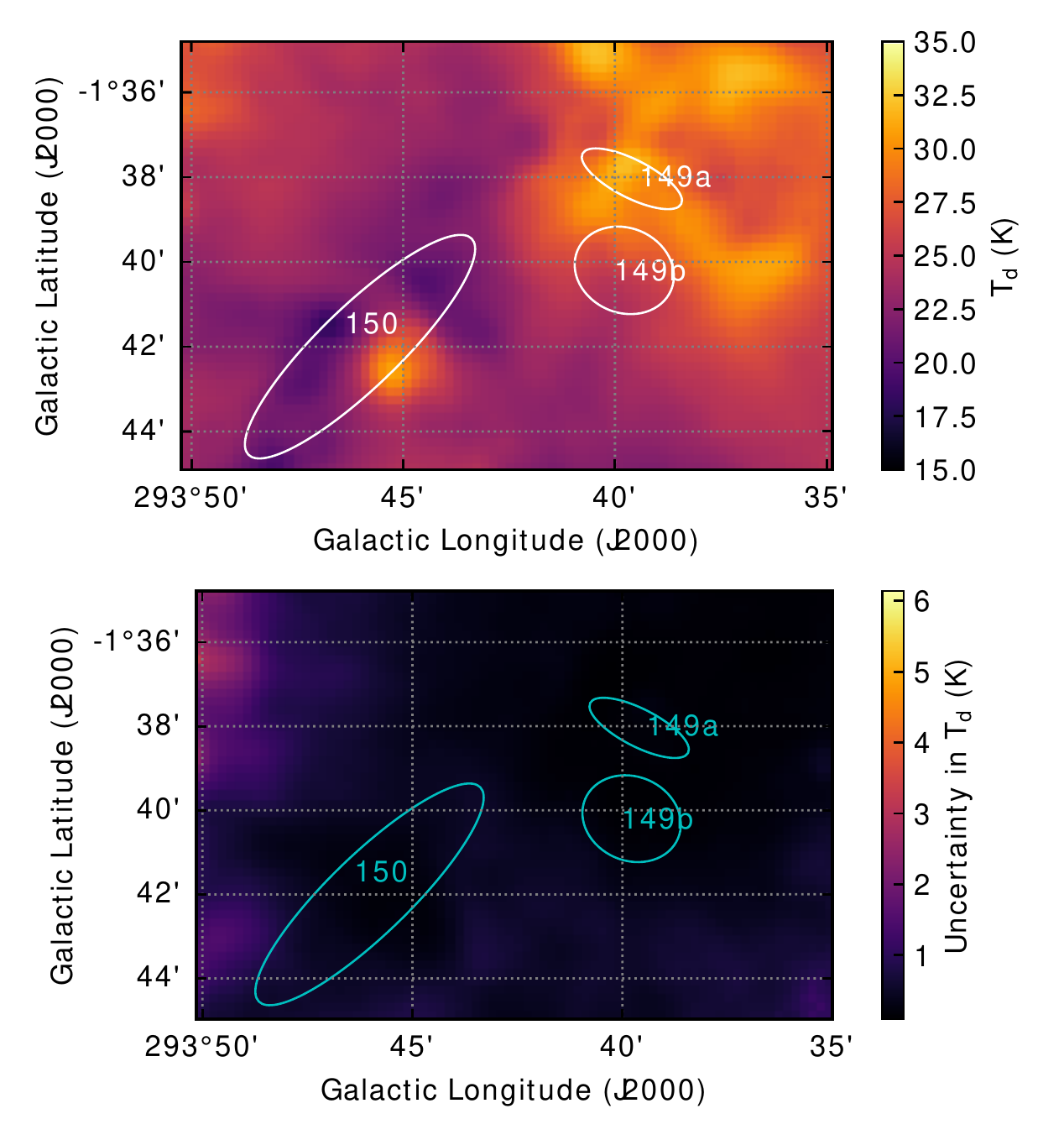}
    \caption{Region 18 \td\, (top) and \td\, uncertainty (bottom) maps.}
  \end{center}
\end{figure}

\begin{figure}
  \begin{center}
    \includegraphics[width=\textwidth]{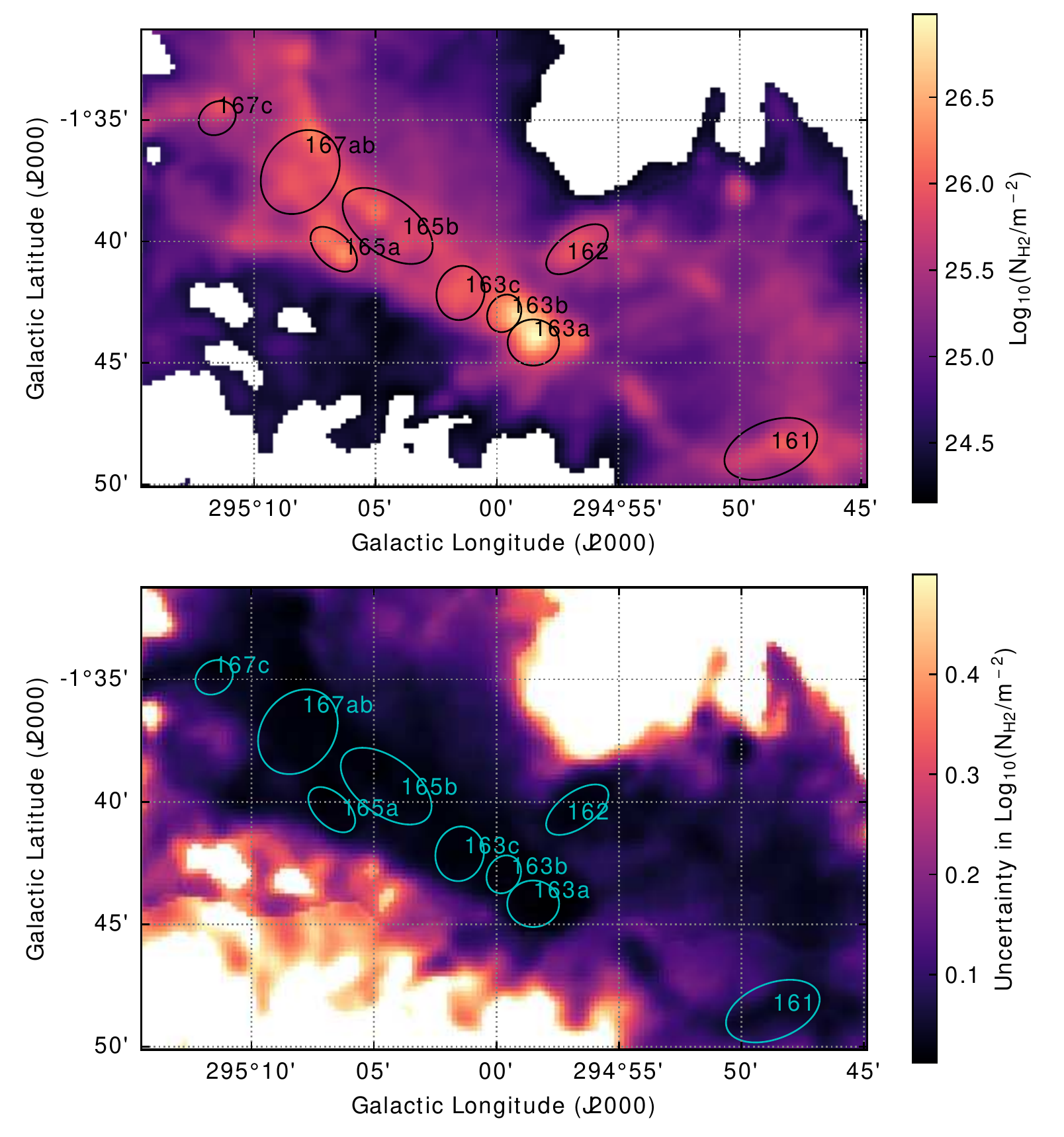}
    \caption{Region 21 \ncol\, (top) and \ncol\, uncertainty (bottom) maps.}
  \end{center}
\end{figure}

\begin{figure}
  \begin{center}
    \includegraphics[width=\textwidth]{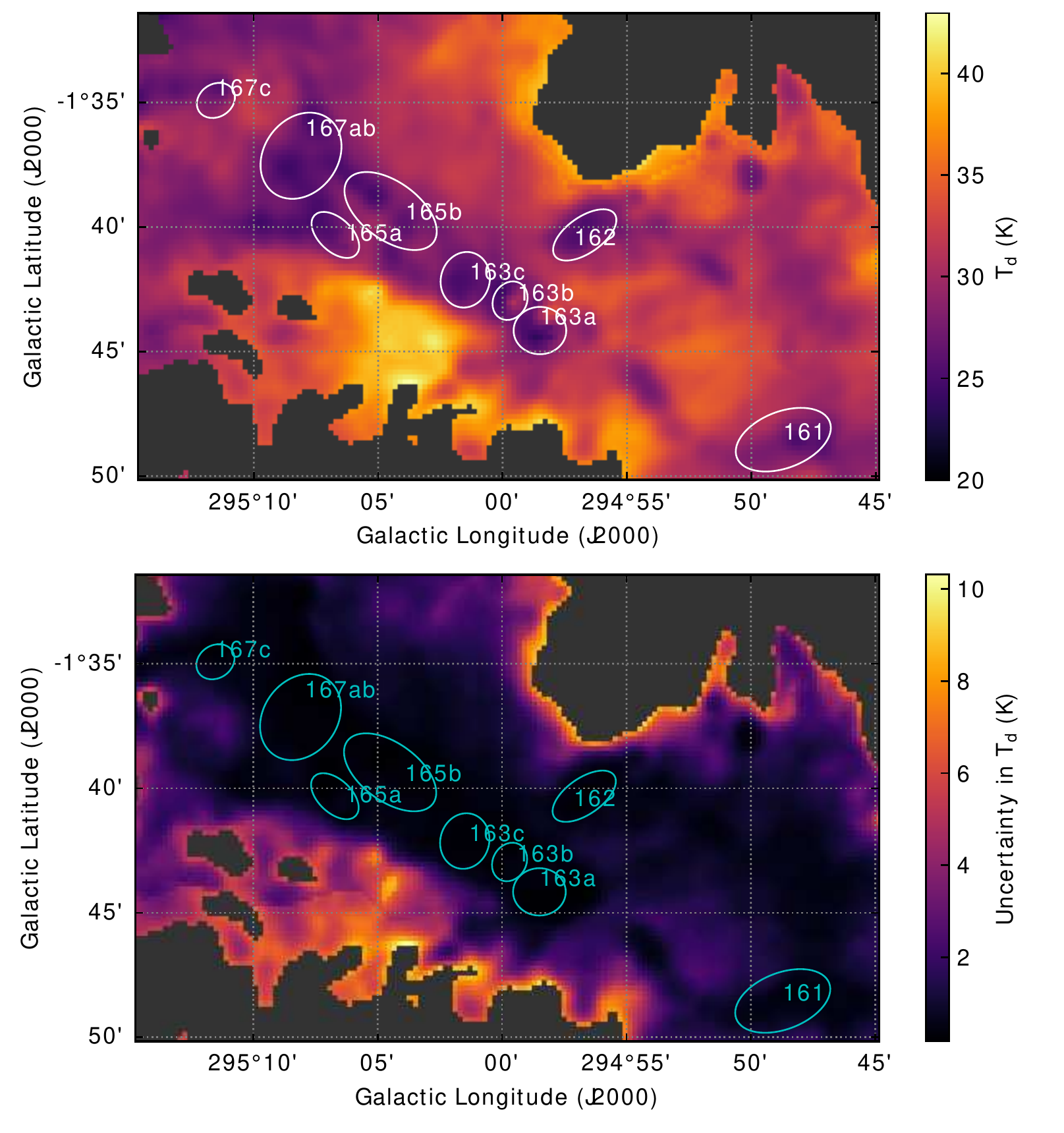}
    \caption{Region 21 \td\, (top) and \td\, uncertainty (bottom) maps.}
  \end{center}
\end{figure}

\begin{figure}
  \begin{center}
    \includegraphics[width=\textwidth]{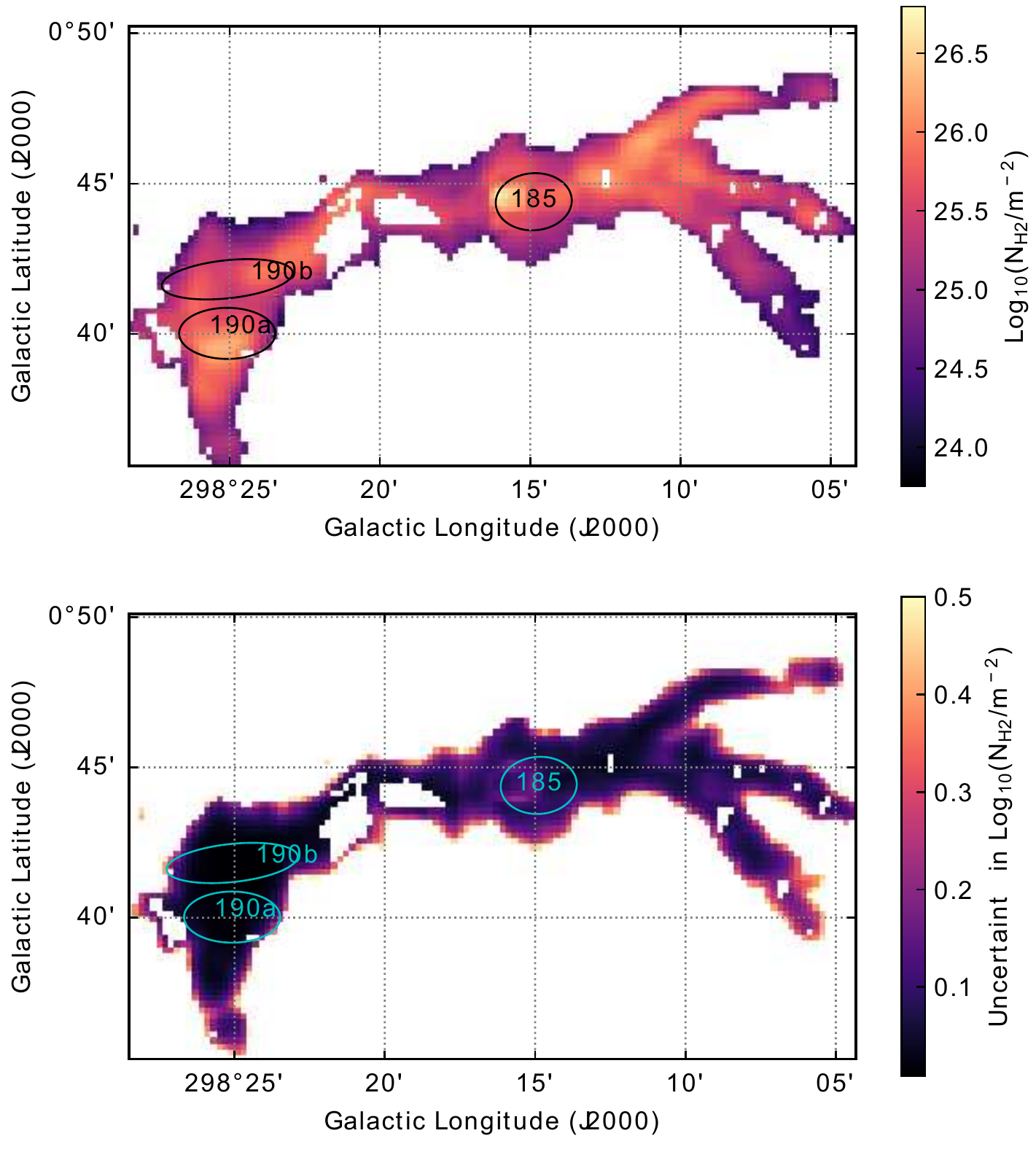}
    \caption{Region 23 \ncol\, (top) and \ncol\, uncertainty (bottom) maps. The gaps in the data are due to the aggressive background subtraction method needed to remove the Dragonfish Nebula.}
  \end{center}
\end{figure}

\begin{figure}
  \begin{center}
    \includegraphics[width=\textwidth]{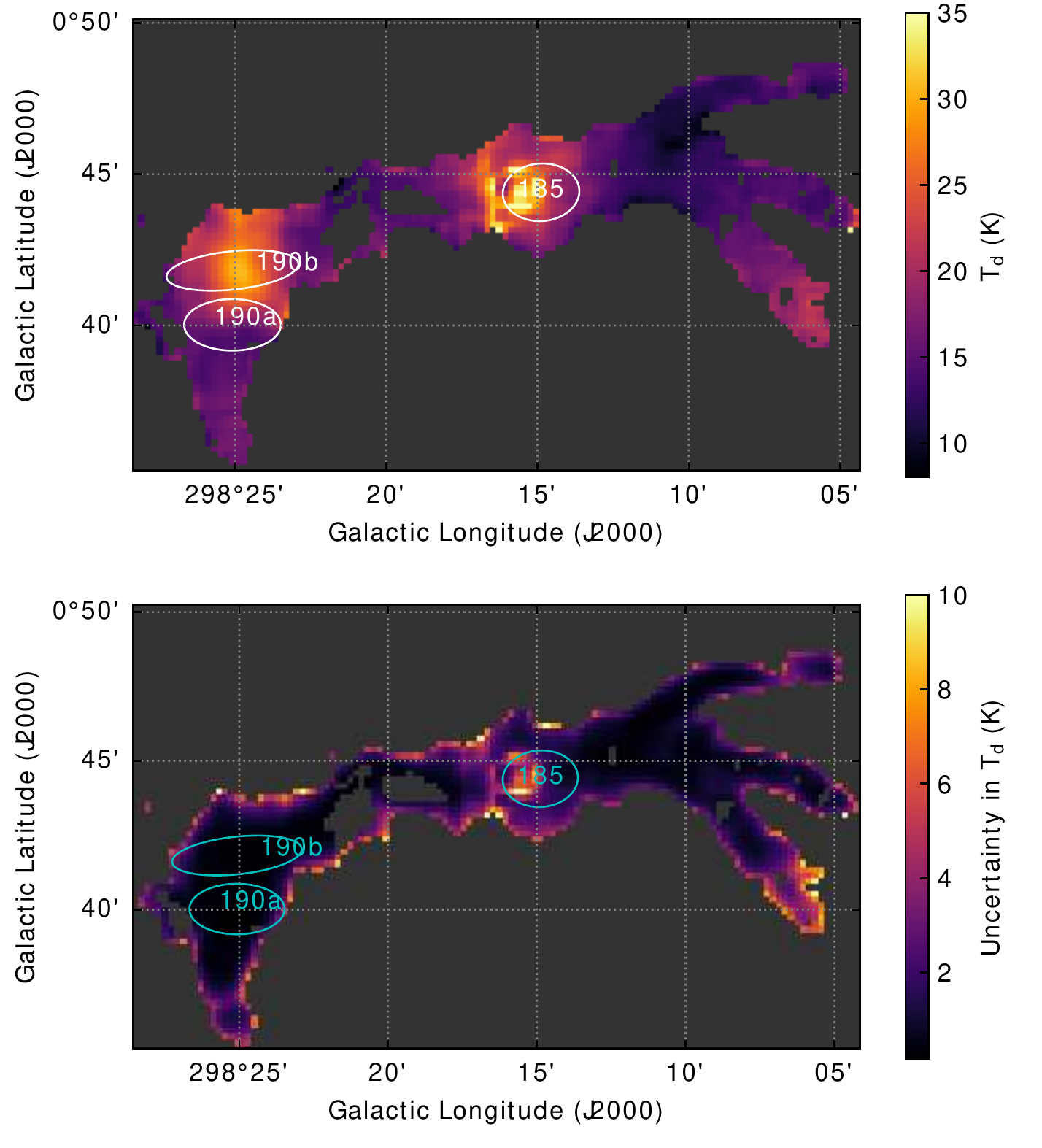}
    \caption{Region 23 \td\, (top) and \td\, uncertainty (bottom) maps.}
  \end{center}
\end{figure}

\begin{figure}
  \begin{center}
    \includegraphics[width=\textwidth]{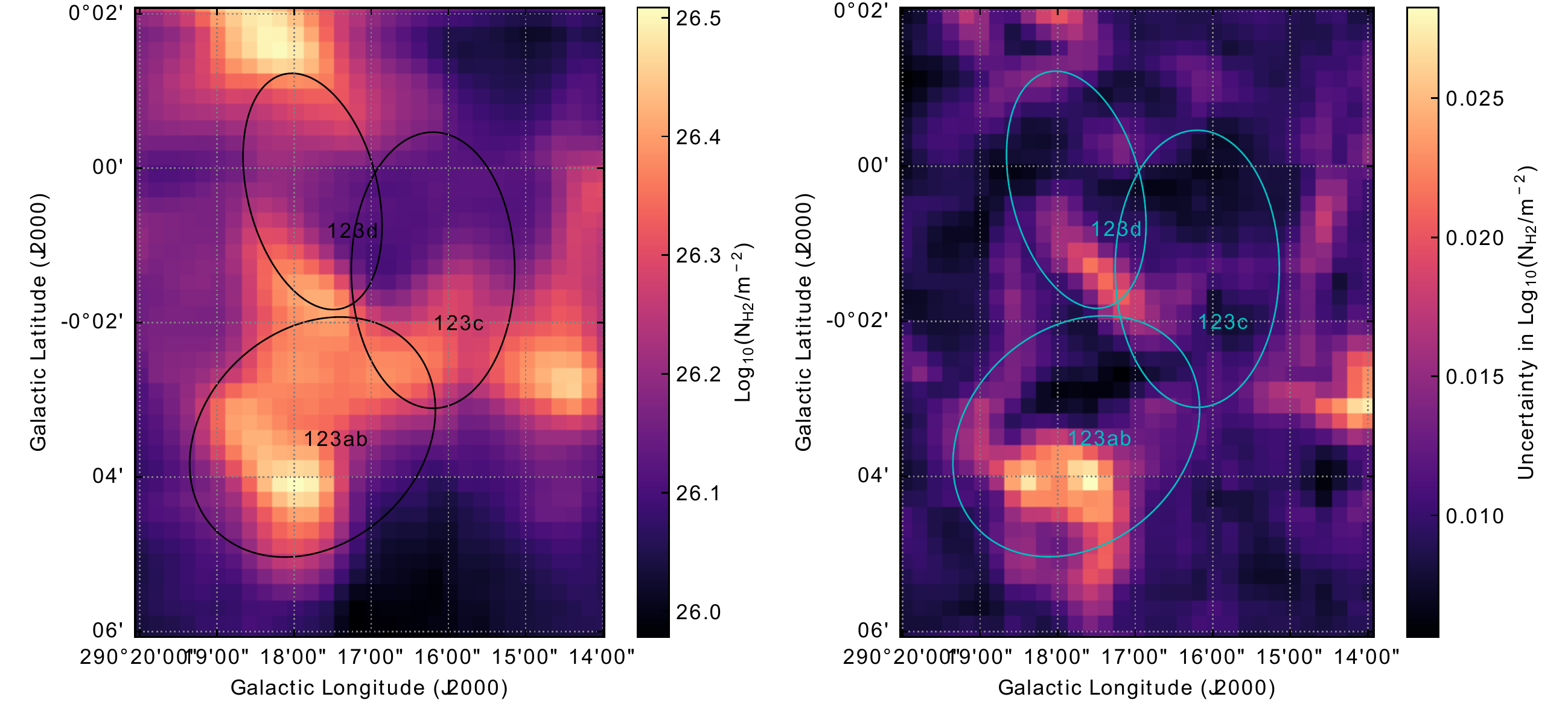}
    \caption{BYF 123 \ncol\, (left) and \ncol\, uncertainty (right) maps.}
  \end{center}
\end{figure}

\begin{figure}
  \begin{center}
    \includegraphics[width=\textwidth]{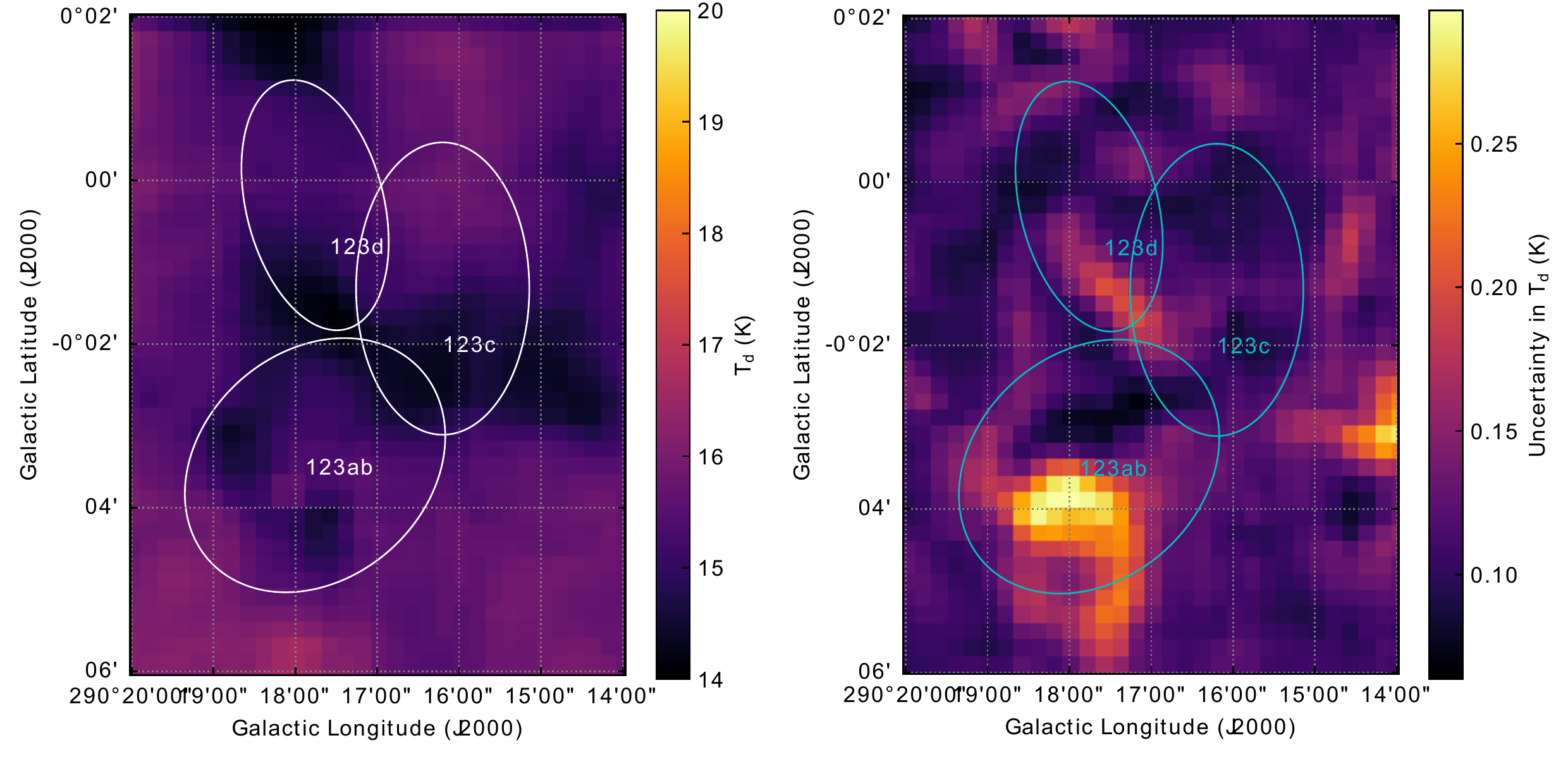}
    \caption{BYF 123 \td\, (left) and \td\, uncertainty (right) maps.}
  \end{center}
\end{figure}

\section{Plots of Dust Temperature vs. H\texorpdfstring{$_2$}{2} Column Density for Each Clump}\label{sec:app2}
Here we plot \td\ versus \ncol\ for the pixels in each clump, masked to the CO emitting area as defined in \citetalias{champ3} and also colored by \td. For completeness, we also include Region-wide plots, masked to all pixels mapped in CO, with translucent points to make coherent structures stand out. The coherent structures are spatially resolved clumps and filaments; note that these generally also show an anti-correlation of \td\ with \ncol.
\begin{figure}
    \centering
    \includegraphics{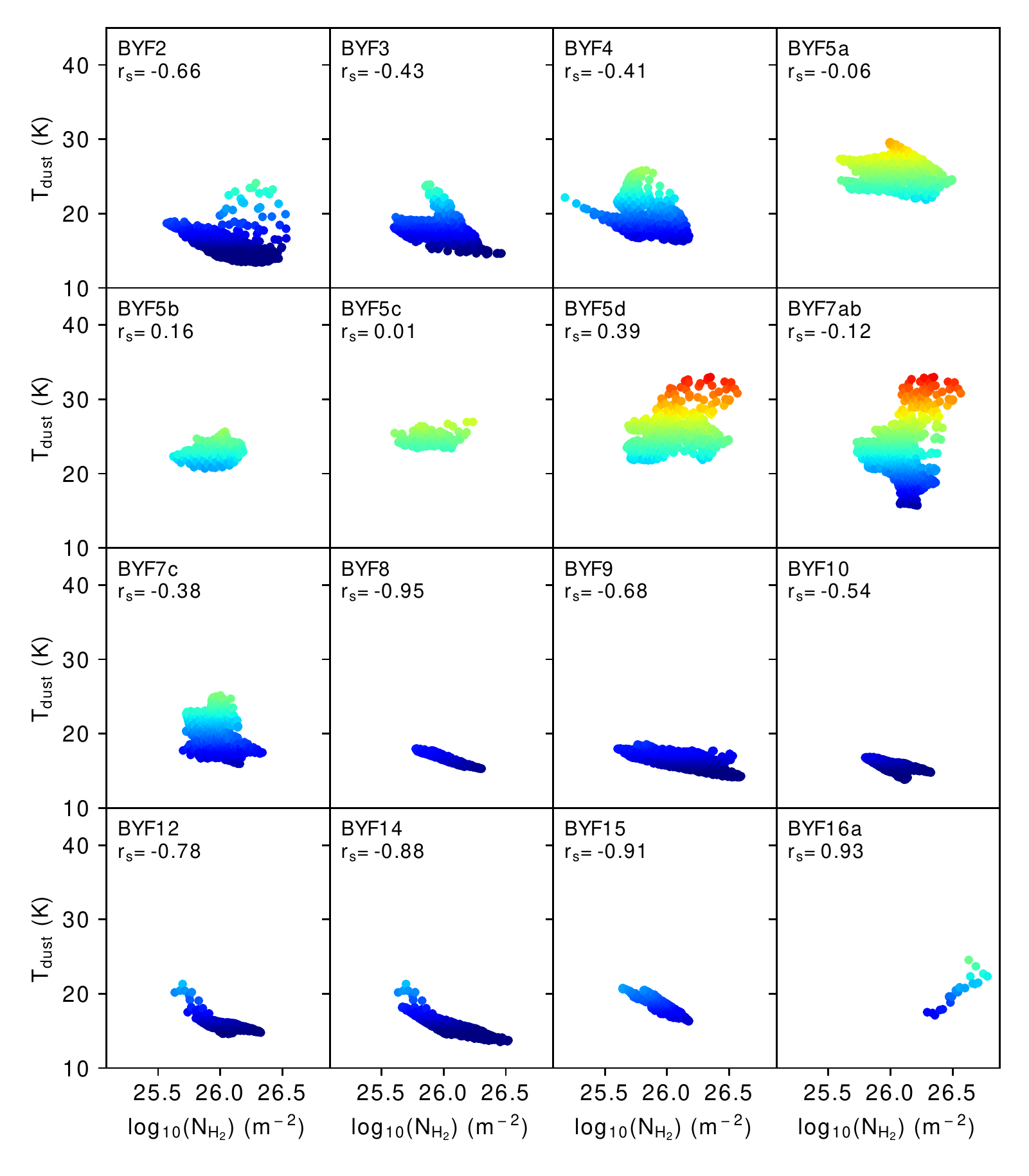}
    \caption{\td\, vs. \ncol\, plots for BYF~2--16a}
    \label{fig:tvn1}
\end{figure}

\begin{figure}
    \centering
    \includegraphics{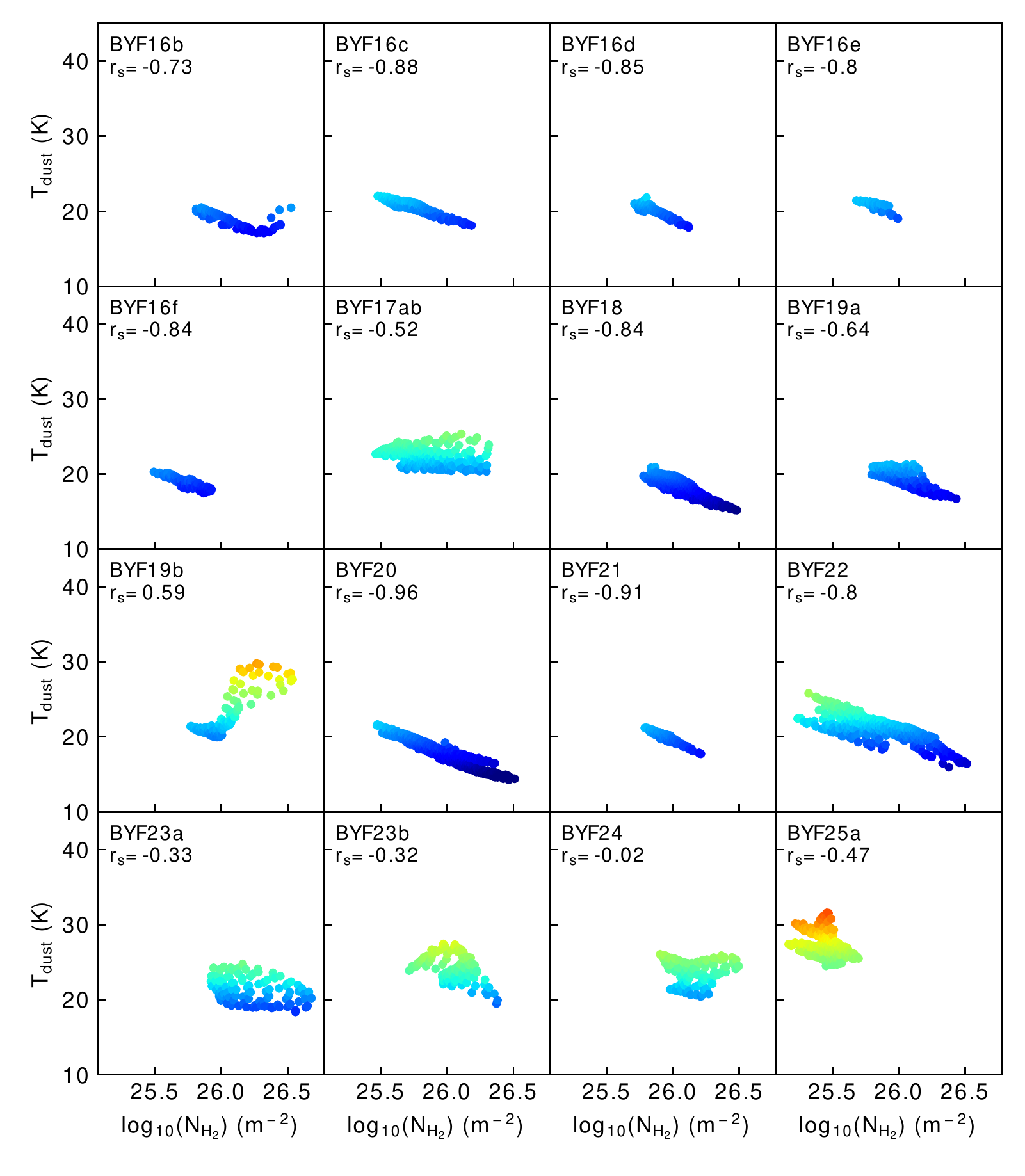}
    \caption{\td\, vs. \ncol\, plots for BYF~16b--25a}
    \label{fig:tvn2}
\end{figure}

\begin{figure}
    \centering
    \includegraphics{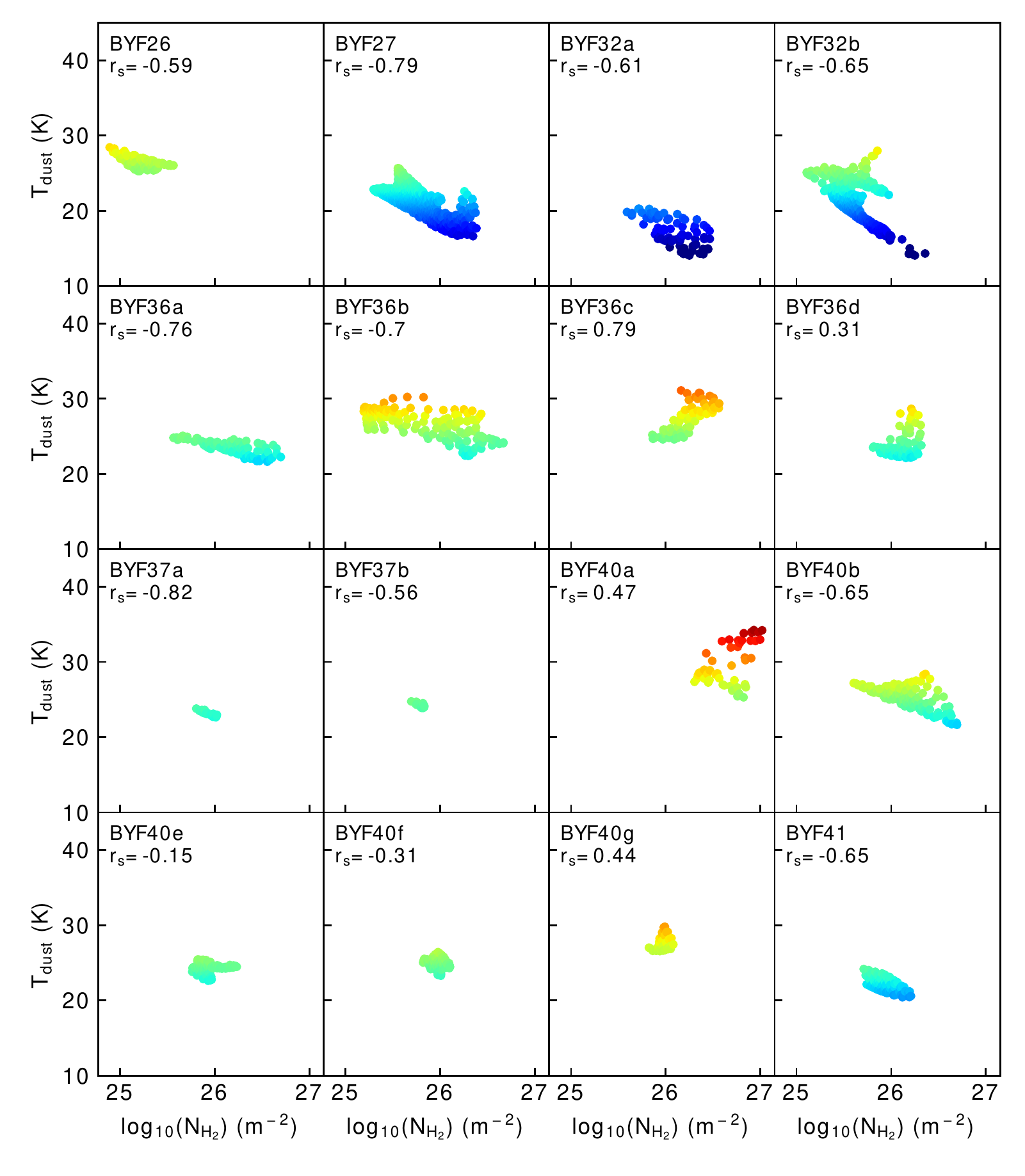}
    \caption{\td\, vs. \ncol\, plots for BYF~26--41}
    \label{fig:tvn3}
\end{figure}

\begin{figure}
    \centering
    \includegraphics{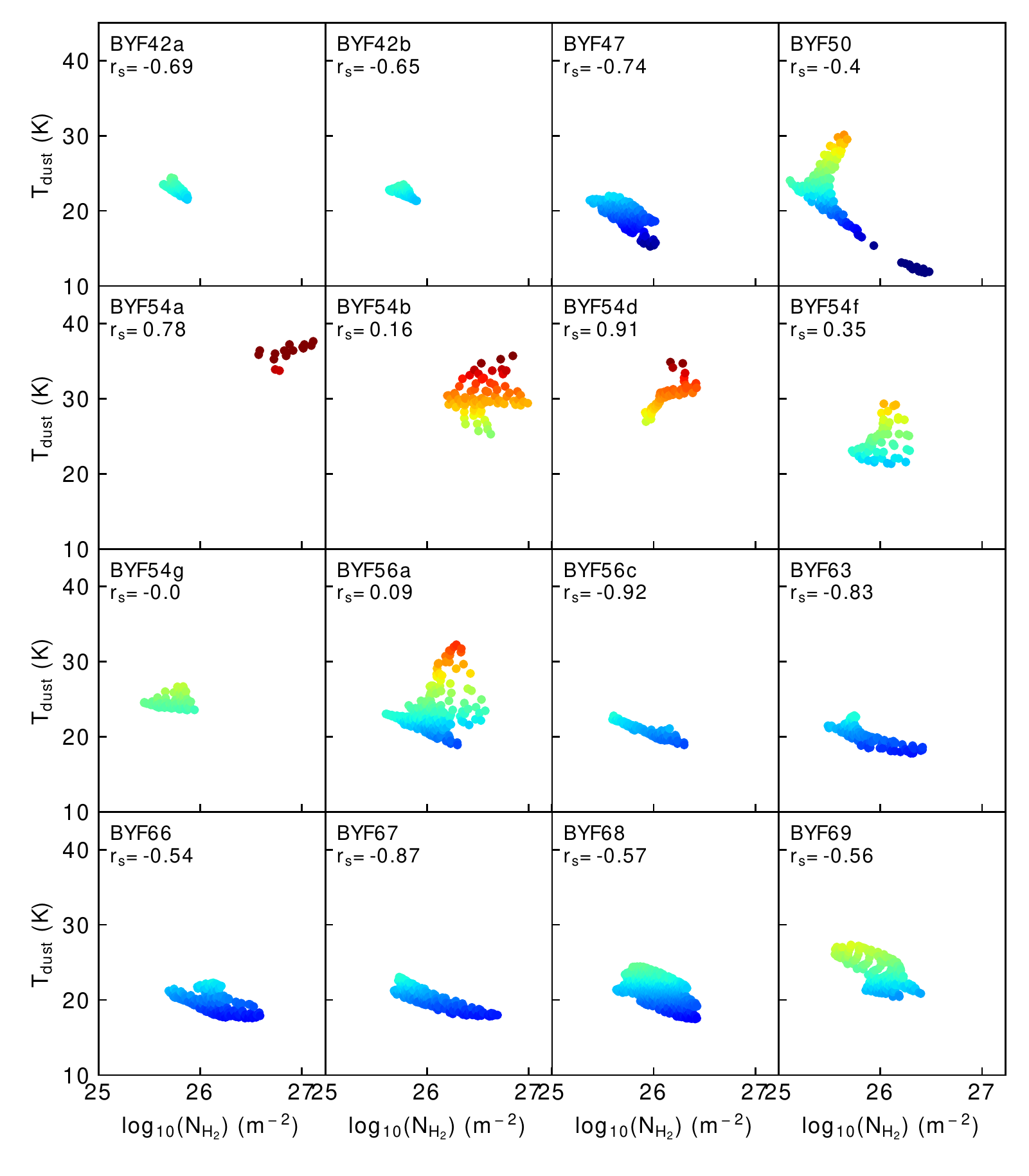}
    \caption{\td\, vs. \ncol\, plots for BYF~42a--69}
    \label{fig:tvn4}
\end{figure}

\begin{figure}
    \centering
    \includegraphics{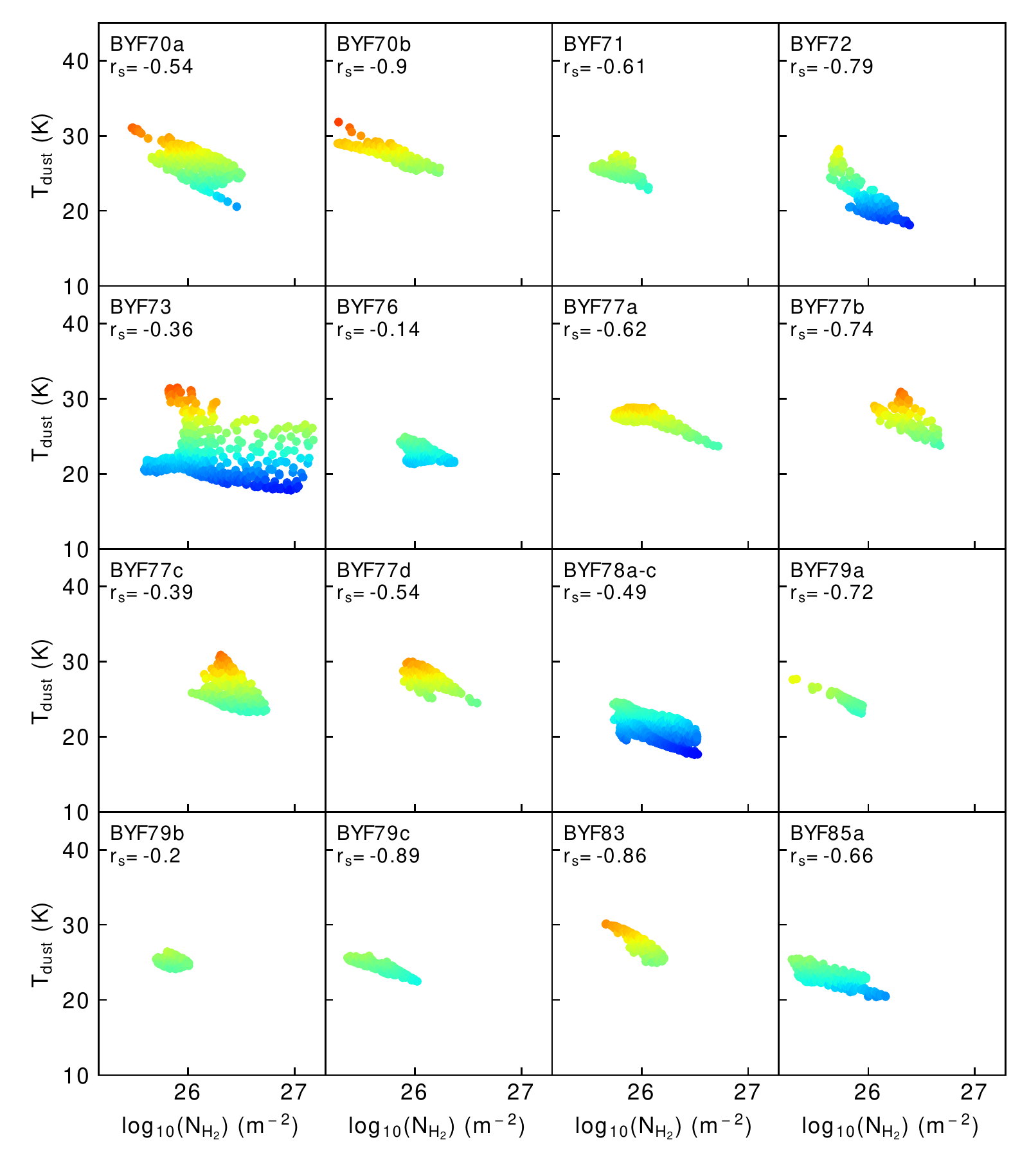}
    \caption{\td\, vs. \ncol\, plots for BYF~70a--85a}
    \label{fig:tvn5}
\end{figure}

\begin{figure}
    \centering
    \includegraphics{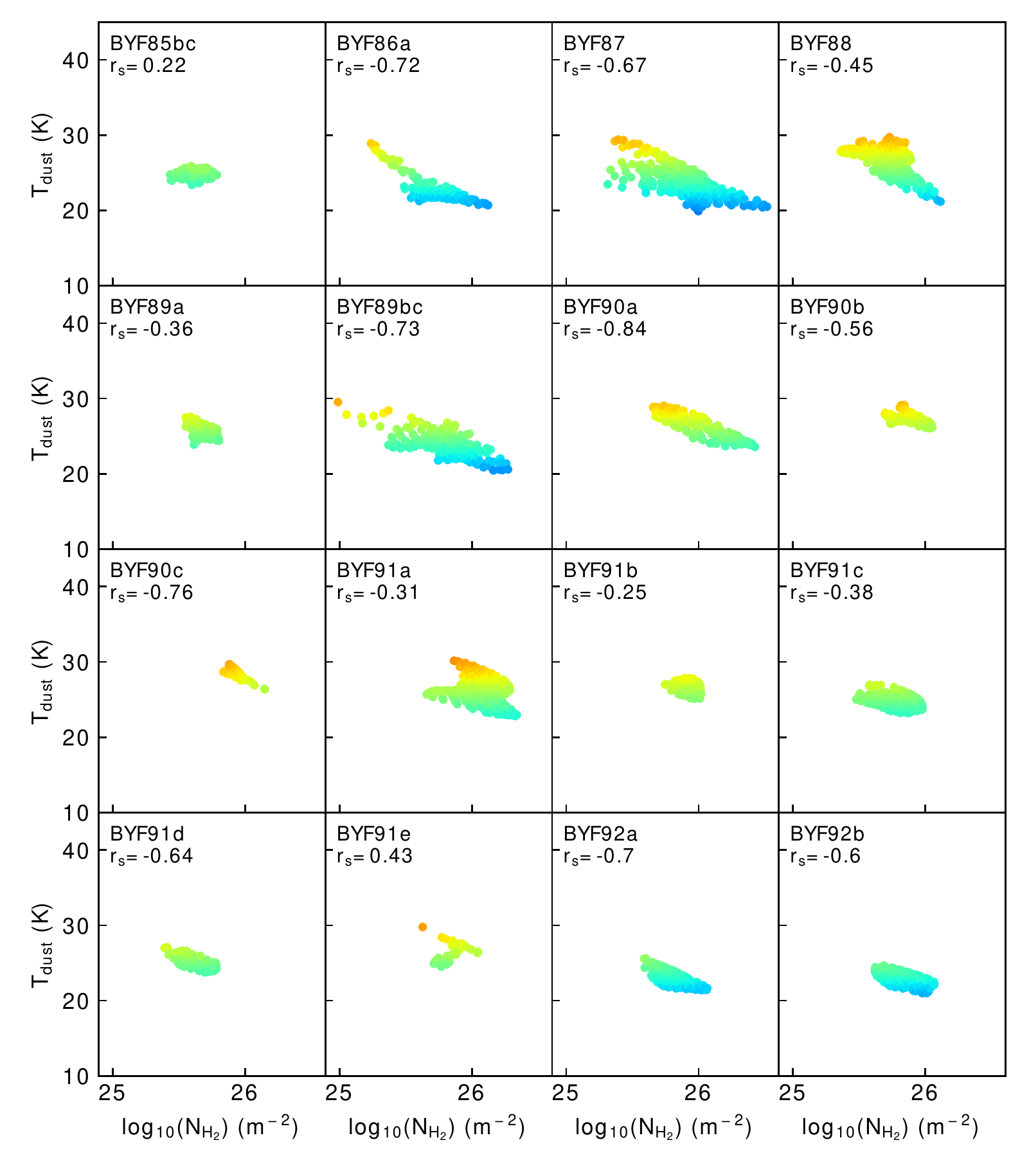}
    \caption{\td\, vs. \ncol\, plots for BYF~85bc--92b. BYF~85b and 85c are not separable without velocity information}
    \label{fig:tvn6}
\end{figure}

\begin{figure}
    \centering
    \includegraphics{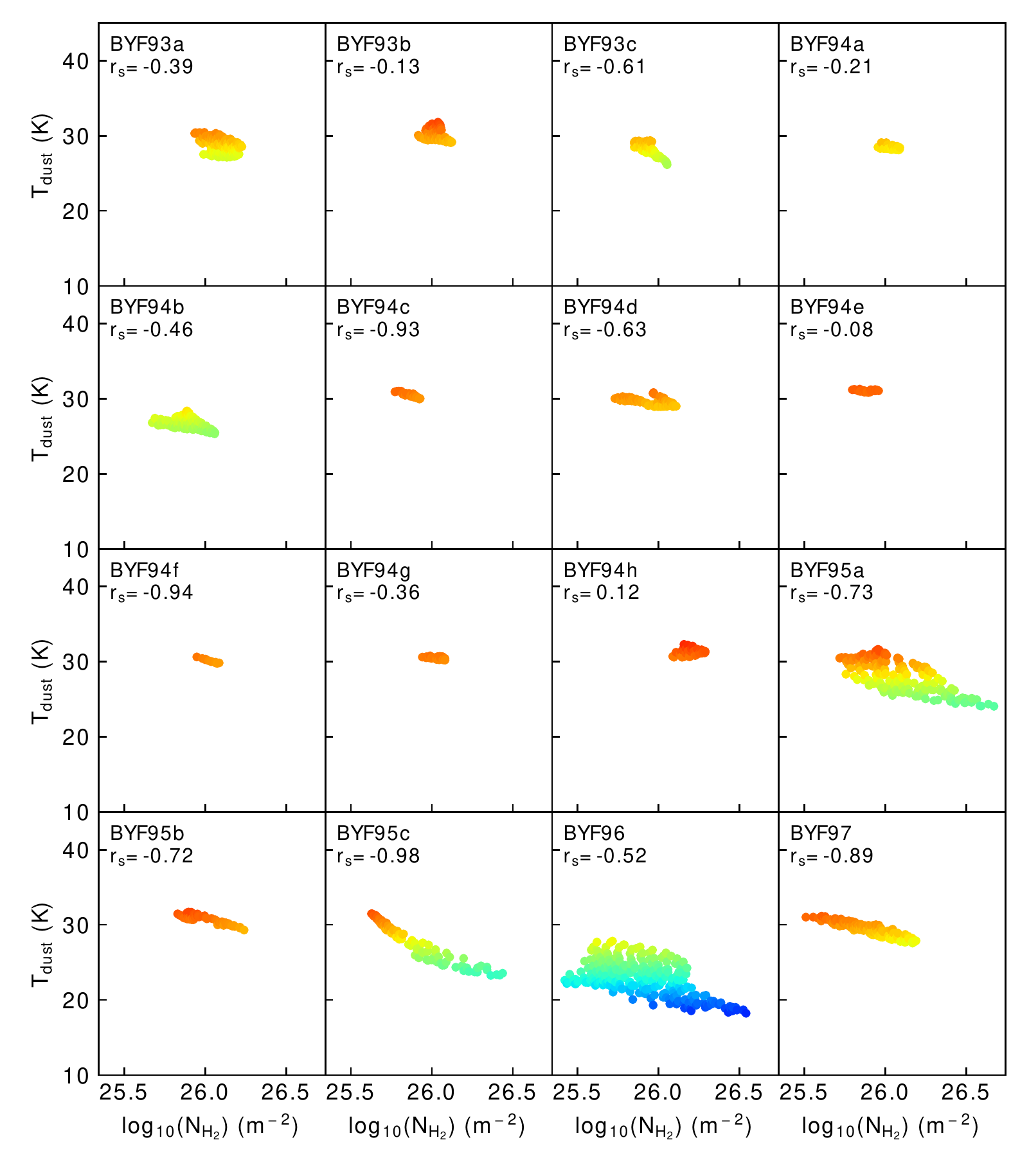}
    \caption{\td\, vs. \ncol\, plots for BYF~93a--97}
    \label{fig:tvn7}
\end{figure}

\begin{figure}
    \centering
    \includegraphics{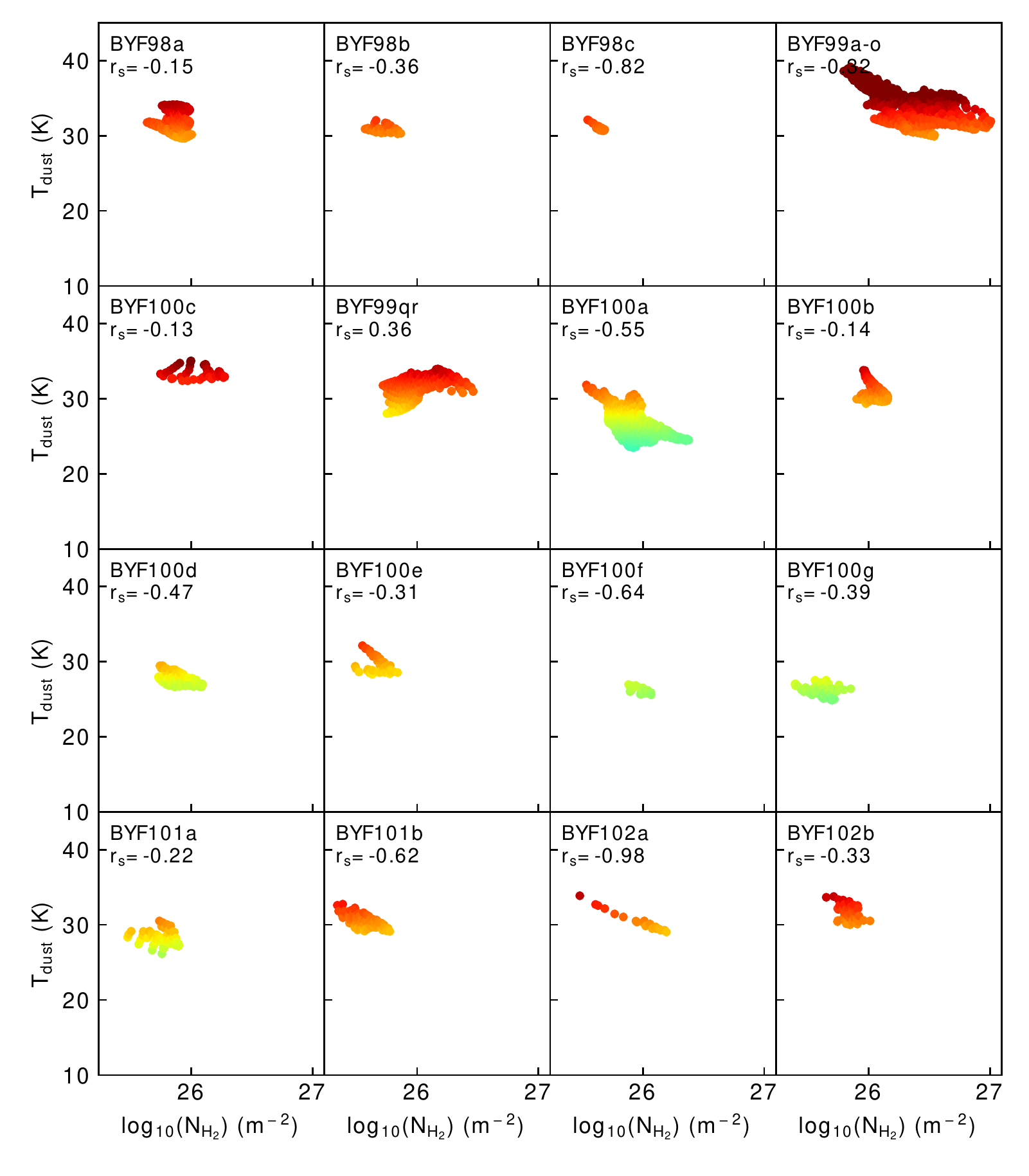}
    \caption{\td\, vs. \ncol\, plots for BYF~98a--102b. Note that most of the components of BYF~99 are stacked along the line of sight, so we had to treat them as only two clumps. BYF~99d lies on top of 100c and so was subsumed by it.}
    \label{fig:tvn8}
\end{figure}

\begin{figure}
\centering
    \includegraphics{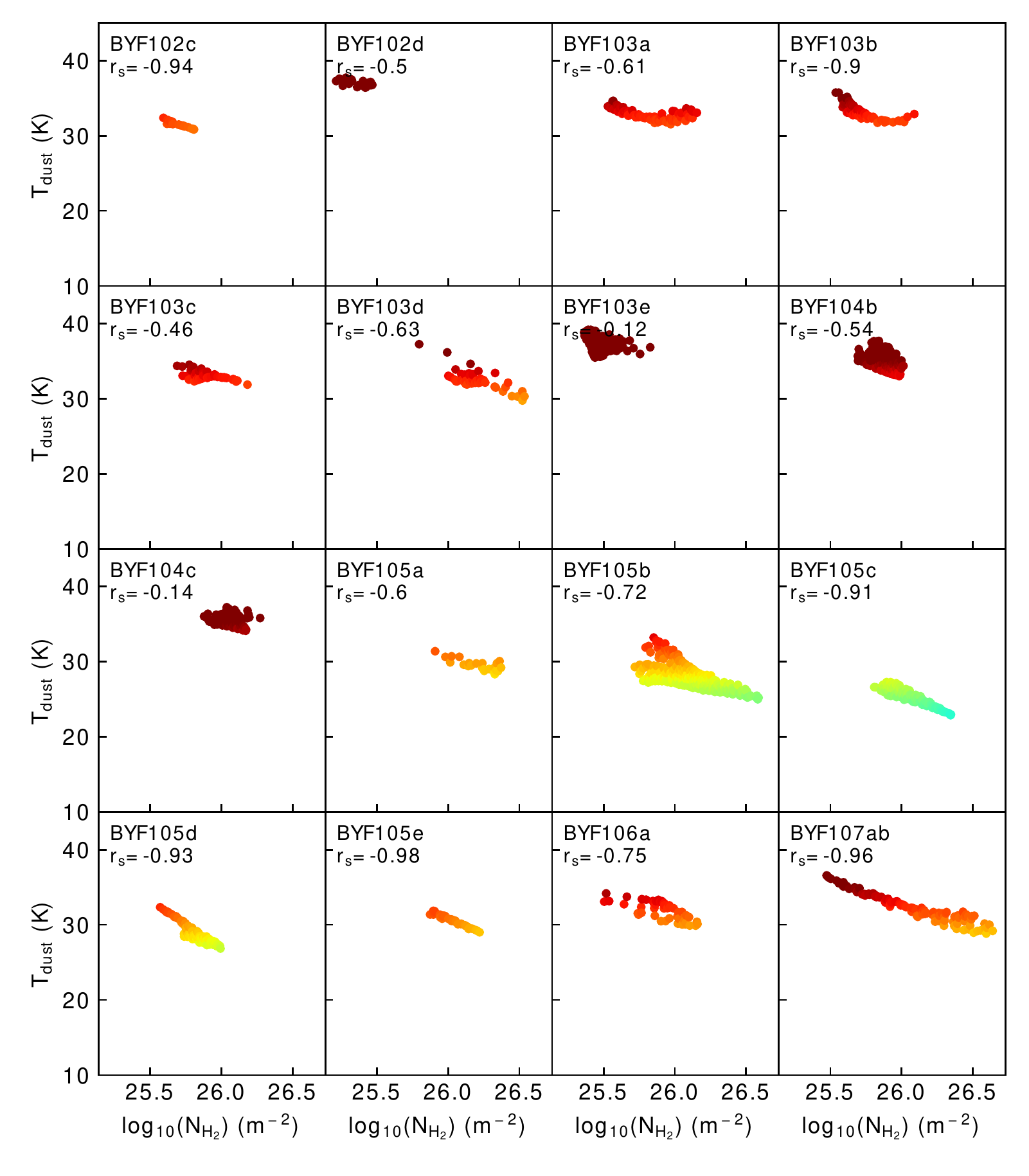}
    \caption{\td\, vs. \ncol\, plots for BYF~102c--107ab. Note the shape of the $T_{\text{d}}$ vs. $N_{\text{H}_2}$ plots for BYF 103a and b; not every cometary globule has this shape, but every clump that has this functional form is a cometary globule.}
    \label{fig:tvn9}
\end{figure}

\begin{figure}
    \centering
    \includegraphics{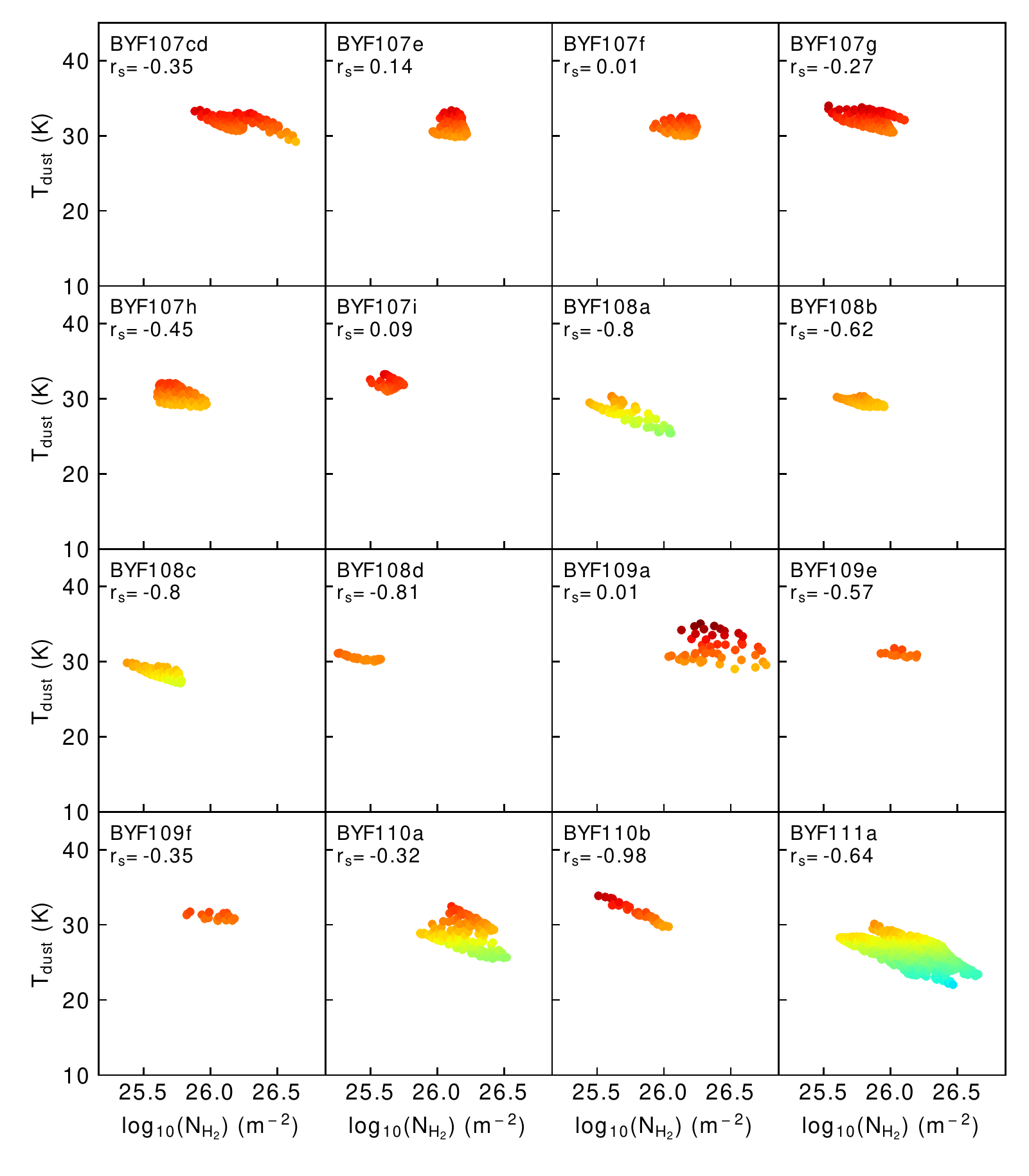}
    \caption{\td\, vs. \ncol\, plots for BYF~107cd--111a}
    \label{fig:tvn10}
\end{figure}

\begin{figure}
    \centering
    \includegraphics{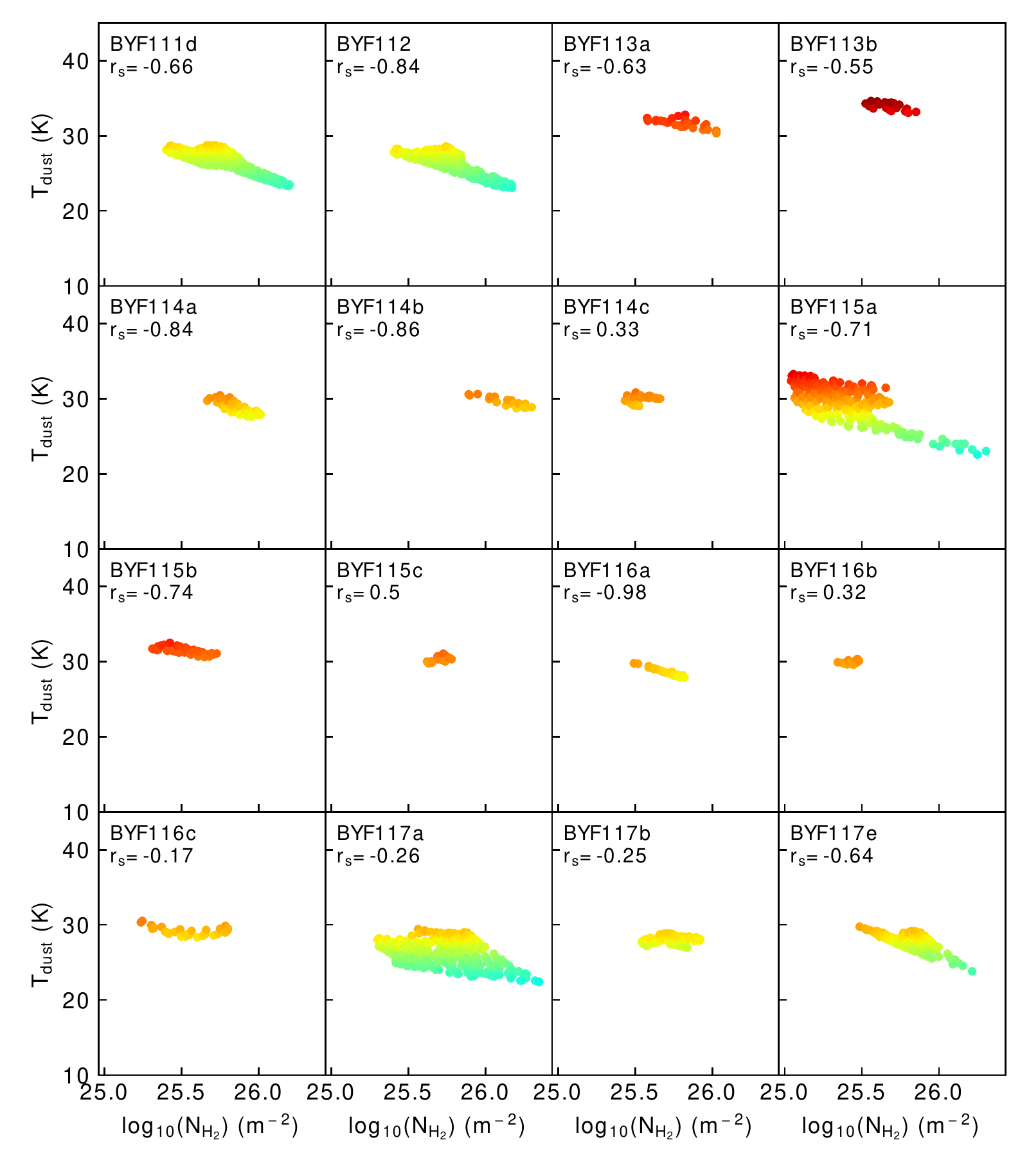}
    \caption{\td\, vs. \ncol\, plots for BYF~111d--117e}
    \label{fig:tvn11}
\end{figure}

\begin{figure}
    \centering
    \includegraphics{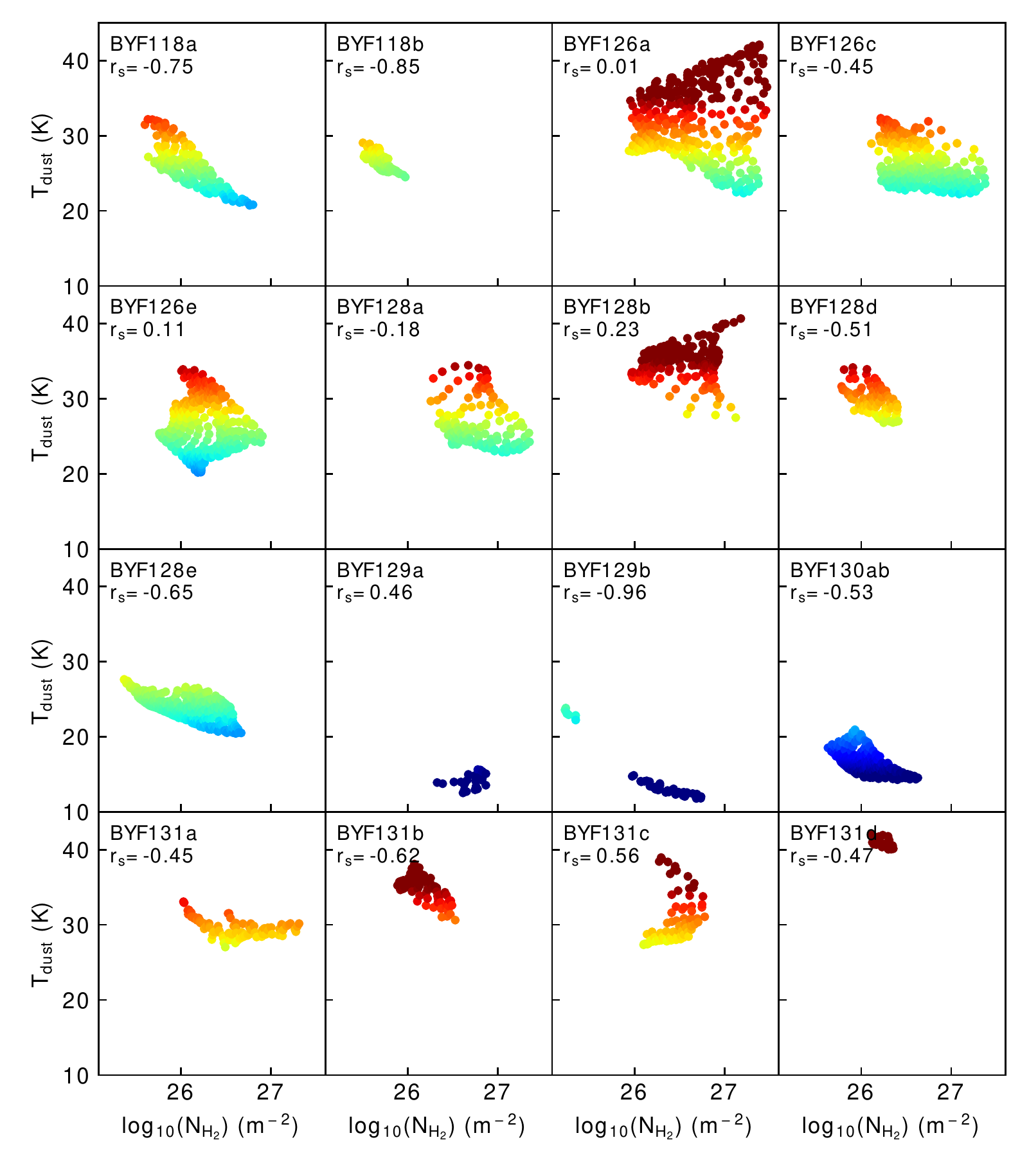}
    \caption{\td\, vs. \ncol\, plots for BYF~118a--131d. Note that the BYF~123 group was moved to the end of this image set because it was not part of any Region.}
    \label{fig:tvn12}
\end{figure}

\begin{figure}
    \centering
    \includegraphics{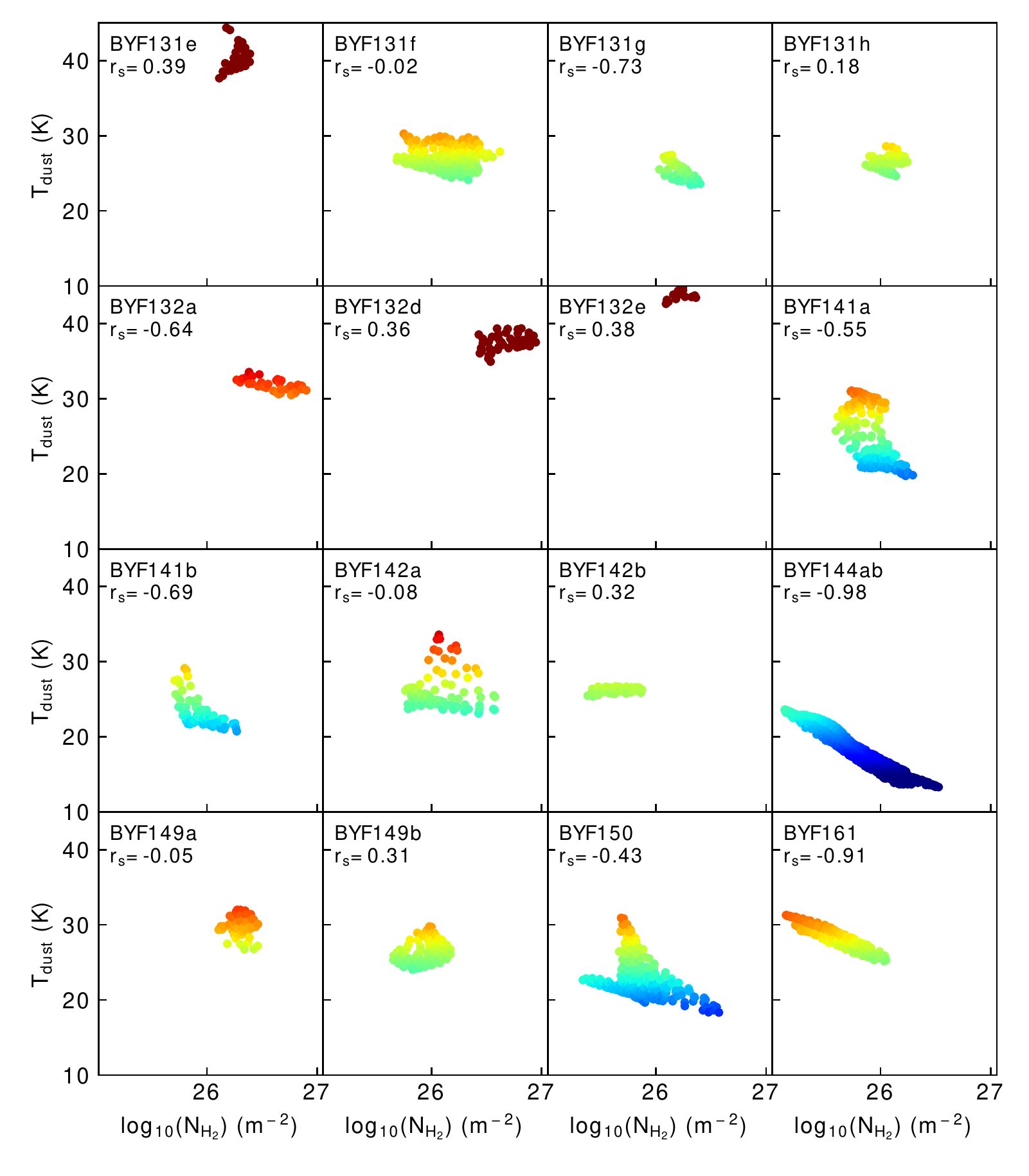}
    \caption{\td\, vs. \ncol\, plots for BYF~131e--161}
    \label{fig:tvn13}
\end{figure}

\begin{figure}
    \centering
    \includegraphics{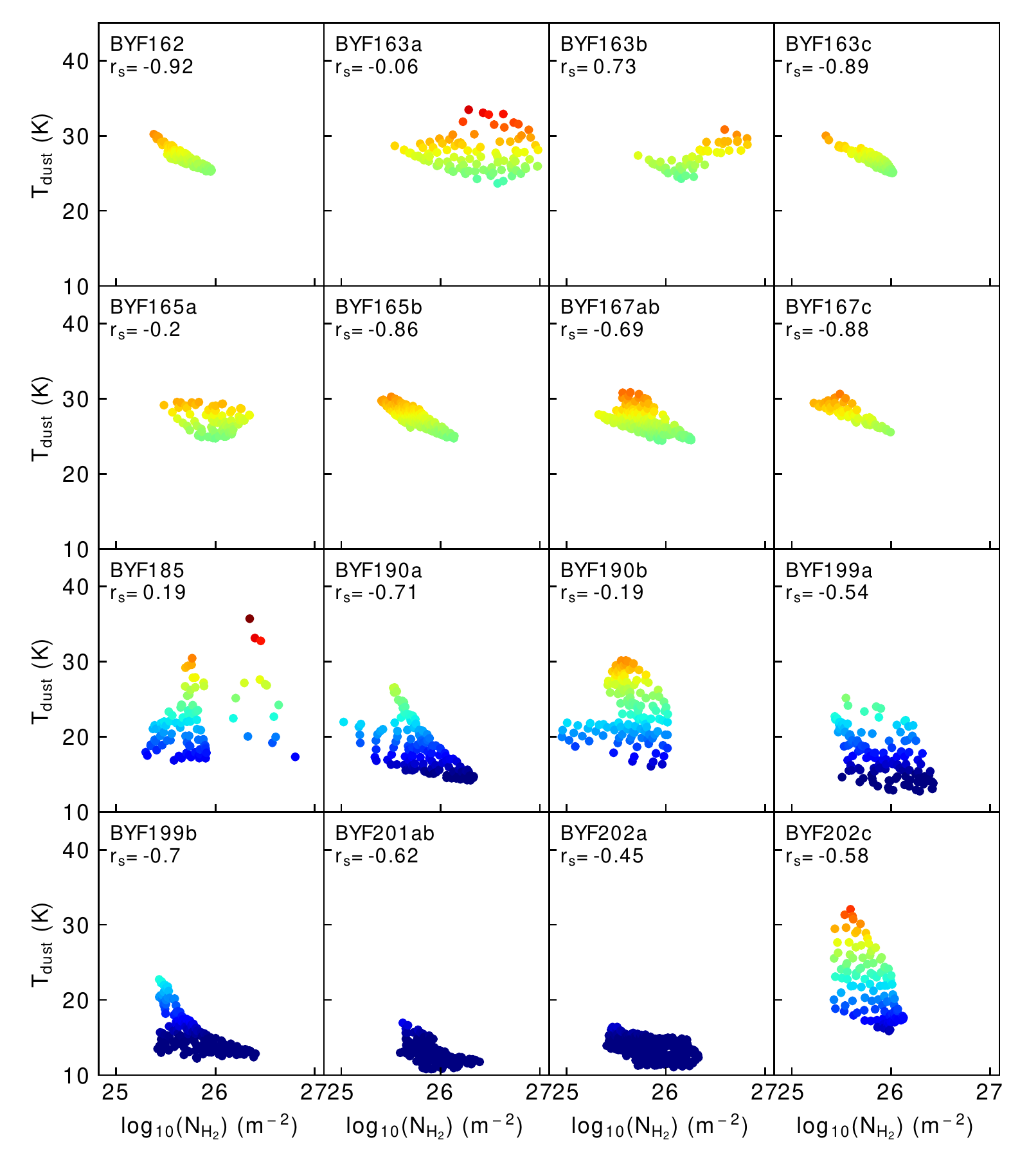}
    \caption{\td\, vs. \ncol\, plots for BYF~162--202c}
    \label{fig:tvn14}
\end{figure}

\begin{figure}
    \centering
    \includegraphics{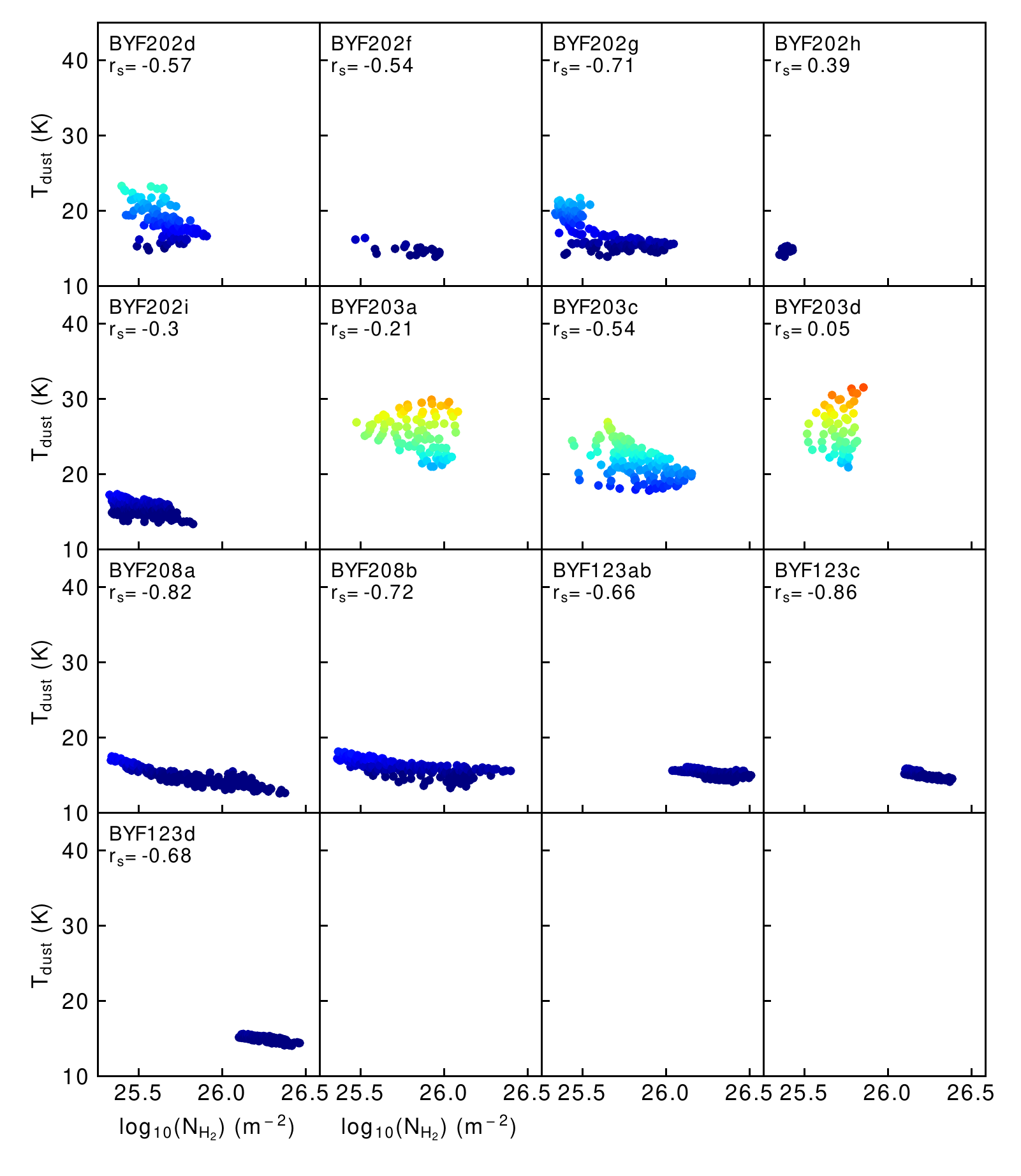}
    \caption{\td\, vs. \ncol\, plots for BYF~202d--208b, plus 123ab, c, and d.}
    \label{fig:tvn15}
\end{figure}

\begin{turnpage}
\begin{figure}
    \centering
    \includegraphics[width=9in]{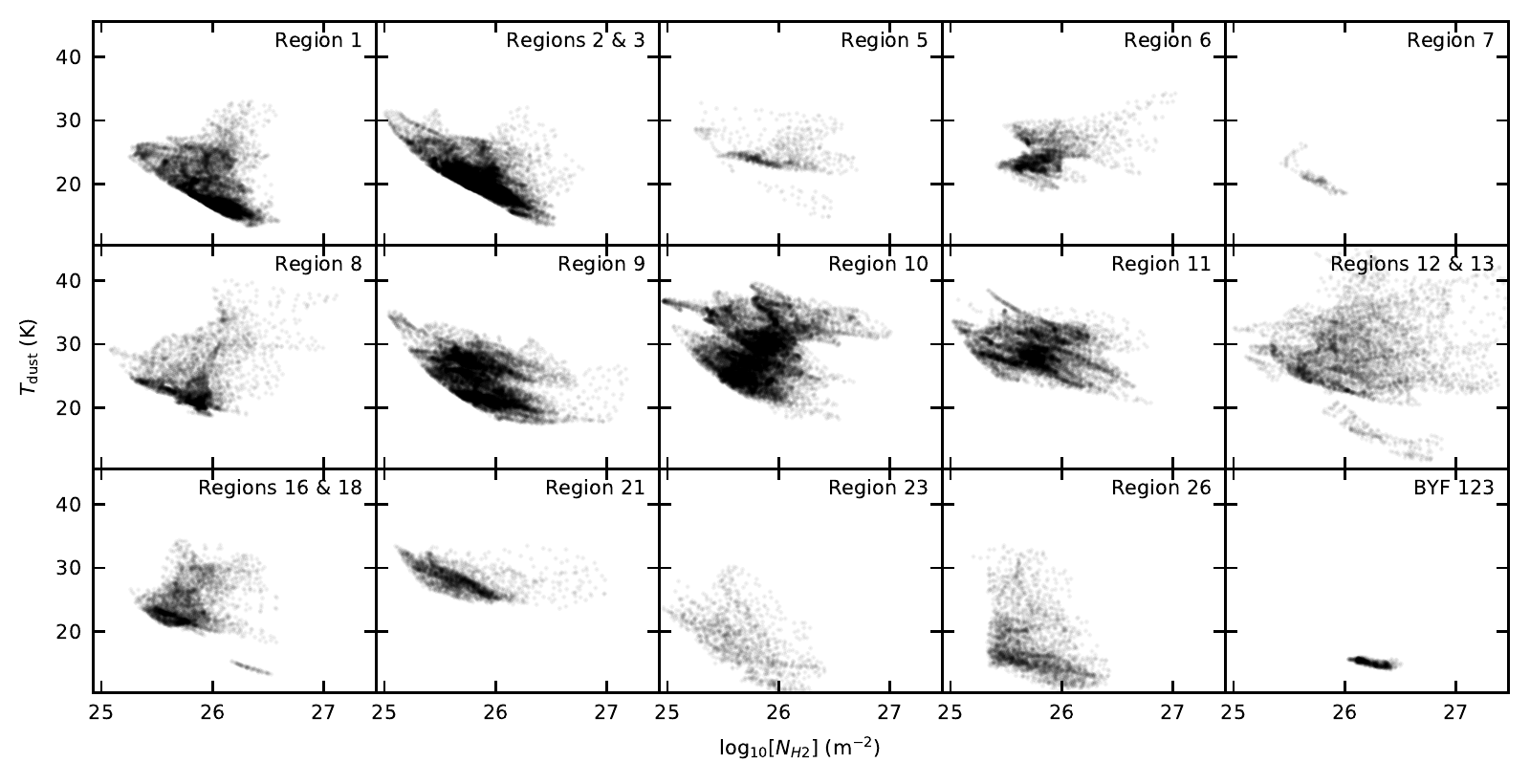}
    \caption{Region-wide pixel-by-pixel plots of \td\ vs. \ncol, masked for where CO data exist. Spatially resolved clumps and filaments structures show up as coherent structures that generally trend downward in \td\ as \ncol\ increases.}
    \label{fig:tvnallregs}
\end{figure}
\end{turnpage}

\section{CO Abundance Maps}\label{sec:app1}
These are the CO abundance maps for CHaMP Regions 1 through 8, 12, 16, 18, 21, and 23. Region 13 is displayed in the main text in figure~\ref{fig:coab13}, and Regions 9 through 11 and 26 were published in \citealt{pittsmn}. Some Regions were split into smaller pieces to eliminate dead space, and in these cases, adjacent sub-Regions may be displayed right-to-left as they would appear on the sky instead of left-to-right in name order.

\begin{figure}
\centering
    \includegraphics[width=\textwidth]{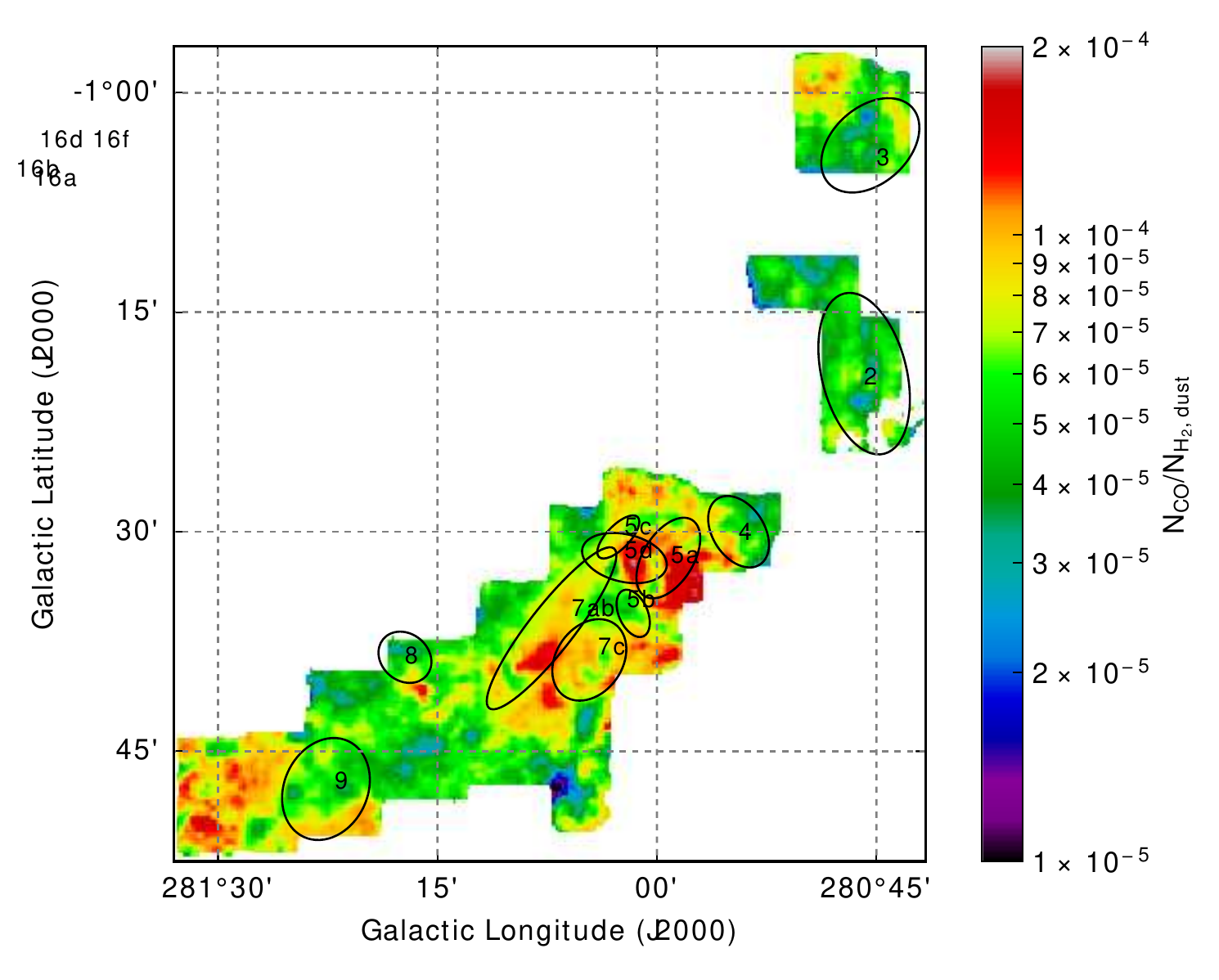}
    \caption{CO abundance map for Region 1.}
    \label{fig:nrat1}
\end{figure}

\begin{figure}
    \centering
        \begin{tabular}{cc}
      \includegraphics[width=0.47\textwidth]{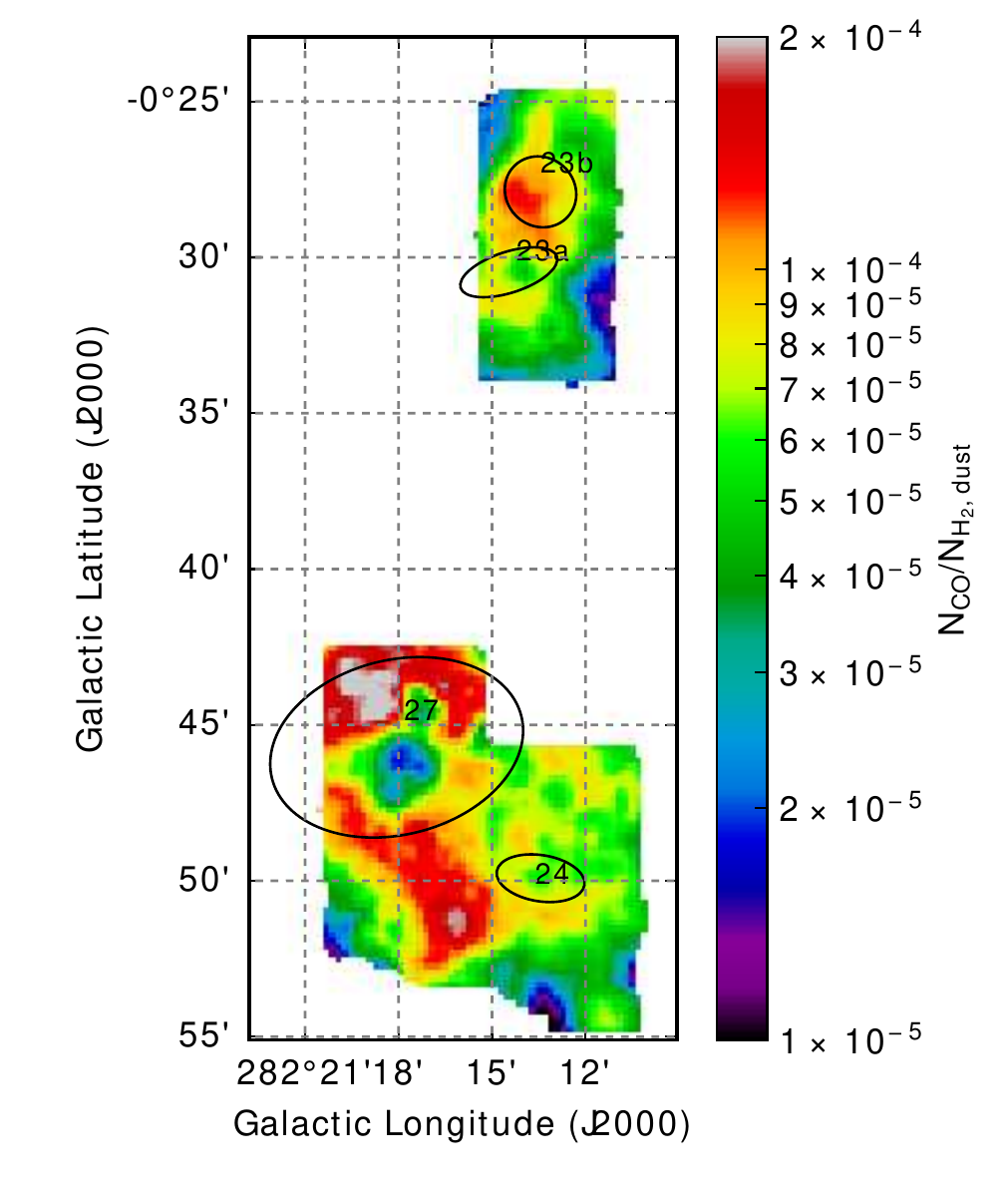}   &  \includegraphics[width=0.5\textwidth]{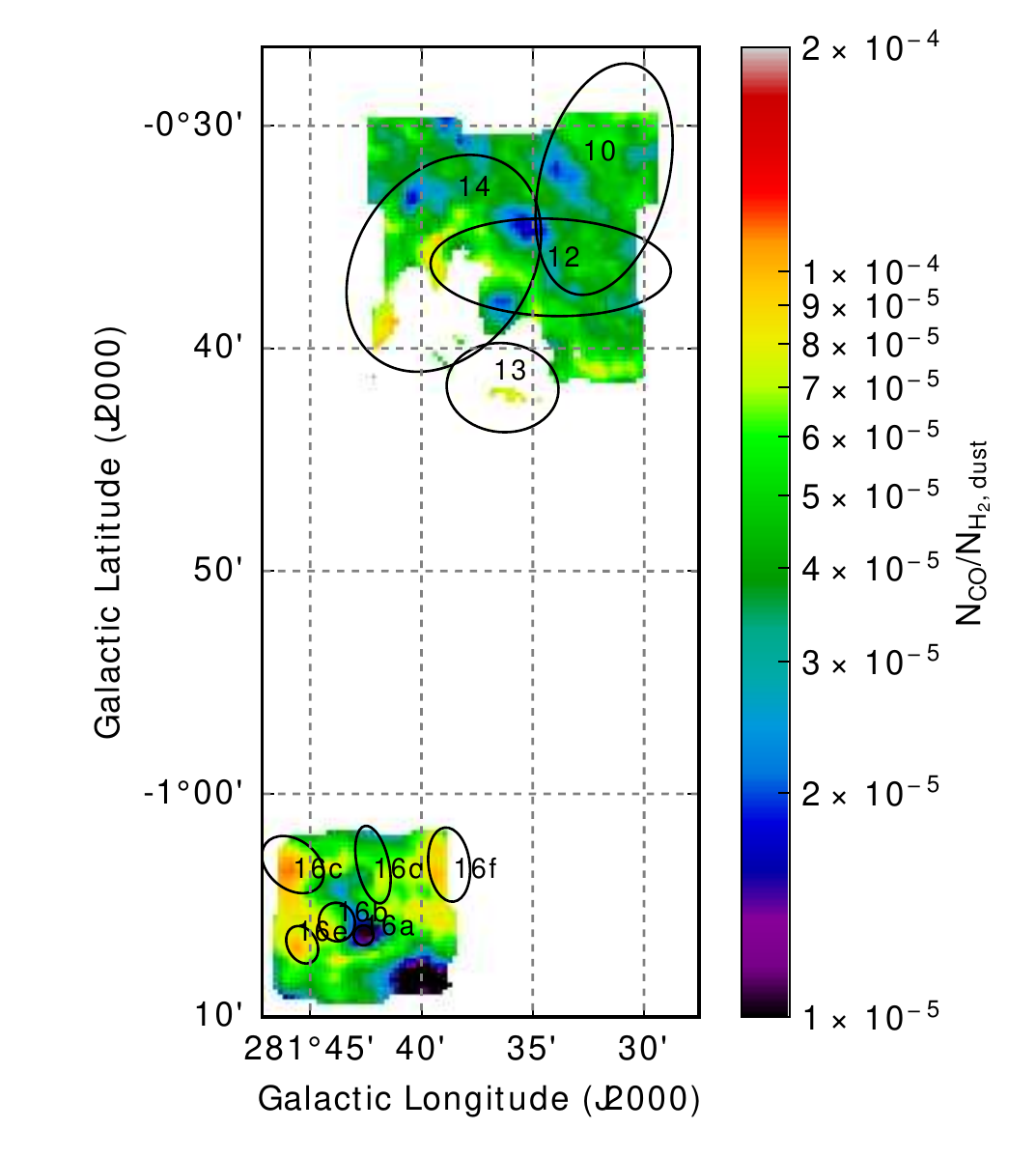}
        \end{tabular}
    \caption{CO abundance maps for Regions 2a (right) and 2b (left)}
    \label{fig:nrat2}
\end{figure}

\begin{figure}
    \centering
    \includegraphics[width=\textwidth]{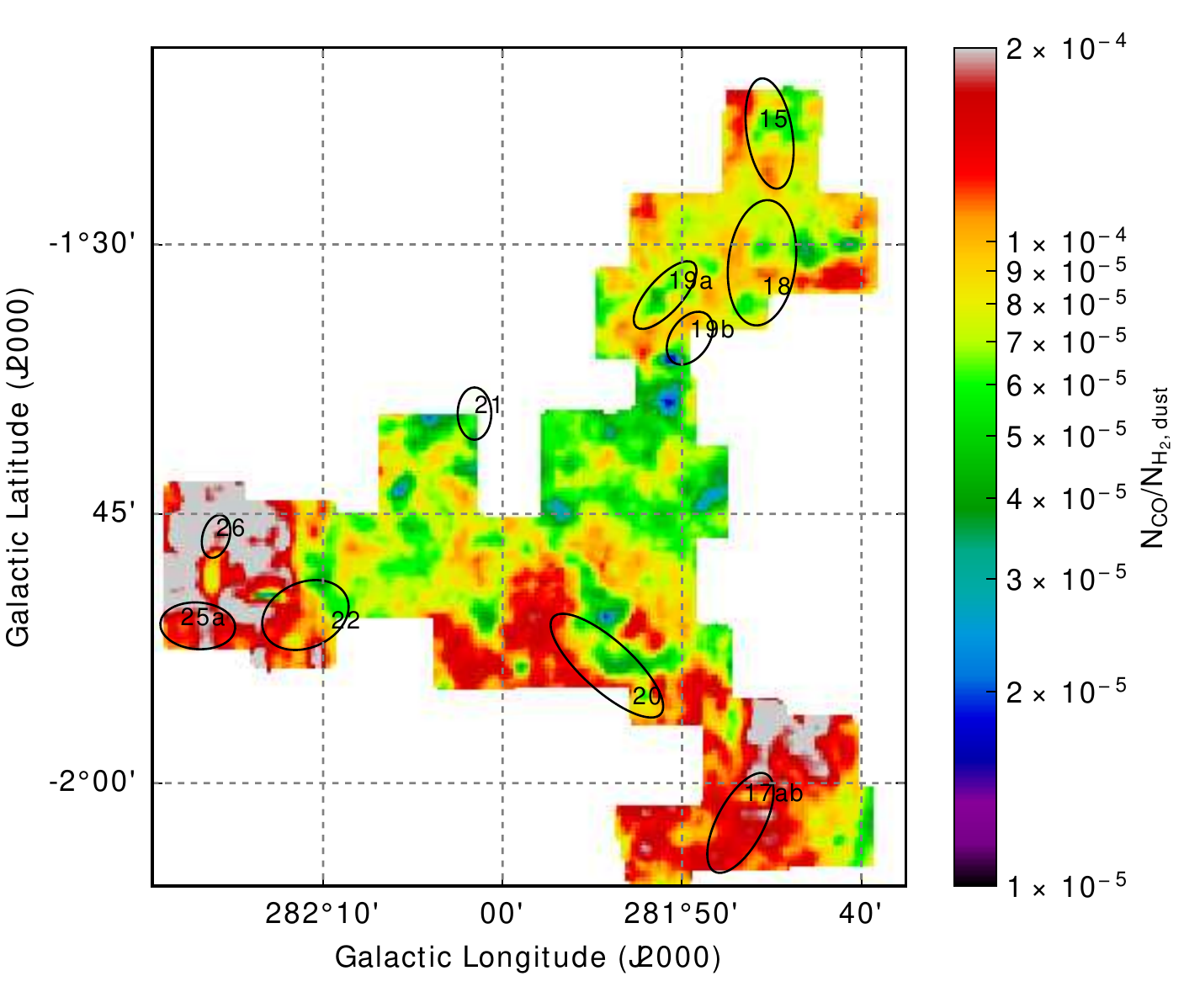}
    \caption{CO abundance map for Region 3}
    \label{fig:nrat3}
\end{figure}

\begin{figure}
    \centering
    \subfloat{\includegraphics[width=0.82\textwidth]{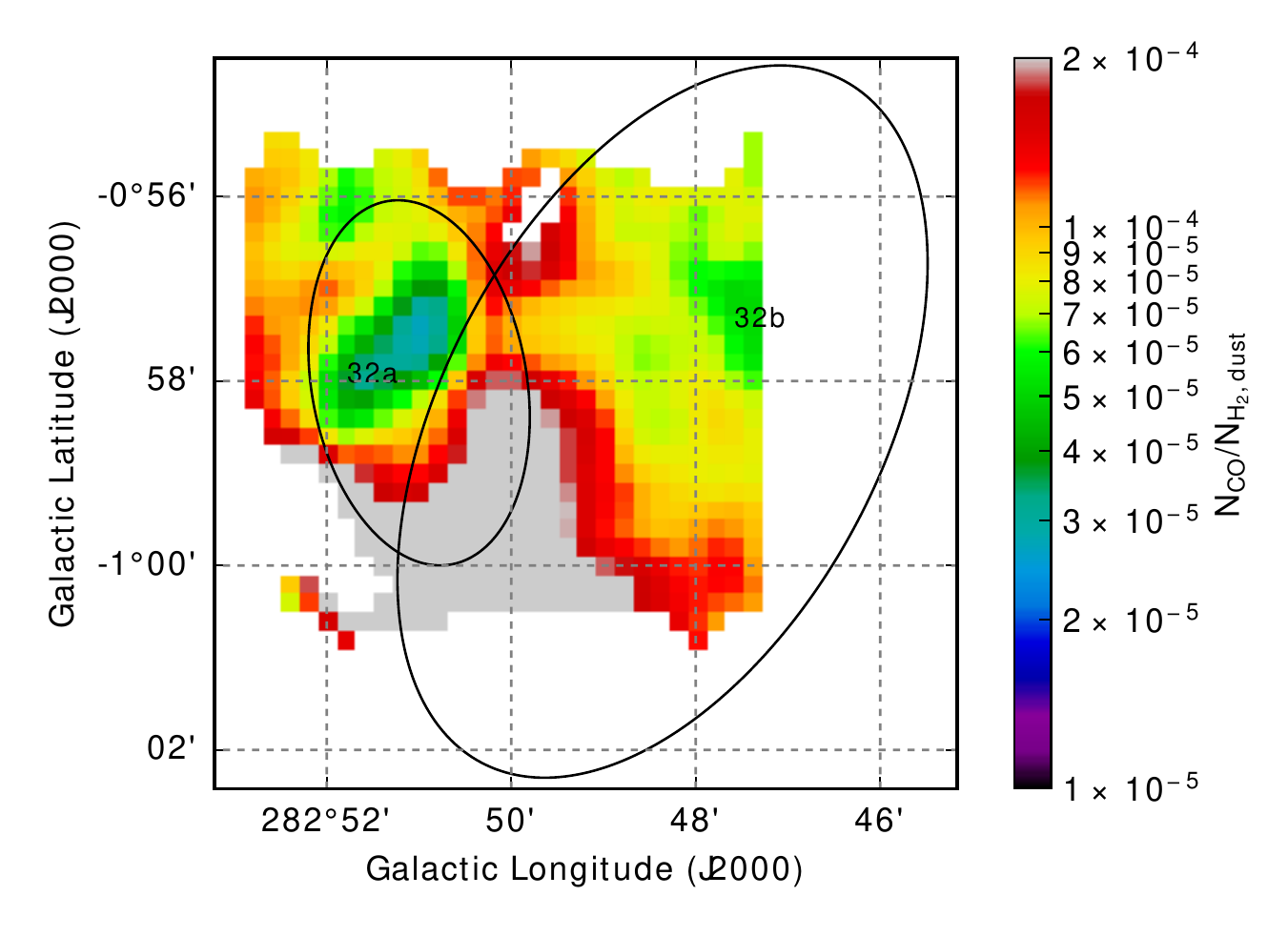}}\\
    \subfloat{\includegraphics[width=0.82\textwidth]{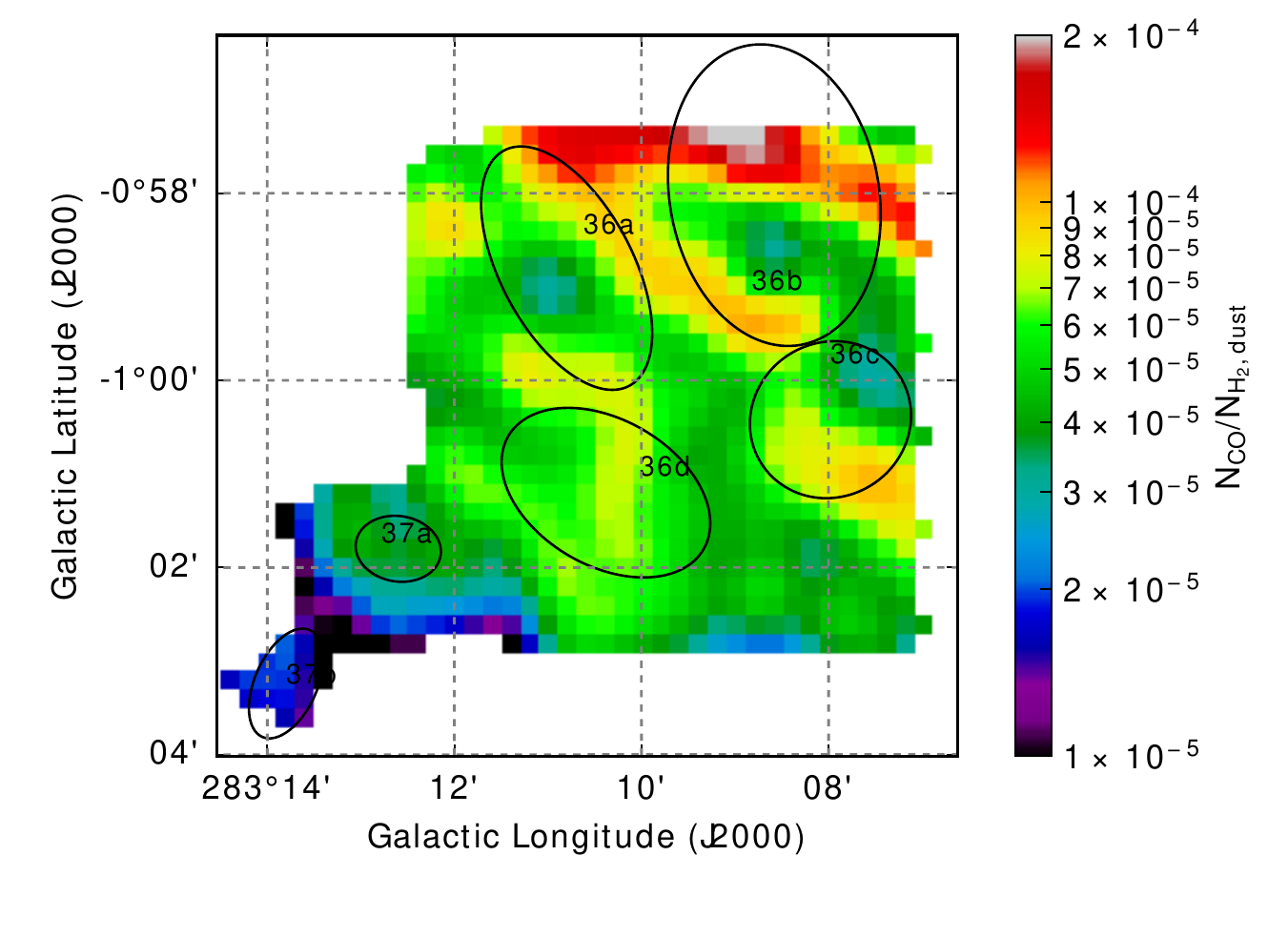}}
    \caption{CO abundance maps for Region 5a (top) and 5b (bottom).}
    \label{fig:nrat5}
\end{figure}

\begin{figure}
    \centering
    \includegraphics[width=\textwidth]{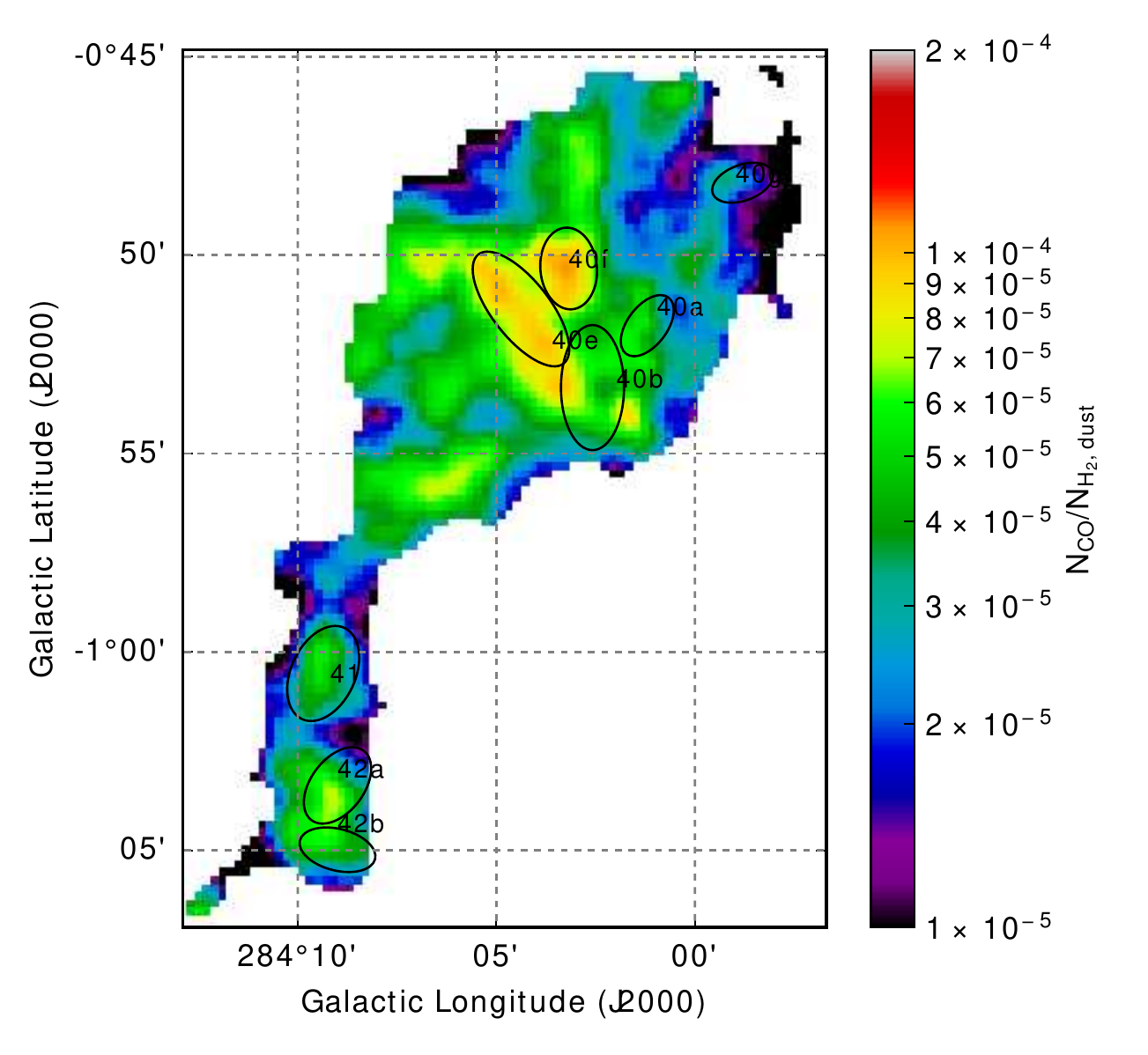}
    \caption{CO abundance map for Region 6}
    \label{fig:nrat6}
\end{figure}

\begin{figure}
    \centering
    \includegraphics{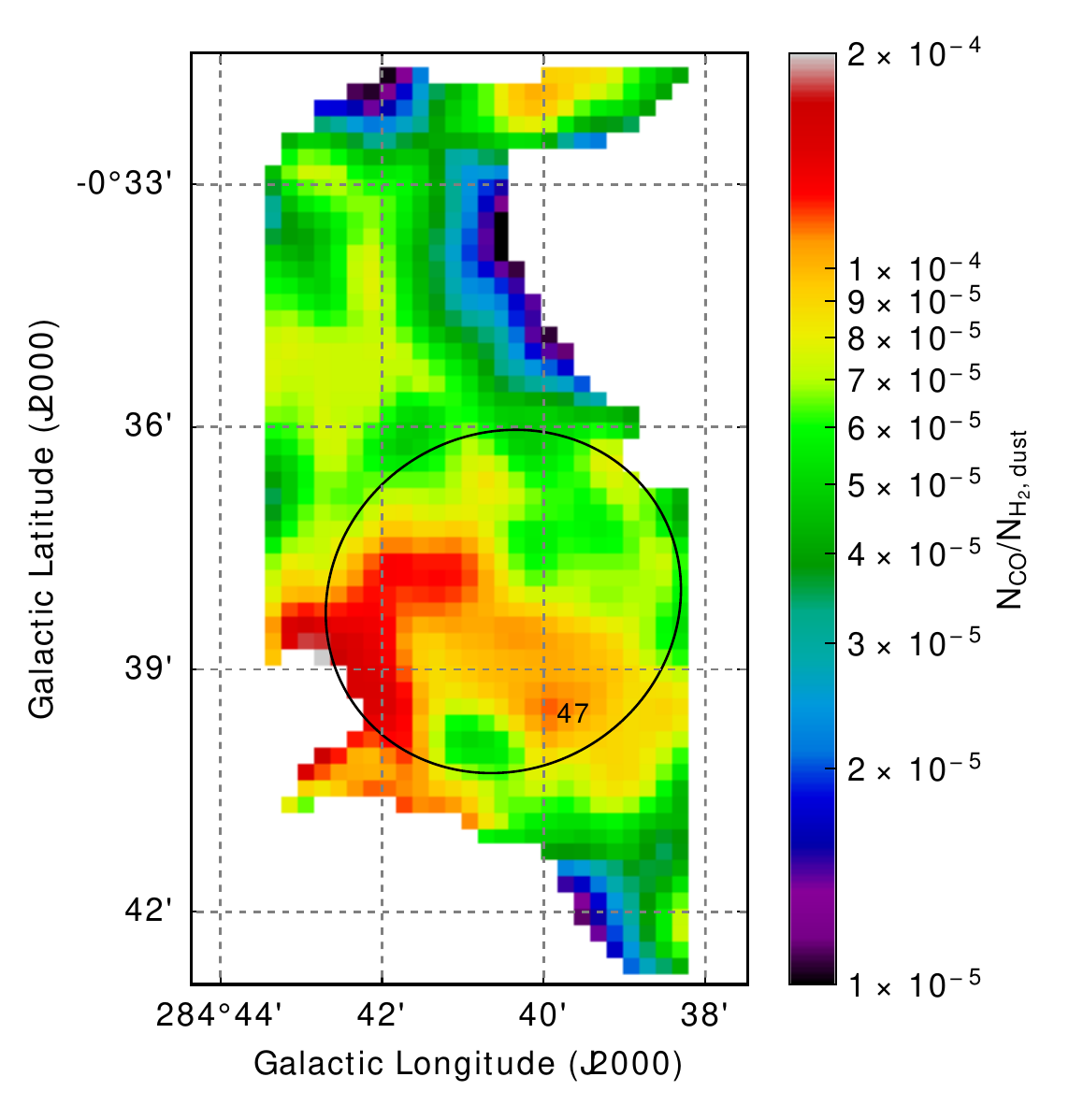}
    \caption{CO abundance map for Region 7}
    \label{fig:nrat7}
\end{figure}
\pagebreak

\begin{figure}
    \centering
     \subfloat{\includegraphics[width=0.8\textwidth]{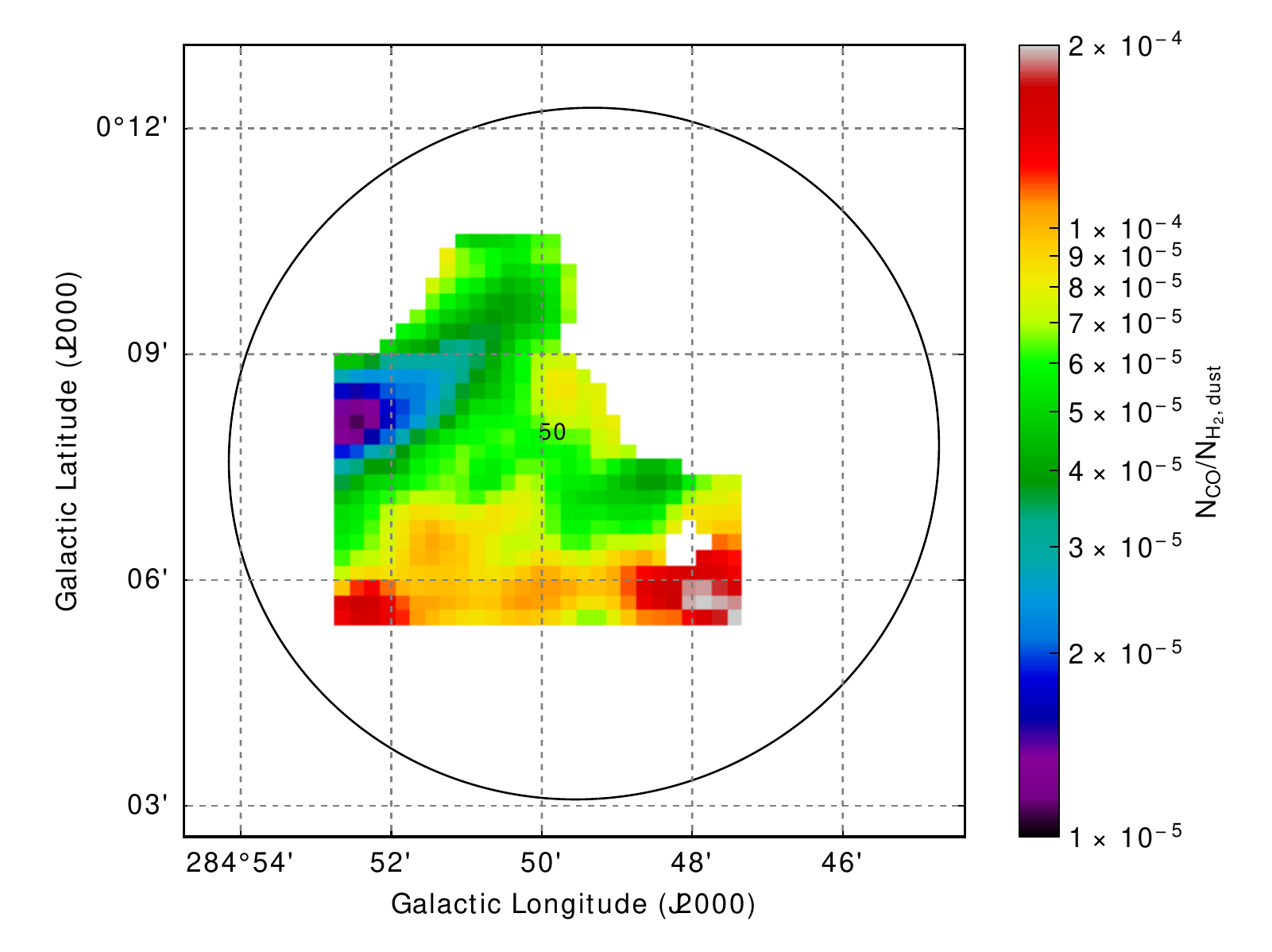}}\\
     \subfloat{\includegraphics[width=0.8\textwidth]{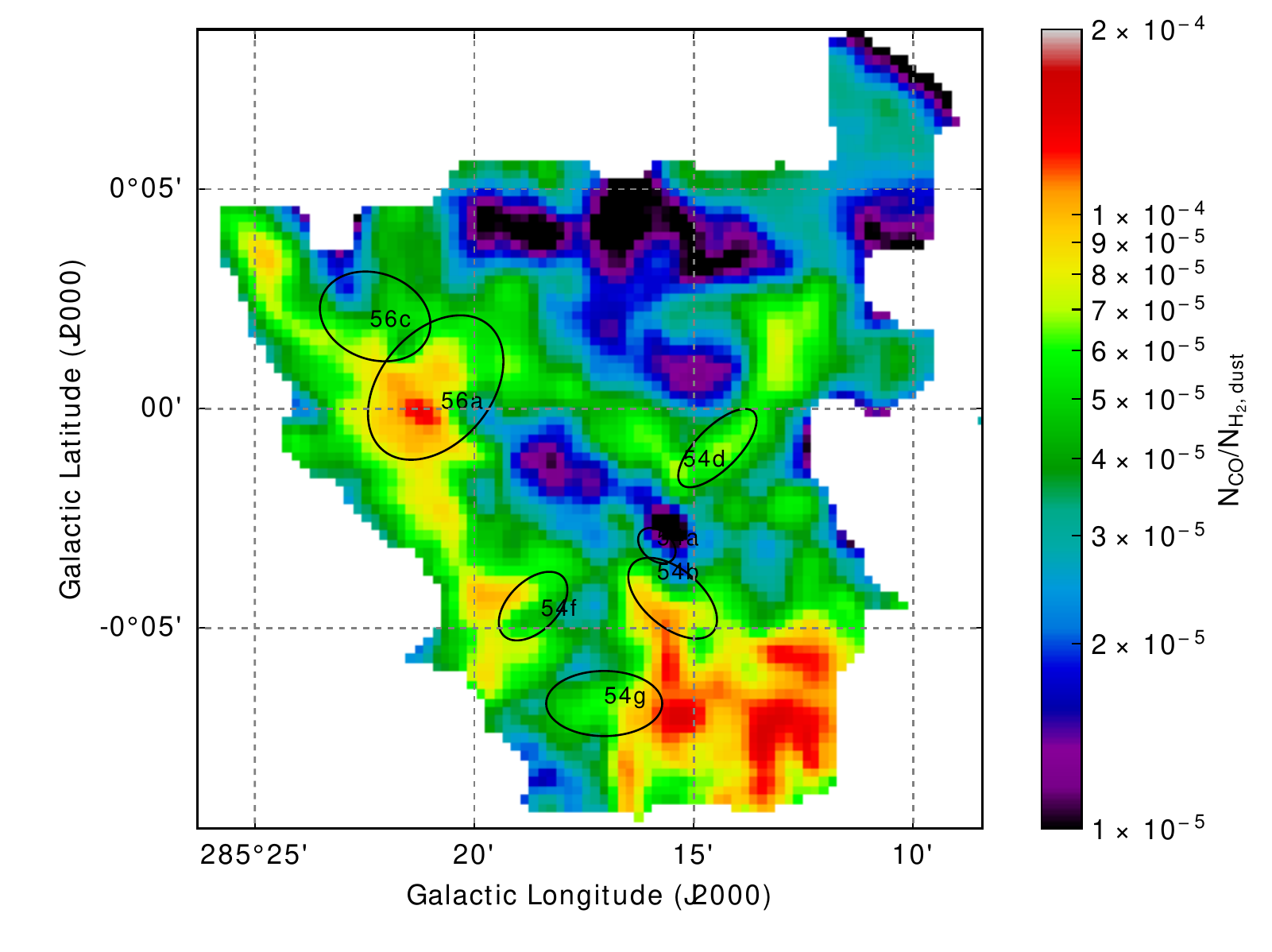}}
    \caption{CO abundance maps for BYF~50 (top) and the rest of Region 8 (bottom)}
    \label{fig:nrat8}
\end{figure}

\begin{figure}
    \centering
    \includegraphics[width=\textwidth]{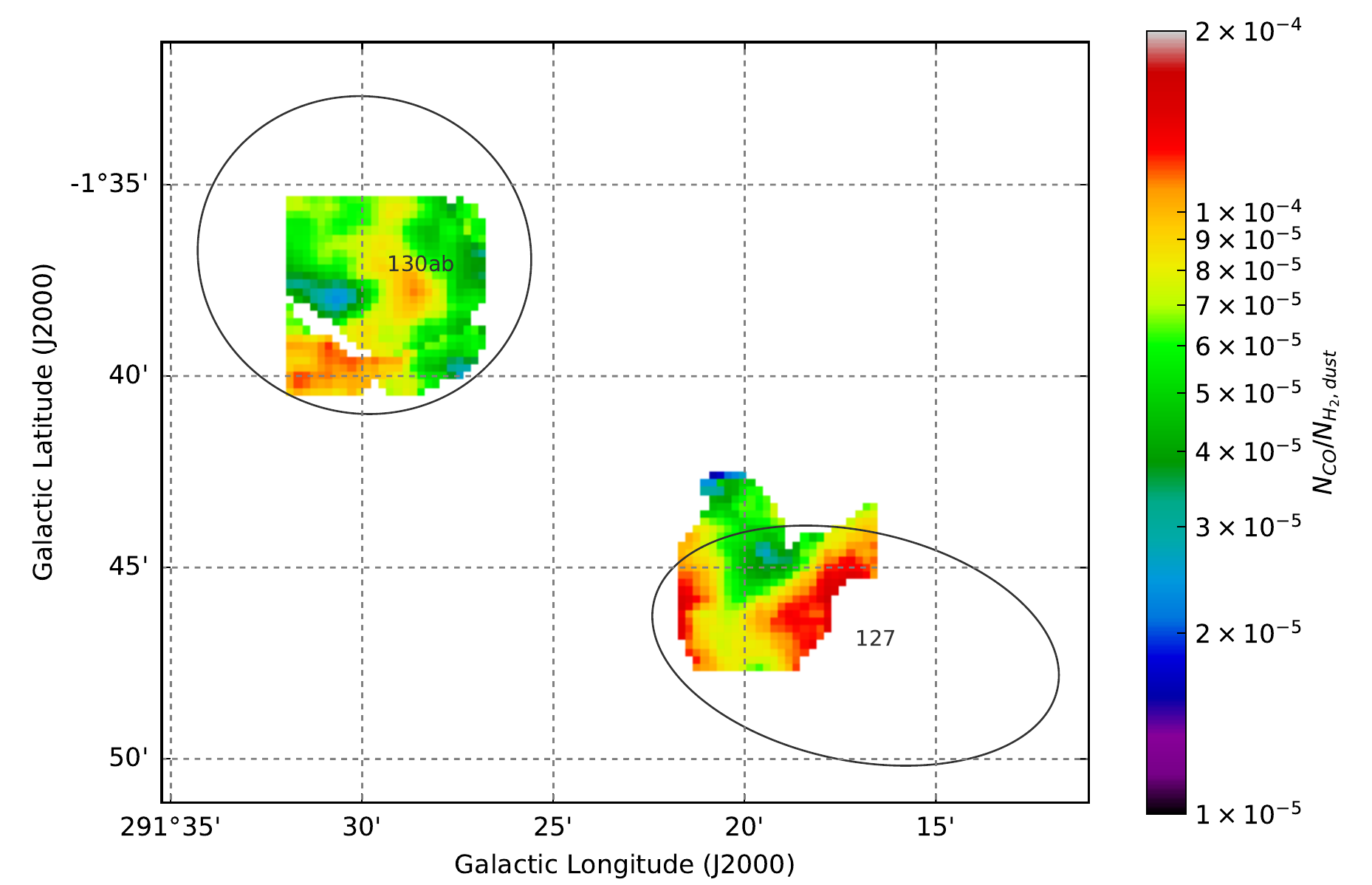}
    \caption{CO abundance map for Region 12}
    \label{fig:nrat12}
\end{figure}

\begin{figure}
    \centering
    \includegraphics[width=\textwidth]{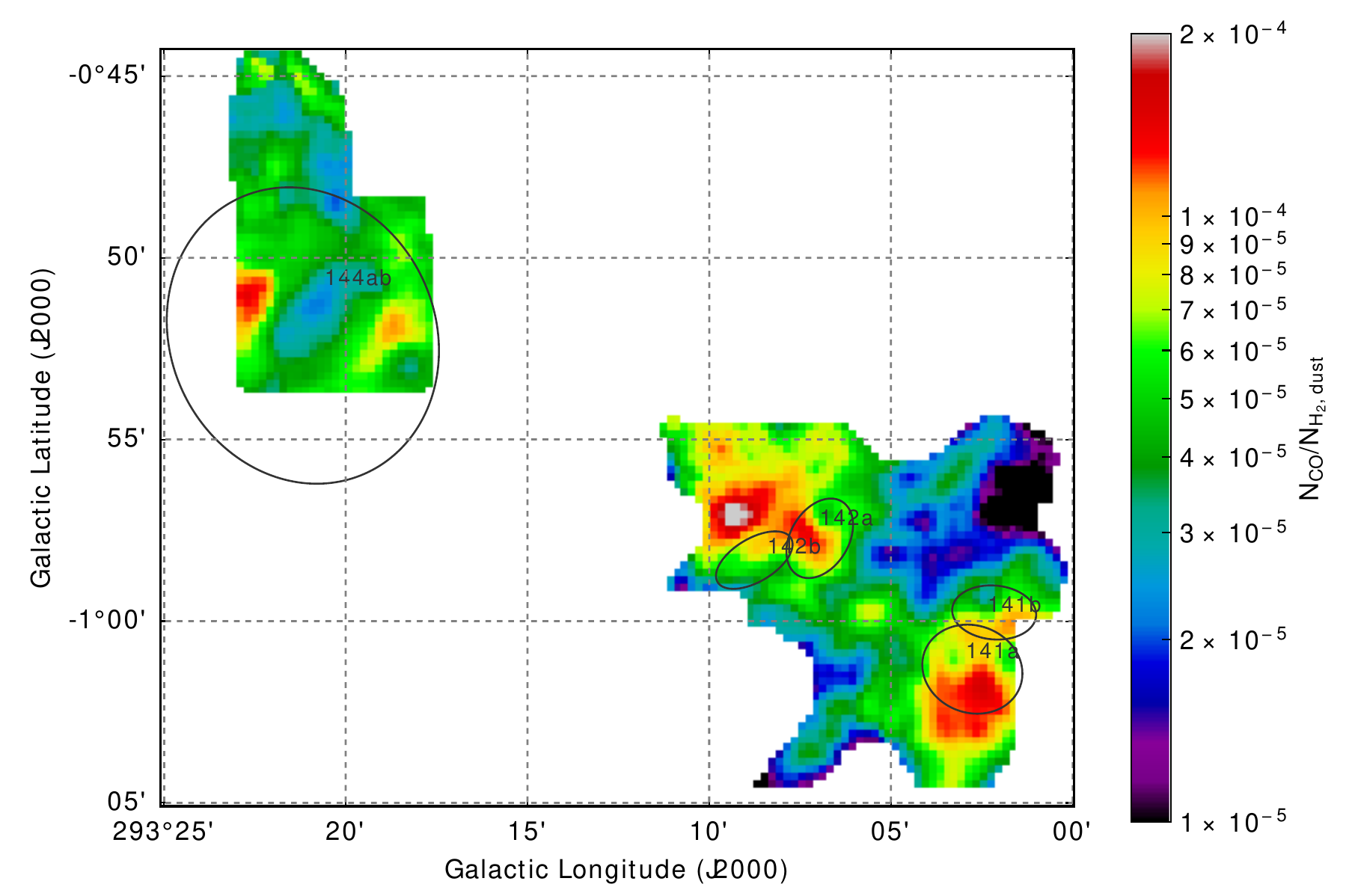}
    \caption{CO abundance map for Region 16}
    \label{fig:nrat16}
\end{figure}

\begin{figure}
    \centering
    \includegraphics[width=\textwidth]{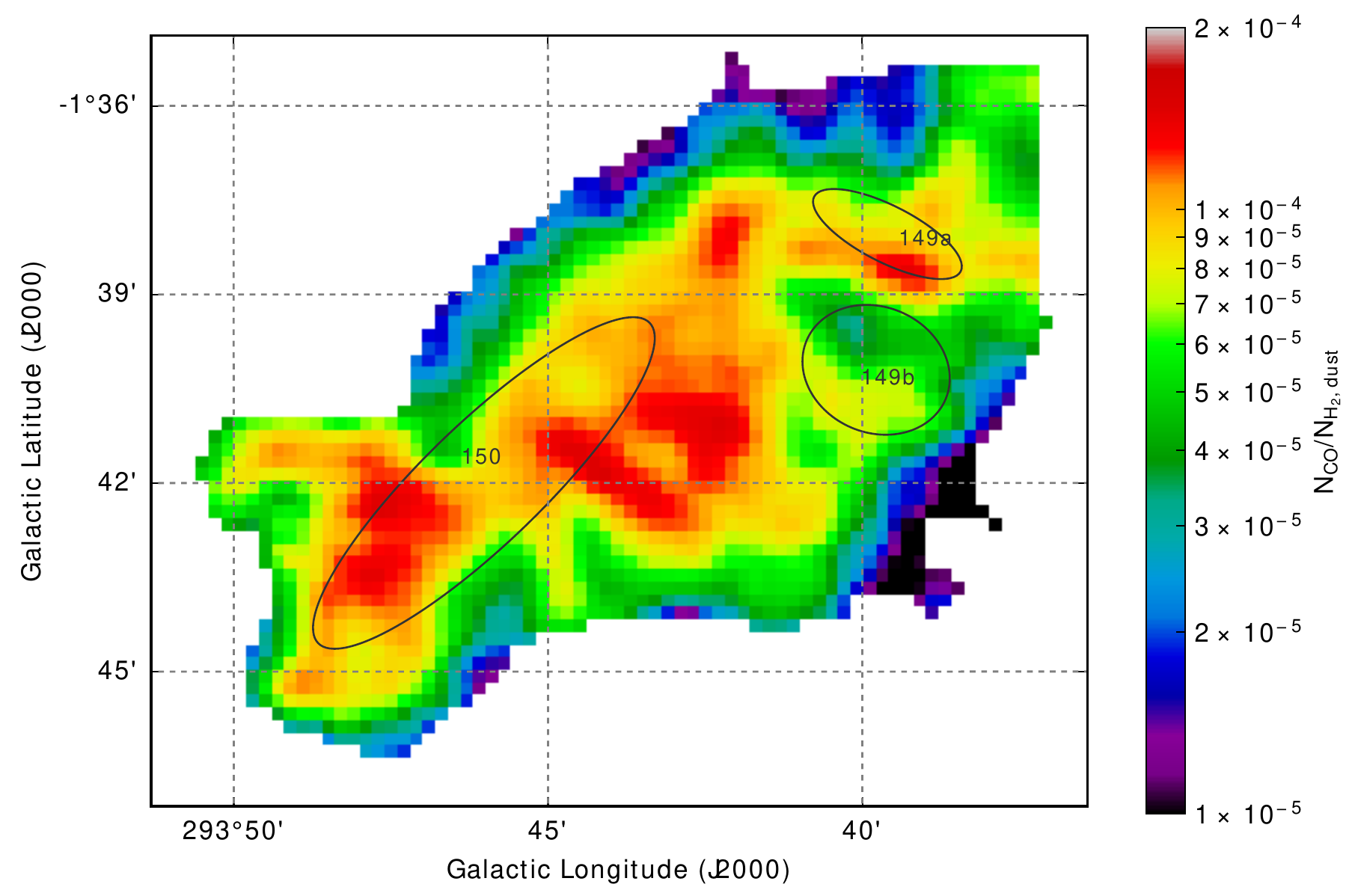}
    \caption{CO abundance map for Region 18}
    \label{fig:nrat18}
\end{figure}

\begin{figure}
    \centering
    \includegraphics[width=1.1\textwidth]{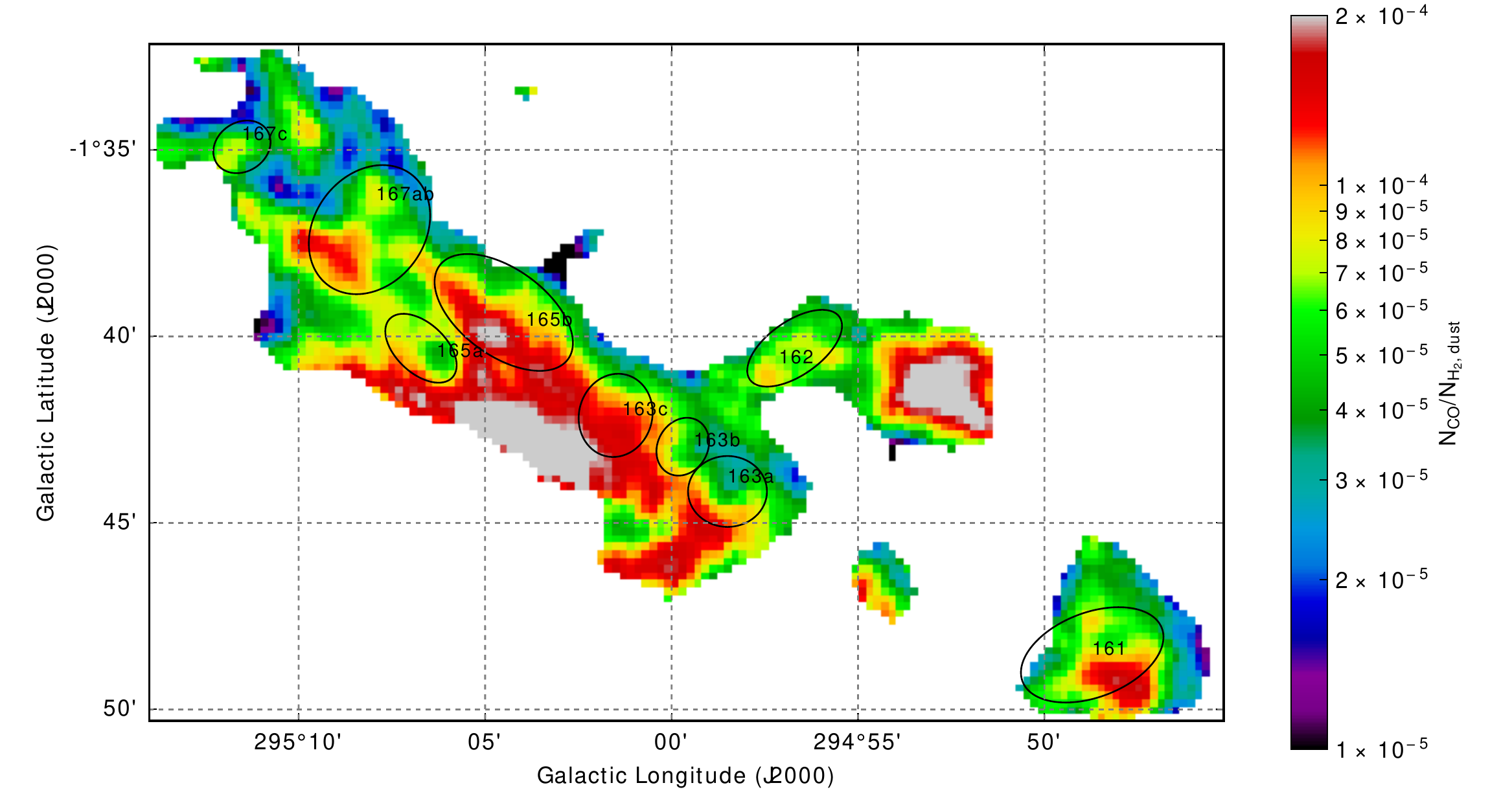}
    \caption{CO abundance map for Region 21}
    \label{fig:nrat21}
\end{figure}

\begin{figure}
    \centering
    \includegraphics[width=\textwidth]{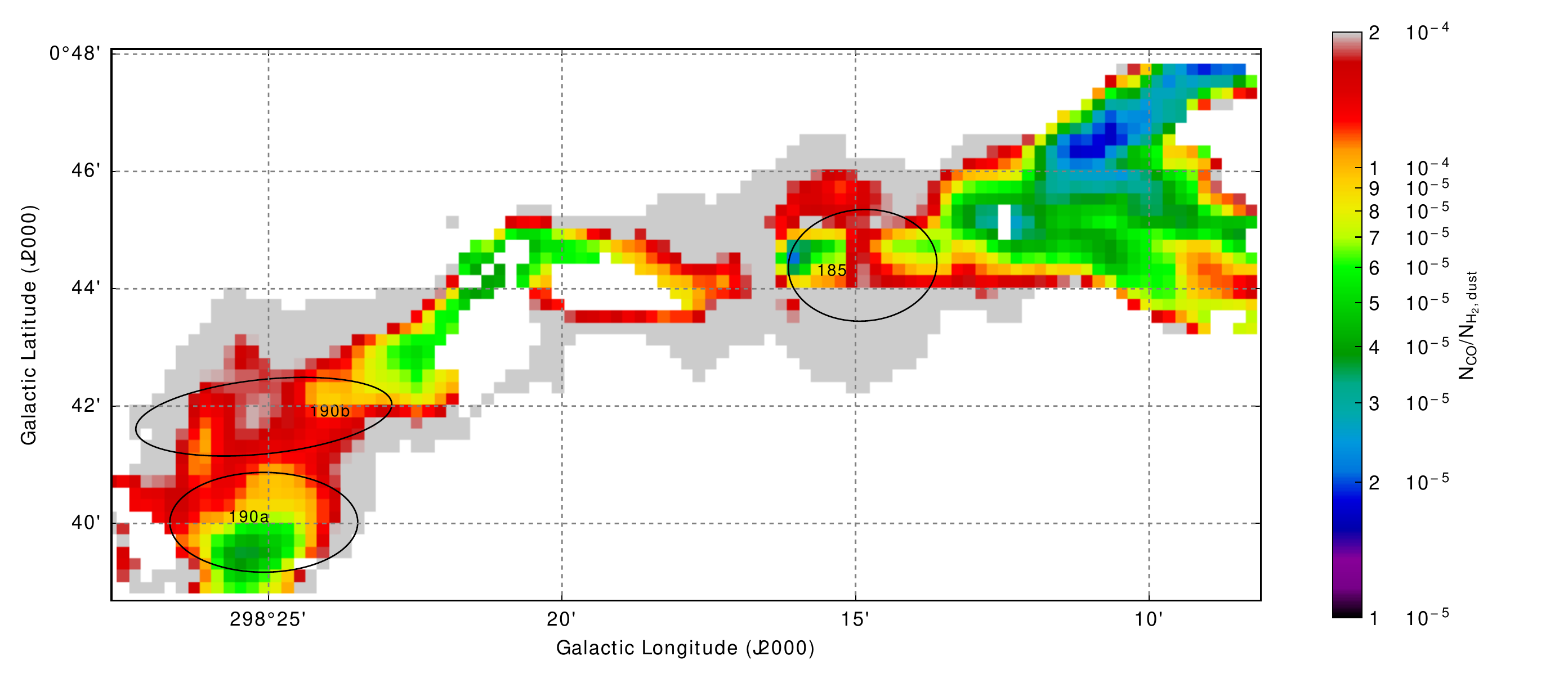}
    \caption{CO abundance map for Region 23, after employing the same method to model and subtract the background Dragonfish Nebula as was used in \protect\citealt{pittsmn} for Region 26. The \emph{Herschel}/PACS maps left a gap in coverage directly over BYF~185, so those data may not be trustworthy.}
    \label{fig:nrat23}
\end{figure}

\begin{figure}
    \centering
    \includegraphics{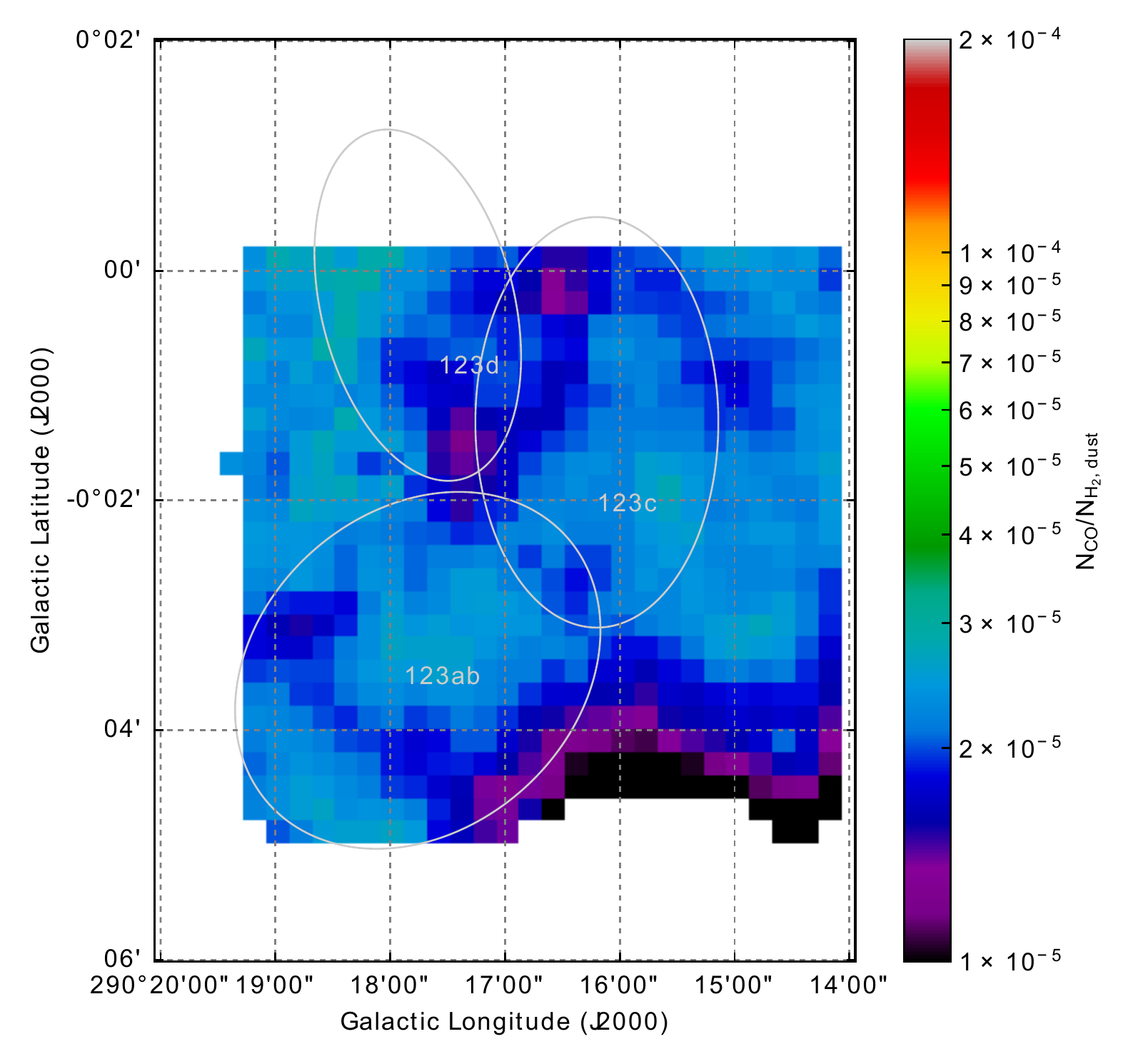}
    \caption{CO abundance map for the isolated clump BYF~123}
    \label{fig:nrat123}
\end{figure}




\end{document}